\documentclass[aps,pra,reprint,amsmath,amssymb,superscriptaddress,onecolumn,longbibliography, notitlepage, floatfix, nofootinbib]{revtex4-2}
\usepackage{mathtools}
\usepackage{siunitx}
\usepackage[hidelinks]{hyperref}
\usepackage{graphicx}
\usepackage{float}
\usepackage{subcaption}
\usepackage[braket, qm]{qcircuit}
\usepackage{xurl}
\usepackage{enumitem}

\newcommand{\im}{\mathrm{i}}

\usepackage{xcolor}

\begin{document}
\setcitestyle{super}
\title{{\fontsize{12.7pt}{18pt}\selectfont Arbitrary control over multimode wave propagation for machine learning}}

\author{
Tatsuhiro~Onodera$^{1,2,\ast}$, 
Martin~M.~Stein$^{1,2,\ast}$, 
Benjamin~A.~Ash$^1$, 
Mandar~M.~Sohoni$^1$, 
Melissa~Bosch$^1$, 
Ryotatsu~Yanagimoto$^{1,2}$, 
Marc~Jankowski$^{2,3}$, 
Timothy~P.~McKenna$^{2,3}$, 
Tianyu~Wang$^{1,\dagger}$, 
Gennady~Shvets$^1$, 
Maxim~R.~Shcherbakov$^{1,\ddagger}$, 
Logan~G.~Wright$^{1,2,\S}$, 
Peter~L.~McMahon$^{1,4}$ \\
\textit{
\newline
\small{$^1$School of Applied and Engineering Physics, Cornell University, Ithaca, NY, USA}\\
\small{$^2$NTT Physics and Informatics Laboratories, NTT Research, Inc., Sunnyvale, CA, USA}\\
\small{$^3$E.\,L. Ginzton Laboratory, Stanford University, Stanford, CA, USA}\\
\small{$^4$Kavli Institute at Cornell for Nanoscale Science, Cornell University, Ithaca, NY, USA}\\
\small{$\dagger$Present address: Department of Electrical and Computer Engineering, Boston University, Boston, MA, USA}\\
\small{$^\ddagger$Present address: Department of Electrical Engineering and Computer Science, University of California, Irvine, CA, USA}\\
\small{$^\S$Present address: Department of Applied Physics, Yale University, New Haven, CT, USA}\\
\small{$^\ast$These authors contributed equally.}\\
\small{Contact: ms3452@cornell.edu, pmcmahon@cornell.edu}
}}

\begin{abstract}
Controlled multimode wave propagation can enable more space-efficient photonic processors than architectures based on discrete components connected by single-mode waveguides. Instead of defining discrete elements, one can sculpt the continuous substrate of a photonic processor to perform computations through multimode interference in two dimensions. Here we design and demonstrate a device whose refractive index can be rapidly reprogrammed across space, allowing arbitrary control of wave propagation. The device, a two-dimensional programmable waveguide, uses parallel electro-optic modulation of the refractive index of a slab waveguide with about $10^4$ programmable spatial degrees of freedom. We implement neural-network inference on benchmark tasks with up to 49-dimensional vectors in a single pass, without digital pre- or post-processing. Theoretical and numerical analysis further indicates that two-dimensional programmable waveguides may offer not only a constant-factor reduction in device area, but also a scaling benefit, with the area required growing as $N^{1.5}$ rather than $N^2$.
\end{abstract}

\maketitle

\section*{Introduction}
Deep neural networks (DNNs) have gained widespread adoption across many domains ranging from computer vision to natural language processing \cite{lecun2015deep}. The size of DNN models has been increasing exponentially over the past decade, leading to exponentially increasing energy costs for running them. Limits to energy costs now impose a practical constraint on how large models can be \cite{patterson2021carbon}, strongly motivating the exploration of alternative, energy-efficient computing approaches for executing DNNs, whose computational cost is typically dominated by that of matrix-vector multiplications (MVMs). Optical neural networks (ONNs), particularly those integrated on photonic chips, that specialize in performing MVMs with optics instead of electronics are one promising candidate approach  \cite{shen2017natphoton, Tait2017neuromorphic, feldmann2019nature, feldmann2021nature, huang2021silicon,  ashtiani2022nature, bandyopadhyay2022single}. 

The dominant paradigm for designing integrated photonic neural networks is to construct networks of discrete, programmable photonic components---such as Mach--Zehnder interferometers, microring resonators, or phase-change-memory cells---connected by single-mode waveguides \cite{shastri2021natphoton}. 
These networks execute a linear-optical operation: an MVM between the vector encoded in the optical input to the chip and the matrix programmed in the discrete components. However, the maximum vector size, $N$, supported by chips using this approach has so far been restricted (Supplementary Table~1) to sizes far below what is necessary for optics to deliver an energy-efficiency advantage ($N \gtrsim 1000$) \cite{hamerly2019large,nahmias2019photonic,anderson2023optical, mcmahon2023natrevphysics}. 
The scale of such chips has been limited by at least two factors: 
First, the co-integration of optical and electrical components poses the challenge of routing $\mathcal{O}(N^2)$ electronic wires carrying the parameters of the matrix multiplication through the perimeter of the chip, often limiting the controllable degrees of freedom to a few hundred parameters.
Second, individual optical components are rather bulky due to the comparably large optical wavelength and the often weak programmability of optical materials. On top of that, substantial portions of the chip's area need to be dedicated to sprawling interconnection regions comprising well-isolated waveguides.

We could achieve far greater spatial efficiency if, instead of building the integrated photonic neural network from discrete components, we treated the entire chip as a blank slate whose refractive-index distribution, $n(x,z)$, we could arbitrarily and reprogrammably sculpt\cite{Hughes-fan2019SciAdv, khoram2019photonres, larocque2021opticsexp, Nakajima-Hashimoto-NTT2022neuralSE, nikkhah2024inverse, wu-feng2023natphoton}. 
Here lies the central challenge that our work tackles: 
How can we make a photonic chip whose refractive-index distribution is programmable, ideally in a way that avoids the integration complexity of introducing electronic wiring?
In conventional nanophotonic chips, $n(x,z)$ is controlled by etching away material in lithographically defined regions---and is fixed at fabrication time. While inverse-designed chips \cite{Molesky-vuckovic-rodriguez2018NatPhoton} realizing MVMs with fixed matrices can be made \cite{nikkhah2024inverse}, we generally would like to be able to program the matrix. Photorefractive crystals were explored several decades ago as a means to implement programmable linear operations with slab waveguides \cite{Psaltis1990holography,Brady1991holographic}, but fell out of favor because the small achievable refractive-index modulation ($10^{-4}$) meant that even centimeter-scale waveguides were unable to perform large-scale operations. Additionally, phase-change materials have recently been demonstrated to realize arbitrary refractive-index distributions \cite{delaney-muskens2021nonvolatile, wu-li2024freeform} but suffer from a limited number of rewrite cycles ($4000$ in ref.\,\cite{delaney-muskens2020family}) and high loss (greater than $\SI{2.8}{dB/mm}$ in ref.\,\cite{wu-li2024freeform, delaney-muskens2020family}). The scale is also currently limited: ref.\,\cite{wu-li2024freeform} reported programming a device with a 3-dimensional input and a 3-dimensional output.

\begin{figure*}[!t]
  \centering
  \includegraphics[width=\textwidth]{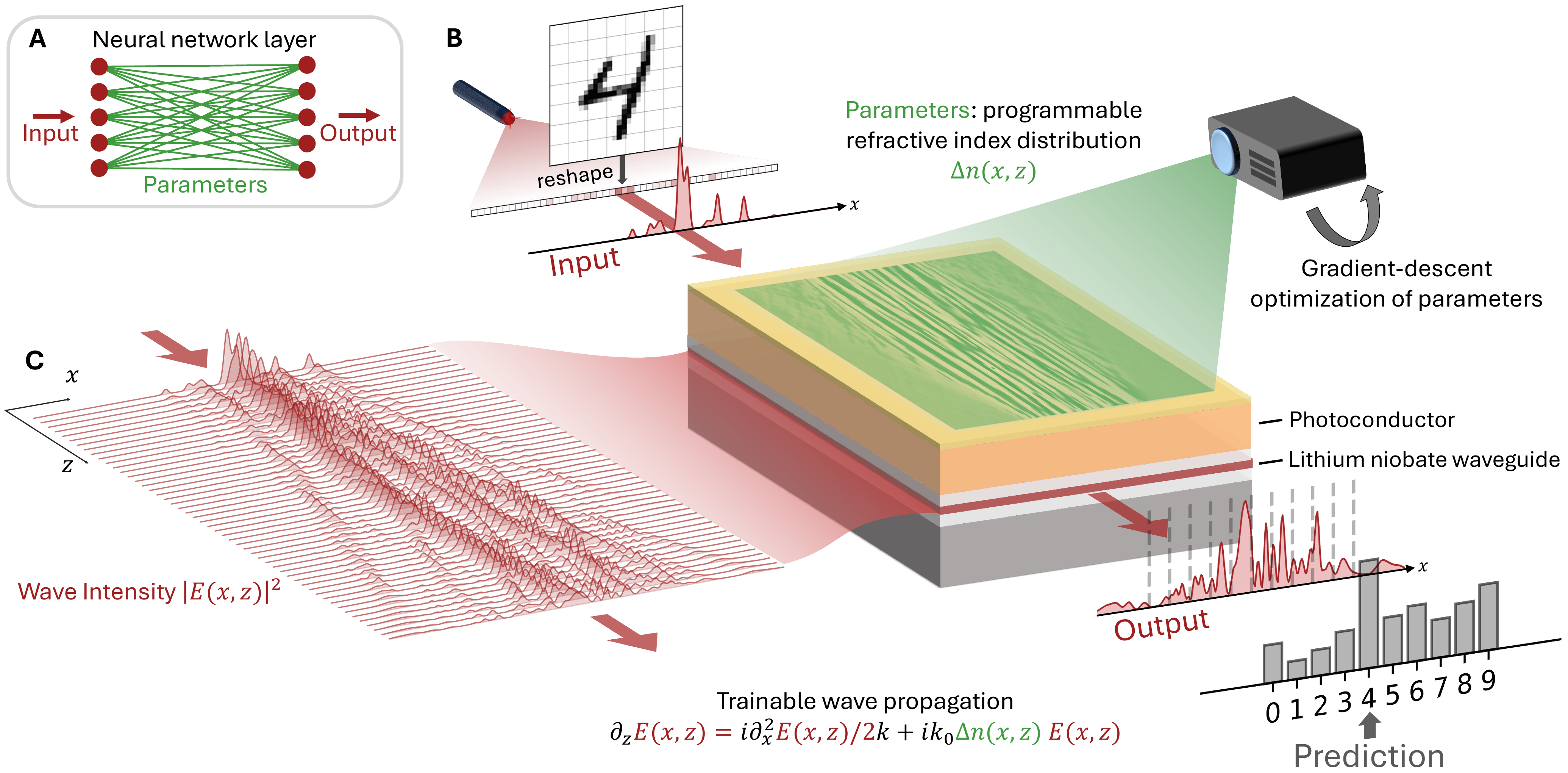}
  \captionsetup{justification=raggedright,singlelinecheck=false}
  \caption{
  \textbf{Machine learning with multimode wave propagation in the 2D-programmable waveguide.} 
  (\textbf{A}) The fundamental unit of an artificial neural network---a \textit{layer}---transforms an input vector into an output vector via a trainable matrix multiplication.
  (\textbf{B}) Analogous to a neural-network layer, the 2D-programmable waveguide linearly transforms an input optical field into an output optical field, via wave propagation through a lithium niobate slab waveguide whose two-dimensional refractive-index modulation $\Delta n(x, z)$ can be continuously and arbitrarily programmed (up to practical limits on resolution and the maximum modulation; see Supplementary Sections  1C and 1D). This refractive-index modulation, which is directly set by the illumination pattern that is projected onto the device (shown in green), is trained so that wave propagation through the waveguide performs machine-learning tasks (handwritten-digit classification shown here as an example). To determine the result of the classification, the output beam's intensity is measured across equally-sized bins; the bin with the highest total power corresponds to the predicted label. 
  (\textbf{C}) Simulated wave intensity in the slab waveguide, which shows that the neural-network computation is performed with complex multimode wave propagation.
  }    
  \label{fig1}
\end{figure*}

In this work, we introduce a photonic chip with a waveguide that is fully programmable in two dimensions: a \textit{2D-programmable waveguide}. The chip uses massively parallel electro-optic modulation to program $n(x,z)$ across  $\sim$10,000 individual regions of a lithium niobate slab waveguide, and we train multimode photonic structures within the chip that perform neural-network inference (Fig.~\ref{fig1}). The structures realized by our 2D-programmable waveguide are similar to inverse-designed nanophotonic devices \cite{Molesky-vuckovic-rodriguez2018NatPhoton,nikkhah2024inverse}: they are computer-optimized, two-dimensional metastructures that control multimode wave propagation. A distinguishing feature of our device is its programmability, setting it apart from typical inverse-designed photonic devices, which are fixed after manufacturing. We achieve programmability optically, decoupling the electronic wiring for programming from the photonic chip: a pattern of light shone on top of our device creates a spatially-varying refractive-index modulation $\Delta n(x,z)$ in the slab waveguide. This is achieved by using the principle of photoconductive gain \cite{chiou-wu2005nature, wu2011natphoton} to induce a refractive-index modulation via the strong electro-optic effect in lithium niobate. In our work we show how we can train the refractive-index distribution so that the complex wave propagation through the device performs a desired neural-network inference (Fig.~\ref{fig1}C). 

\section*{Operating principle of the device}

\begin{figure}[!t]
  \centering
  \includegraphics[width=\textwidth]{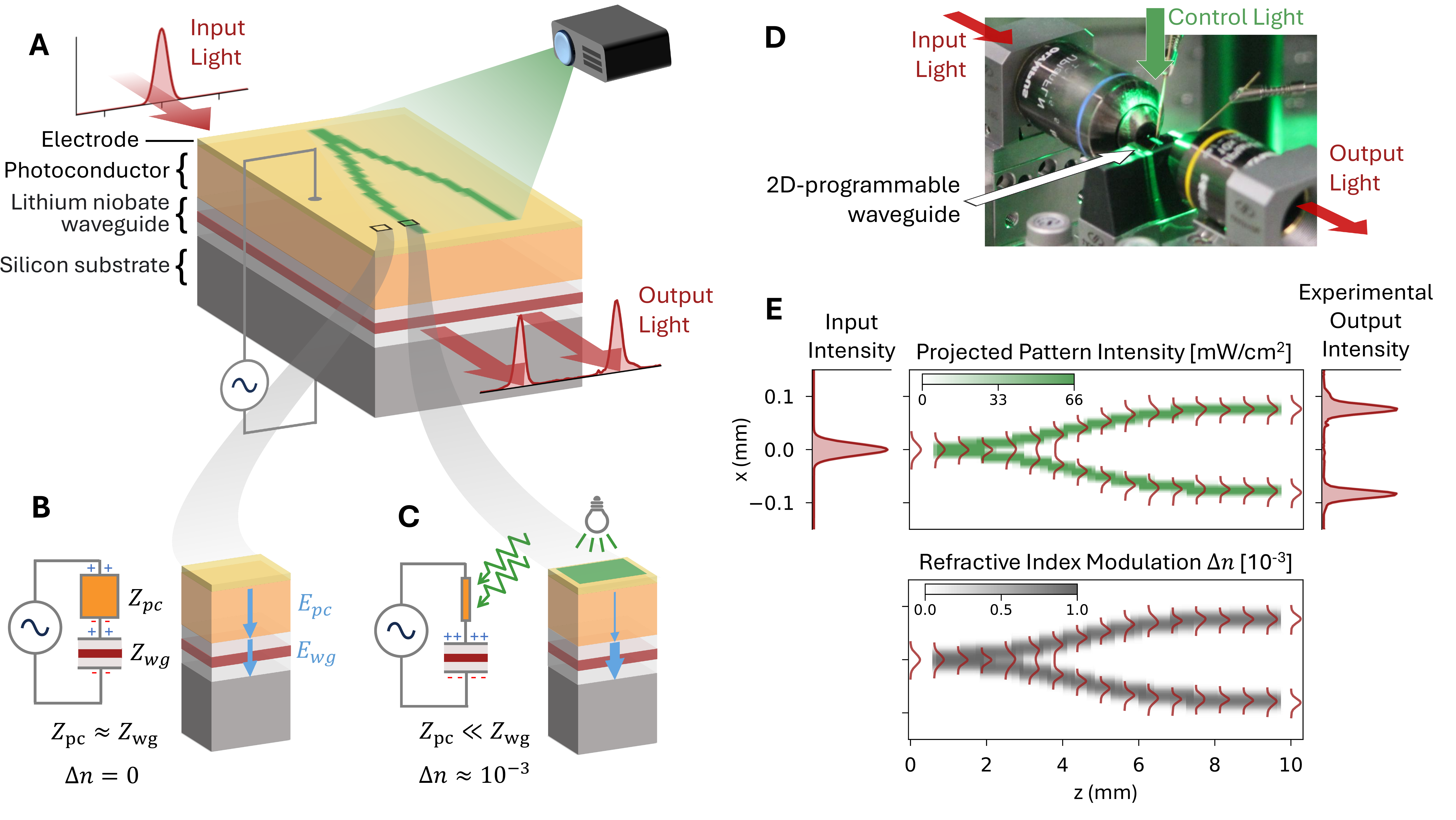}
  \captionsetup{justification=raggedright,singlelinecheck=false}
  \caption{
  \textbf{Operating principle of the 2D-programmable waveguide.}
  (\textbf{A}) The 2D-programmable waveguide consists of a nanophotonic stack of four layers: (1) A conductive silicon substrate that doubles as the ground electrode, (2) a Z-cut lithium niobate (in red) slab waveguide with silicon dioxide cladding (in white), (3) a photoconductive layer for optical control of the refractive index, and (4) a gold electrode. 
  (\textbf{B-C}) Electrical-circuit models of the 2D-programmable waveguide in regions with and without illumination. There is a voltage division between the photoconductor and the slab waveguide, with impedances $Z_{\text{pc}}$ and $Z_{\text{wg}}$, respectively. 
  (\textbf{C}) Upon illumination, the resistance of the photoconductor decreases, leading to an increase in the electric field (blue arrows) inside the waveguide, which induces a refractive-index modulation via the electro-optic effect.
  (\textbf{D}) A photograph of our prototype 2D-programmable waveguide in our experimental setup.
  (\textbf{E}) Experimental realization of a Y-branch splitter on the 2D-programmable waveguide, which splits the input light into two equal output beams. The projected pattern (in green) directly corresponds to the induced refractive-index modulation (in gray). A simulation of the wave propagating through the device is overlaid with the patterns (in red).}
  \label{fig2}
\end{figure}

In Fig.~1, we show a schematic for how to perform machine learning with the device. In machine learning, data is often encoded in vector-form, $\vec{x}_l$, and processed via programmable matrices $W$ and activation functions $f$ in ``layers'': $\vec{x}_{l+1} = f(W\vec{x}_l)$ (illustrated in Fig.~1A). Here, we amplitude-encode the machine-learning input data $\vec{x}_l$ in the 1D input optical field distribution (Fig.~1B), $E(x, z=0)$, which serves as the initial condition for programmable wave propagation described by the partial differential equation: 
\begin{align}
\label{eq:PDEwavepropagation}
\frac{\partial E(x, z)}{\partial z} = \frac{\im}{2k}\frac{\partial^2 E(x, z)}{\partial x^2} + \im k_0 \Delta n(x, z) E(x, z),
\end{align}
a solution of which is shown in Fig.~1C.
Here, $z$ denotes the propagation direction and $x$ the transverse direction, while $k_0$ and $k$ are the wavenumbers in free space and the slab waveguide, respectively. The refractive-index distribution of the slab waveguide, $n(x,z) = n_\text{0} + \Delta n(x,z)$, has two contributions: a spatially uniform part, $n_0$, that is the refractive index of the waveguide when no programming light is impinging on it, and a programmable part that is induced by electro-optic modulation, $\Delta n(x,z)$, loosely corresponding to the programmable matrix weights $W$.  After the optical field has propagated through the device, we measure its intensity, $I_\text{out}(x) \propto |E(x, z=L)|^2$, at the output facet and bin it to produce the machine-learning task's output vector, corresponding to $\vec{x}_{l+1}$.

In Fig.~2, we show how patterns of light program the refractive-index modulation $\Delta n(x,z)$.
First the device converts the projected pattern of light into a spatially-varying quasi-static electric-field. Second, the electro-optic waveguide changes its refractive-index $\Delta n(x,z)$ in response to this electric field. 
Our device is an electro-optic waveguide coated by a photoconductive film and sandwiched between a pair of electrodes with an oscillating bias voltage applied across them.
For regions of the chip that are illuminated at intensities of tens of $\SI{}{mW/cm^2}$, the photoconductor's impedance drops substantially, increasing the bias electric field within the slab waveguide.
Then the lithium niobate waveguide locally changes its refractive-index via the Pockels effect, due to the spatially varying bias electric field.
In our device, the largest refractive-index modulation is approximately $10^{-3}$, limited by the geometry of the material stack and a safety margin to prevent dielectric breakdown (see Supplementary Section 1C). We note that the refractive-index modulation in our device can take on continuous values by continuously varying the intensity of the projected pattern.

We projected illumination patterns across a $\SI{9}{mm} \times \SI{1}{mm}$ area, achieving a pixel resolution of $\SI{9}{\micro\meter} \times \SI{9}{\micro\meter}$ (Limits to the spatial resolution are discussed in Supplementary Section 1C).  This configuration enabled us to control the refractive-index distribution $n(x,z)$ with 10,000 degrees of freedom (see Supplementary Section 3A for our calculation of the parameter count) and update the entire distribution at a rate of $\SI{3}{Hz}$. To maximize the refractive-index modulation, we set the amplitude of the oscillating bias voltage across the electrodes to be up to $\SI{1000}{V}$. Given that CMOS electrode backplanes can only support spatially programmable voltages of around $\SI{10}{V}$, our approach of using photoconductive gain is crucial to achieving large electro-optic modulation: it allows us to apply a large voltage to a single unpatterned electrode, and realize controllable high voltages at \textit{virtual electrodes} via the patterned illumination \cite{chiou-wu2005nature, wu2011natphoton}. Due to the large impedance of the device, at the highest voltages less than \SI{1}{mW} electrical power is dissipated across the active area of the device.

To illustrate the operating principle of our device, we projected a pattern in the shape of a Y-branch splitter onto the 2D-programmable waveguide (Fig.~2A). The refractive-index modulation was approximately proportional to the projected pattern $\Delta n(x,z) \propto I(x, z)$ (Fig.~2E), up to spatial smoothing and a weak nonlinearity due to voltage division. We coupled a single input Gaussian beam into the device using a beamshaper and measured the intensity of the output light with a camera, as shown in Fig.~2E.  We note that since we use an oscillating bias voltage to drive the device, the induced refractive-index pattern also oscillates in time. Therefore, measurements of the output intensity (and optical inputs, if not continuous-wave) need to coincide with the driving voltage peaking. We discuss approaches to overcome this limitation and further details on the experimental setup in the Methods.

\section*{Machine-learning demonstrations with the 2D-programmable waveguide}

\begin{figure*}[t!]
  \centering
  \includegraphics[width=\textwidth]{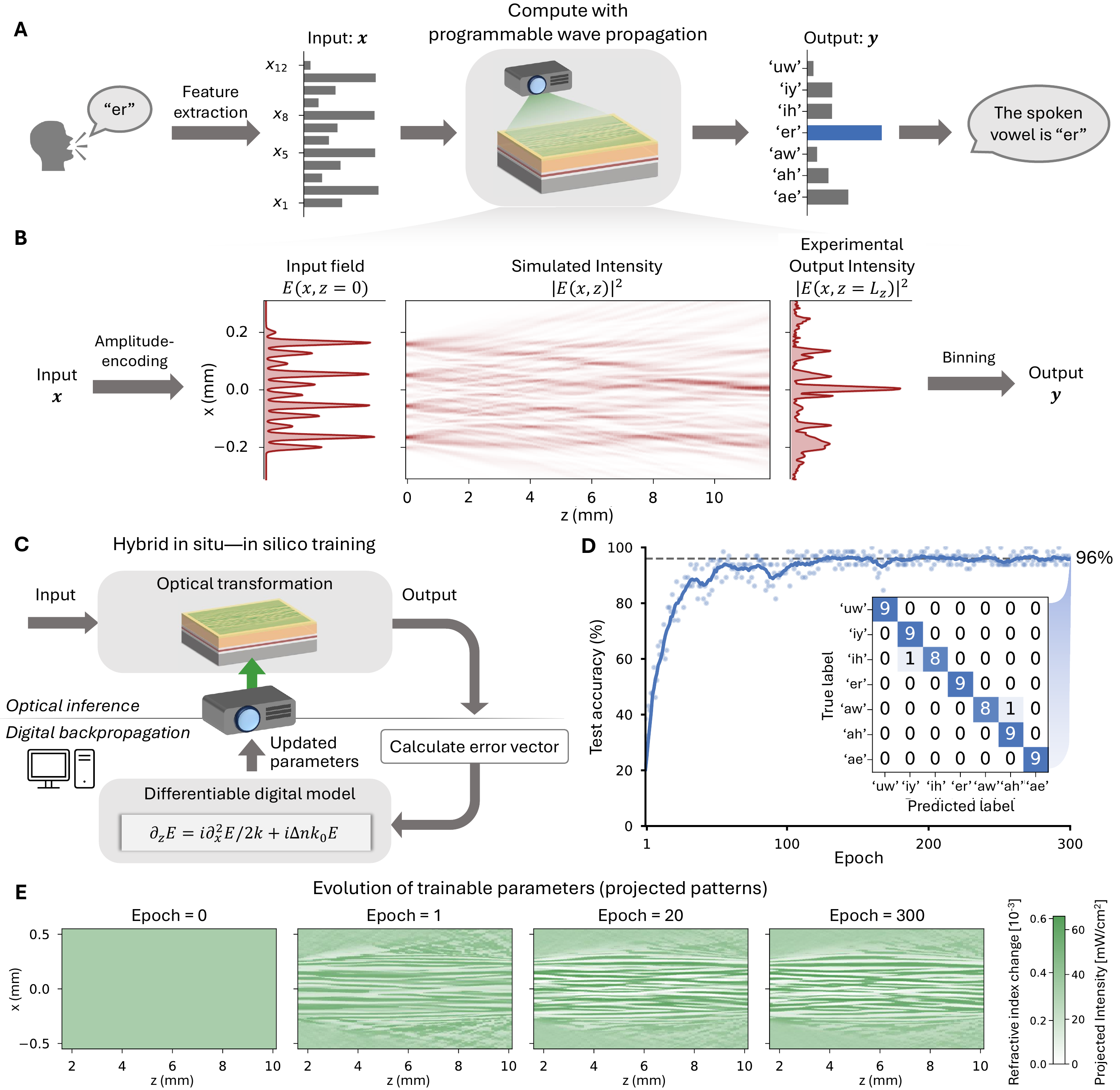}
  \captionsetup{justification=raggedright,singlelinecheck=false}
  \caption{
  \textbf{Vowel classification with the 2D-programmable waveguide.}
  (\textbf{A}) Overview of approach: The task involves predicting a spoken vowel, here ``er'', from a 12-dimensional input vector representing formant frequencies extracted from audio recordings. The 2D-programmable waveguide was trained to take in this input vector and output a 7-dimensional vector with a one-hot encoding format that indicates the predicted vowel.
  (\textbf{B}) Left: The input vector was amplitude-encoded into 12 Gaussian spatial modes to produce the initial optical field distribution. Center: Simulated wave propagation in the chip after training of the projected pattern. Right: The experimentally measured output intensity. It was binned, i.e., the total power within equally-spaced spatial bins was calculated to produce the 7-dimensional output vector.
  (\textbf{C}) Illustration of physics-aware training, a hybrid in-situ--in-silico backpropagation algorithm, which we used to train the parameters of the 2D-programmable waveguide. 
  (\textbf{D}) Test accuracy as a function of epoch. The inset shows the confusion matrix on the test dataset after training.
  (\textbf{E}) Evolution of the trainable parameters, the projected patterns, at different stages of training.\newline}
  \label{fig3}
\end{figure*}

We next applied the 2D-programmable waveguide to performing machine-learning tasks: vowel classification \cite{hillenbrand1995acoustic} and MNIST handwritten-digit classification \cite{lecun1998mnist}. Both tasks are used as benchmarks in studies of similar on-chip optical neural networks, providing useful points of comparison \cite{wu-feng2023natphoton, shen2017natphoton, feldmann2021nature}.

The vowel-classification dataset \cite{hillenbrand1995acoustic} comprises formant frequencies extracted from audio recordings of spoken vowels by various speakers. The task is to predict which of the 7 vowels is spoken, given a 12-dimensional input vector of formant frequencies. We divided the dataset into a training set and a test set, comprising 196 (75\% of the dataset) and 63 (25\%) samples, respectively. 

In Fig.~3, we present our experimental results on performing vowel classification with the 2D-programmable waveguide.  
As shown in Fig.~3A and 3B, we encoded the 12-dimensional input vectors in the amplitudes of 12 equidistant spatial Gaussian modes that were simultaneously projected onto the device's input facet. For readout, we measured the intensity at the output facet with a camera and binned the camera pixels into 7 different regions, with each region corresponding to a specific vowel. The predicted vowel for an input is given by the region that receives the most optical power. 
Thus, as shown in the simulated intensity distribution $|E(x,z)|^2$ in Fig.~3B, the device learned to use complex multimode wave propagation to direct most power toward the region corresponding to the correct vowel. 
For more details on the output decoding and the overall computational model of our ONN demonstrations, see Supplementary Section 5A.

The refractive-index distribution to implement vowel classification was learned using \textit{physics-aware training} \cite{wright-onodera-mcmahon2022nature}, a modified backpropagation algorithm (Fig.~3C). In this algorithm, the forward pass is performed by the experimental setup, while the backward pass is computed with a digital model of the experiment. The hybrid in-situ--in-silico nature of the algorithm allows for efficient training even in the presence of both imperfect models and experimental noise (Supplementary Section 3C). 
The digital model was challenging to construct due to the large number of parameters and the complexity of the wave propagation. Initially, a purely physics-based model (using Eq.~\ref{eq:PDEwavepropagation}) provided qualitative but not quantitative agreement with the experimental results. 
The remaining discrepancies were largely removed with data-driven refinements to the physics-based model (Supplementary Section 4). 

Using physics-aware training, we trained the 2D-programmable waveguide for a total of 300 epochs, which took approximately one hour on the experimental setup (see Fig.~3D). 
Fig.~3E shows the evolution of the initially uniform illumination pattern into a complex pattern that resembles the refractive-index distributions found in inverse-designed photonic devices.
Fig.~3D shows that despite the complexity of the projected pattern, the 2D-programmable waveguide successfully performed the vowel-classification task, and achieved a test accuracy of 96\% after training.

\begin{figure}[t!]
  \centering
  \includegraphics[width=\textwidth]{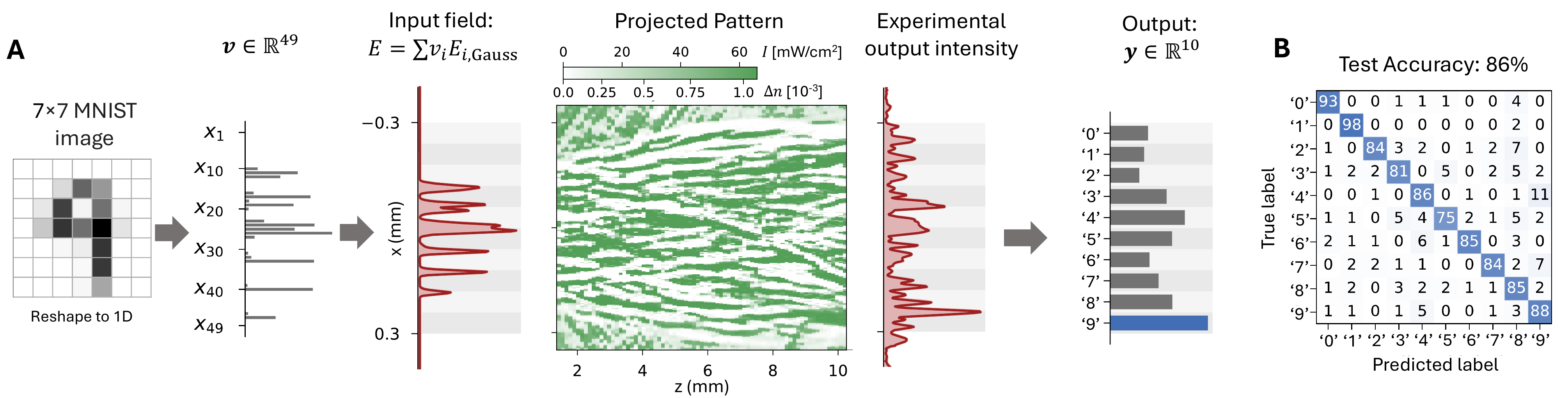}
  \captionsetup{justification=raggedright,singlelinecheck=false}
  \caption{
  \textbf{MNIST handwritten-digit classification with the 2D-programmable waveguide: neural-network inference with high-dimensional input vectors.}
  (\textbf{A}) We performed MNIST handwritten-digit classification with the 2D-programmable waveguide. Each image from the MNIST dataset was electronically downsampled and reshaped to a 49-dimensional vector. We trained the device to perform machine learning on this high-dimensional input vector with the same procedure as the vowel-classification task (see Fig.~3). (\textbf{B}) The confusion matrix, derived from evaluating on the test dataset of 10,000 images. After 10 epochs of training, the system achieved 86\% accuracy on the test dataset. As a baseline, a single-layer digital neural network with a $49\times10$ matrix achieves 90\% accuracy on this same task.
  }
  \label{fig4}
\end{figure}

In Fig.~4, we present our experimental results on MNIST handwritten-digit classification. The task consists of classifying 14-by-14-pixel images of handwritten digits from 0 to 9. We divided the MNIST dataset in the standard manner into 60,000 training images and 10,000 test images. We down-sampled each MNIST image to 7-by-7 pixels, then flattened it to a 49-dimensional input vector. 

To train the refractive-index distribution to perform MNIST classification, we followed the same procedure used for the vowel-classification task (Fig.~3): the 2D-programmable waveguide processed the 49-dimensional input vector to produce a 10-dimensional output vector that corresponds to the 10 possible digits (more details in Supplementary Section 5C). As shown in Fig.~4B, the system achieved 86\% accuracy on the test dataset after 10 epochs of training, which took about 10 hours on the experimental setup. This falls 4 percentage points short of the 90\% accuracy that a one-layer digital neural network achieves on this downsampled MNIST classification task, likely due to imperfect modeling and experimental drifts. Nevertheless, this result demonstrates that complex wave propagation in our device can be harnessed to perform computations comparable to that of a single-layer neural network with a $49\times10$ matrix of trainable parameters.

\section*{Length (and area) scaling for 2D-programmable waveguides}
\label{sec:scaling-argument}

\begin{figure*}[!t]
  \centering
  \includegraphics[width=\textwidth]{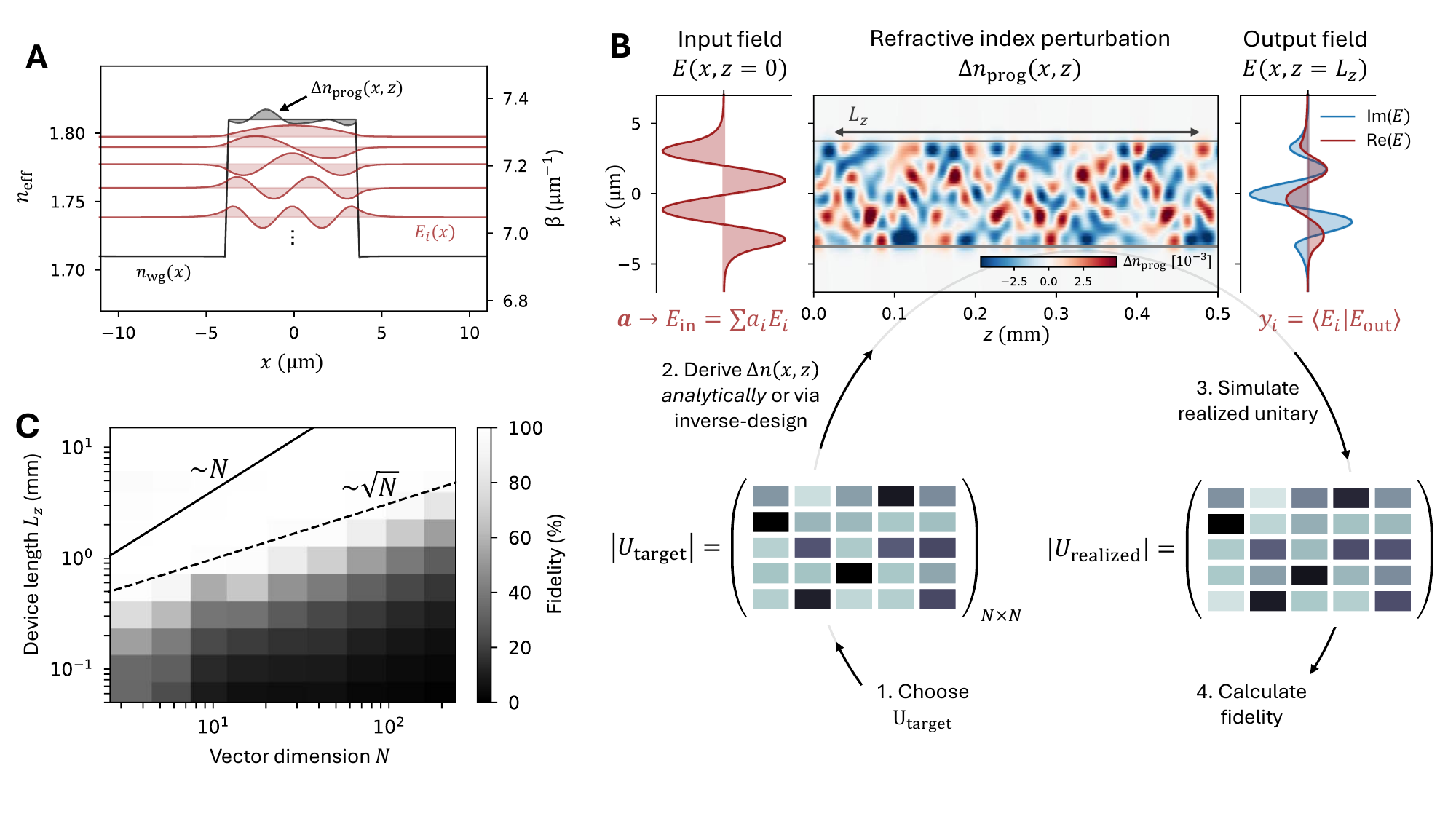}
  \captionsetup{justification=raggedright,singlelinecheck=false}
  \caption{
  \textbf{Simulation study showing the required length for devices realizing arbitrary photonic $N\times N$ unitaries scales as $\sqrt{N}$} 
  (\textbf{A}) We consider a weak refractive-index profile $\Delta n_\text{prog}(x,z)$ embedded into a step-index multimode waveguide with profile $n_\text{wg}(x)$. As light propagates through the perturbed waveguide, its $N$ eigenmodes interact with each other as described by coupled-mode theory. These interactions generate an $N\times N$ unitary matrix-vector multiplication (MVM). 
  (\textbf{B}) Vectors $\vec{a}$ are encoded in the modal amplitudes of the input electric field. Given a unitary matrix $U_\text{target}$, we determine the refractive-index perturbation $\Delta n_\text{prog}(x,z)$ that is embedded in the step-index waveguide analytically (Supplementary~Section~8B) or via in silico inverse design (Supplementary~Section~8C). We verify that after propagation through the multimode waveguide, the modal amplitudes have transformed according to $U_\text{realized}\approx U_\text{target}$.
  (\textbf{C}) Fidelities of $N\times N$ Haar-random unitary matrices realized in waveguides of various lengths $L_z$ with an inverse-designed $\Delta n_\text{prog}(x,z) < \Delta n_\text{max} = 5\cdot 10^{-3}$. The separation between unitaries realized with high versus low fidelity is well described by the line $L_z = \sqrt{N}\lambda_0/\Delta n_\text{max}$.
  }   
  \label{fig5}
\end{figure*}

We turn to a discussion of the theoretical size-scaling of different on-chip photonic processors.
Integrated photonics can be used to implement matrices of size $N \times N$ using circuits of width much wider than ${\sim}N \lambda_0$ ($N$ well-isolated waveguides) and length greater than ${\sim}N\lambda_0 / \Delta n_\text{prog}$ ($N$ $\pi$-phase shifters).\cite{Wu2024, clements2016optimal} 
Since the programmable refractive-index $\Delta n_\text{prog}$ is often very small, this limits circuits to be either very long or operate on low-dimensional inputs.
In the 2D-programmable-waveguide paradigm presented in this paper, the required device width scales as ${\sim}N\lambda_0$, imposed by the diffraction limit. 
One might intuitively expect that the length $L_z$ of 2D-programmable waveguides would also need to scale as ${\sim}N\lambda_0 / \Delta n_\text{prog}$---just as it does for the aforementioned circuit approaches for universal linear transformations on chip~\cite{Wu2024}. 
In this section, we present analytical and numerical results showing that, beyond the \emph{constant-factor} improvement in spatial footprint by avoiding single-mode waveguides, 2D-programmable waveguides may offer a different size-\emph{scaling}.

Our analytical argument, presented in detail in Supplementary Section 8A\&B, investigates the amount of phase-shift necessary to optically implement a given unitary transformation of dimension $N$ in a programmable multimode waveguide.
As the unitary dimension increases, the phase-shift per element, linked to the \textit{generator} of the unitary, becomes smaller (more precisely: for a given unitary one can always find a generator whose elements have a root-mean-square smaller than $\pi/\sqrt{N}$).
This suggests an exciting possibility: High-dimensional optical matrix-vector multipliers based on interference may require much less propagation distance through phase shifters than commonly assumed.
This property of unitary matrices has recently\cite{hamerly2024towards} been exploited to show that in \mbox{3-MZI} meshes, the average length of phase shifters can scale as $1/\sqrt{N}$---but universal MZI meshes by construction have a circuit depth that scales as $N$, independent of how short the individual phase-shifter elements are. However, in 2D-programmable waveguides this insight can potentially lead to large practical benefits because the devices effectively consist entirely of phase shifters, with nothing else constraining the total system length. The intuitive core of the mathematical argument is that if one needed a full $\pi$ phase-shift per phase-shifter, the required device length would scale as $N\lambda_0/\Delta n_\text{prog}$, but since the required phase shift goes as $\pi/\sqrt{N}$, the total required length scales as $N\lambda_0/(\sqrt{N}\Delta n_\text{prog})=\sqrt{N}\lambda_0/\Delta n_\text{prog}$.

We analytically find a refractive index distribution, $\Delta n_\text{prog}(x,z)$, whose magnitude (or equivalently: length) only scales as $\sqrt{N}$ (as measured by the root-mean-square of the distribution). Our construction is similar to the one presented by Larocque\cite{larocque2021opticsexp} in that it couples the modes of a multimode waveguide, but crucially differs by implementing parallel global couplings rather than a sequence of pairwise mode couplings, thereby achieving better scaling.

Our analytical argument relies on strong approximations: we used coupled-mode theory under a rotating wave approximation to calculate the propagation of unidirectional, scalar waves in a perturbed multimode waveguide. 
To validate our theory, we present results from numerical simulations that make far fewer assumptions: 
we simulated unidirectional, scalar waves in a perturbed multimode waveguide (Supplementary~Section~8D). 
Our analytical argument also only shows that the \textit{root-mean-square} of the refractive-index distribution scales as ${\sim}\sqrt{N}$. However, in practical devices, it is usually the \textit{maximum} value of the refractive-index change that is limited. Our simulations show that even with a strictly imposed maximum value of the programmable refractive-index strength, unitaries can be implemented accurately over a propagation distance that scales as ${\sim}\sqrt{N}$. 
We emphasize that our analytical argument permits completely general dense unitary matrices. 

This length-scaling is surprising in part because fundamental geometric considerations in linear optical devices suggest that the length required to perform arbitrary operations on $N$ modes is proportional to $N\lambda_0$.\cite{Miller2023, Li2022WadeHsu} 
Our theory, showing that $L_z{\sim}\sqrt{N}\lambda_0/\Delta n_\text{prog}$ is sufficient to perform universal optical operations on $N$ modes, implies a different length-scaling but does not contradict the bound by Miller\cite{Miller2023} unless $N$ is very large, often $\gg1000$ using common values (see Supplementary Section 8D).
Therefore, our result suggests that there is a large practical regime in which the better-than-linear length-scaling can be exploited to create universal optical processors with modest programmability or short propagation distances.

However, we should also note that the length-scaling argument does not directly apply to the device we used in the reported experiments, since our scaling result assumes guided modes in the $x$-direction, while our experiments were performed in a slab waveguide wide enough that light was effectively unguided.
It is nonetheless the case that one could---as far as we can tell without having explicitly performed these experiments---realize devices and experiments that do satisfy the assumptions of the theory (see discussion in Supplementary Section 8D).
For example, we simulated a multimode waveguide with a modest width of \SI{0.3}{mm}, length of \SI{5}{mm}, a refractive-index programmability no larger than ${\Delta n_\text{prog}=5\cdot 10^{-3}}$ with \SI{500}{nm} resolution, and showed that such a waveguide can realize a $200 \times 200$-dimensional unitary with high fidelity (see Fig.~\ref{fig5} and Supplementary Section 8D). A device of width $\SI{1.5}{mm}$ and length $\SI{11}{mm}$ with the same programmable refractive-index magnitude and resolution should be able to realize arbitrary unitaries with dimension as high as $1000\times 1000$.

\section*{Discussion and Outlook}
We have introduced and demonstrated a 2D-programmable photonic processor comprising a lithium niobate slab waveguide whose refractive-index distribution, $n(x,z)$, can be continuously programmed. 
The device design enables programming by parallel electro-optic modulation with approximately 10,000 degrees of freedom. We used our chip to perform neural-network inference by training the refractive-index distribution and consequently the multimode wave propagation through the chip. 
To train the device, we developed a physics-based model of the chip's behavior, along with a data-driven refinement, allowing the model to be sufficiently accurate that it supports backpropagation-based training \cite{wright-onodera-mcmahon2022nature}.

The predominant approach to building integrated photonic neural networks is to fabricate large arrays of discrete components connected by single-mode waveguides \cite{shastri2021natphoton}. In contrast, we adopted the conceptual approach of using wave propagation in distributed spatial modes \cite{Brady1991holographic, Hughes-fan2019SciAdv, khoram2019photonres, larocque2021opticsexp, Nakajima-Hashimoto-NTT2022neuralSE, gu-pan2023m3icro, nikkhah2024inverse}, and experimentally validated the theoretical predictions\cite{larocque2021opticsexp, khoram2019photonres, gu-pan2023m3icro} that this approach will be more space-efficient. Our prototype chip was able to perform neural-network inference with input vectors of dimension up to 49, which is larger than the capability of the neural-network photonic chips reported in Refs.\,~\cite{shen2017natphoton,bandyopadhyay2022single,ashtiani2022nature,feldmann2021nature,huang2020demonstration, wu-feng2023natphoton, Zhang-Liu2021NatureComms, pai2023experimentally}, and more space efficient than any of these chips based on networks of discrete components (see Supplementary Section 6 for a detailed comparison). This large input dimension enabled us to use our chip to perform MNIST handwritten-digit classification with a single pass through the chip, and without using any digital-electronic parameters.

One of the most promising applications of our approach is in reducing the energy cost of neural network inference, which remains (and is likely to remain) dominated by linear matrix--vector multiplication. There exists a break-even point beyond which optical devices could significantly outperform electronic hardware in this task, owing to the more favorable energy scaling with dimension $N$ (optics: $E\sim N$; electronics: $E\sim N^2$)\cite{hamerly2019large,nahmias2019photonic,anderson2023optical, mcmahon2023natrevphysics}. Reaching this regime requires high-dimensional operations: at low $N$, the overhead cost from analog-to-digital conversion and optical--electronic transduction outweighs the benefits, with estimates placing the break-even point around $N = 1000$, far beyond what is currently possible on a single chip. 
We derived a theoretical scaling law (Supplementary Section 8) describing how the dimension of possible matrix--vector multiplications in a 2D-programmable waveguide scales with the device dimensions and the refractive-index change. 
Surprisingly, the device length only needs to scale as $\sqrt{N}$, better than the most common approaches of designing photonic circuits~\cite{clements2016optimal, Wu2024}.
This result may enable all-optical matrix--vector multipliers with dimension exceeding the break-even point of energy efficiency. The development of such devices would make hybrid neural network architectures, in which analog optics performs the linear operations and electronic circuits implement the nonlinearities, energy-competitive, changing the energy-scaling of neural network inference.

To conclude, we believe that our device concept, with its ability to programmably control multimode wave propagation, may create new opportunities in the broader fields of optical computing and optical information processing \cite{wetzstein2020nature, shastri2021natphoton, mcmahon2023natrevphysics}. Although our work in this paper has focused on machine learning, our device could also be used to solve integral equations \cite{mohammadi2019science} and combinatorial-optimization problems \cite{RoquesCarmes2020heuristic}. More broadly, our chip is essentially an arbitrary (passive) photonic device that can be reconfigured on demand: any photonic device that can be specified as an inhomogeneous refractive-index distribution can be realized. Such devices can even be learned directly---effectively by performing inverse design \cite{Molesky-vuckovic-rodriguez2018NatPhoton}, but in situ in real time. Our concept will potentially enable the development of reprogrammable photonic simulators supporting novel studies of bound states in the continuum \cite{hsu2016bound} and topological photonics \cite{price2022roadmap}, as well as applications in engineering. It may ultimately even be possible to make a device that combines programmable linear wave propagation (this work), programmable nonlinear wave propagation\cite{Hughes-fan2019SciAdv, Marcucci-Conti2020PhysRevLetters, Tegin2021scalable,Nakajima-Hashimoto-NTT2022neuralSE} (a natural extension of this work to having programmable $\chi^{(2)}(x, z)$\cite{yanagimoto2025programmableonchipnonlinearphotonics}), and programmable gain/loss (demonstrated in Ref.\,~\cite{wu-feng2023natphoton}), giving rise to a reconfigurable on-chip platform capable of realizing almost every functionality we have in free-space optics.

\section*{Acknowledgements}
We gratefully acknowledge the Air Force Office of Scientific Research for funding under Award Number FA9550-22-1-0378, and the National Science Foundation for funding under Award Number CCF-1918549. We thank NTT Research for their financial and technical support. This work was performed in part at the Cornell NanoScale Facility, a member of the National Nanotechnology Coordinated Infrastructure (NNCI), which is supported by the National Science Foundation (Grant NNCI-2025233). P.L.M. acknowledges financial support from a David and Lucile Packard Foundation Fellowship. We acknowledge helpful discussions with 
Chris Alpha, 
Nicholas Bender,
Jeremy Clark, 
Anthony D'Addario, 
Noah Flemens, 
John Grazul,
Ryan Hamerly,
David Heydari,
Phil Infante, 
Oscar Jaramillo,
Vladimir Kremenetski,
Mario Krenn,
Kangmei Li,
George McMurdy, 
Roberto Panepucci, 
Carl Poitras, 
Sridhar Prabhu,
Aaron Windsor,
Fan Wu,
and Yiqi Zhao.

\section*{Author contributions}
T.O., L.G.W. and P.L.M. conceived the project.
M.M.Stein, T.O.,  L.G.W. and P.L.M. designed the devices and experiments.
T.O., M.M.Stein, B.A.A., and R.Y. performed the device fabrication with aid and recipe development from M.R.S., M.B., M.J., and T.P.M.. G.S. supervised M.R.S. and M.B..
M.M.Stein, T.O., M.M.Sohoni and T.W. designed and built the imaging setup to program the refractive-index patterns.
T.O., M.M.Stein built the high-voltage and beamshaper setups, performed the experiments, and analyzed the results.
T.O. derived the scaling law.
M.M.Stein, T.O., L.G.W. and P.L.M. wrote the manuscript with input from all authors.
P.L.M. supervised the project.

\section*{Competing interests}
T.O., M.M.Stein, M.R.S., L.G.W., and P.L.M. are listed as inventors on a patent application \\(WO2023220401A1) on the 2D-programmable waveguide. The other authors declare no competing interests.

\clearpage

\captionsetup[figure]{name=Extended Data Fig.}
\setcounter{figure}{0} 

\section*{Methods}
\label{methods}
\subsection{Device fabrication and design}

\begin{figure*}[!htbp]
  \centering
  \includegraphics[width=\textwidth]{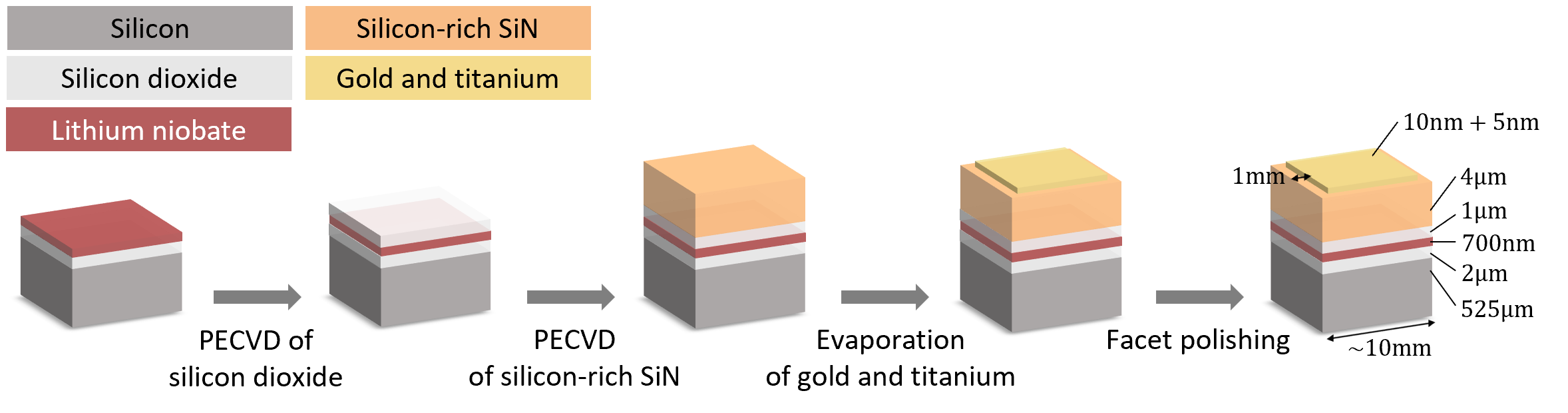}
  \captionsetup{justification=raggedright,singlelinecheck=false}
  \caption{\textbf{Fabrication process and device geometry of the 2D-programmable waveguide.}
  Devices were fabricated from thin-film lithium niobate on insulator wafers consisting of a heavily doped p-type silicon substrate with a resistivity of \SIrange{0.01}{0.05}{\ohm\centi\metre}, a \SI{2}{\micro\metre} PECVD silicon dioxide layer, and \SI{700}{\nano\metre} Z-cut MgO-doped lithium niobate bonded by ion-cutting. After dicing, we deposited \SI{1}{\micro\metre} of SiO\textsubscript{2} cladding and \SI{4}{\micro\metre} of silicon-rich SiN (photoconductive layer) by PECVD with alternating plasma pulses to minimize stress. Electrodes were formed by evaporating \SI{10}{\nano\metre} Ti and \SI{5}{\nano\metre} Au, with tape masking to prevent breakdown at chip edges. Waveguide facets were polished using progressively finer abrasives to reduce coupling loss. The final device has facet dimensions of approximately \SI{10}{\milli\metre}~\texttimes~\SI{10}{\milli\metre}.}

  \label{fig:fabrication}
\end{figure*}

As shown in Extended Data Fig. 1, we started our fabrication processes from a thin-film lithium niobate wafer purchased from NanoLN. It was a p-type silicon wafer with a substrate conductivity of 0.01-0.05$\,$$\si{\ohm\cm}$, \SI{2}{\micro\meter} of silicon dioxide deposited via plasma-enhanced chemical vapor deposition (PECVD), and \SI{700}{nm} of Z-cut $\mathrm{MgO}$-doped lithium niobate that is wafer-bonded with the ion-cut technique. We diced small pieces from the wafer using a Disco Dicing Saw for further processing. We deposited an additional \SI{1}{\micro\meter} of silicon dioxide via PECVD as a cladding, followed by another deposition of \SI{4}{\micro\meter} of silicon-rich silicon nitride (SRN), which is the photoconductive layer, via PECVD. The SRN layer was deposited in an Oxford Plasmalab 100 by flowing \SI{40}{sccm} of $\mathrm{SiH_4}$, \SI{10}{sccm} of $\mathrm{N_2O}$, and \SI{1425}{sccm} of $\mathrm{N_2}$ into the deposition chamber at a temperature of \SI{350}{\celsius} and a pressure of \SI{1900}{mTorr}. We alternated pulses of high and low frequency power during deposition to minimize film stress, with \SI{160}{W}, \SI{12}{s} low frequency pulses and \SI{200}{W}, \SI{8}{s} high frequency pulses.

Next, we evaporated electrodes onto the chip using a CVC SC4500 E-gun Evaporation System. We first evaporated \SI{10}{nm} of titanium as an adhesion layer, then \SI{5}{nm} of gold. To prevent dielectric breakdown between the top electrode and the conductive substrate through air at the edges of the chip, we covered the perimeter of the chip with tape before evaporation. The tape acted as a mask, preventing deposition closer than around \SI{1}{mm} to the edges, thereby increasing the path length between the top electrode and substrate through air. To minimize coupling losses into the waveguide, we used an Allied Multiprep Polisher to polish the waveguide facets. We polished using silicon carbide paper of successively finer grain size, starting at \SI{3}{\micro\meter}, then moving to \SI{1}{\micro\meter}, and \SI{0.5}{\micro\meter} roughness.

\subsection*{Device design and characterization}
We used Z-cut lithium niobate for the slab waveguide, so that the crystal axis of the lithium niobate is parallel to the strong electric field that is induced by the bias voltage, which points out of the waveguide plane (in the $y$ direction). As shown in Supplementary Fig.~1, we used the transverse magnetic (TM) mode of the slab waveguide, whose optical electric field is also oriented in the $y$ direction. Since the $r_{33}$ electro-optic coefficient is largest in lithium niobate, this configuration maximized the strength of the electro-optic modulation.

The thickness of the lithium niobate layer is chosen for single-mode operation (see Supplementary Section 1A). To maximize the refractive-index modulation, it is beneficial to have a thicker photoconductor and a thinner silicon dioxide cladding (see Supplementary Section 1B). The silicon dioxide cladding is chosen to be sufficiently thick to ensure low propagation loss. Thus, we balanced these tradeoffs to arrive at the device geometry shown in Extended Data Fig.~1. The device has a propagation loss of less than $\SI{1}{dB/cm}$ at a wavelength of $\SI{1550}{nm}$ (see Supplementary Section 1E).

In order to maximize the refractive-index contrast of the slab waveguide between the bright (illuminated) and dark regions, it is important to design the photoconductor to have high dark resistance and low bright resistance. Thus, to optimize the material properties of the photoconductor, we swept the silicon to nitrogen ratio in the silicon-rich silicon nitride photoconductor (by varying the amount of $\mathrm{SiH_4}$ gas we flowed into the PECVD) and chose the material with the largest photoconductive contrast. We characterized the refractive-index modulation as a function of the intensity of the projected pattern with an off-axis holography setup (see Supplementary Section 1C). The maximum refractive-index modulation that we achieved in this work is approximately $10^{-3}$. We show in Supplementary Fig.~3 that this can be increased to beyond $4\times 10^{-3}$ by using a photoconductor layer that is twice as thick (\SI{8}{\micro\meter}) and by further optimizing the photoconductive properties. In Supplementary Section 7B, we also discuss how the refractive-index modulation can be further increased by switching to a different material for the waveguide core.

\subsection{Experimental setup}

\begin{figure*}[!t]
  \centering
  \includegraphics[width=\textwidth]{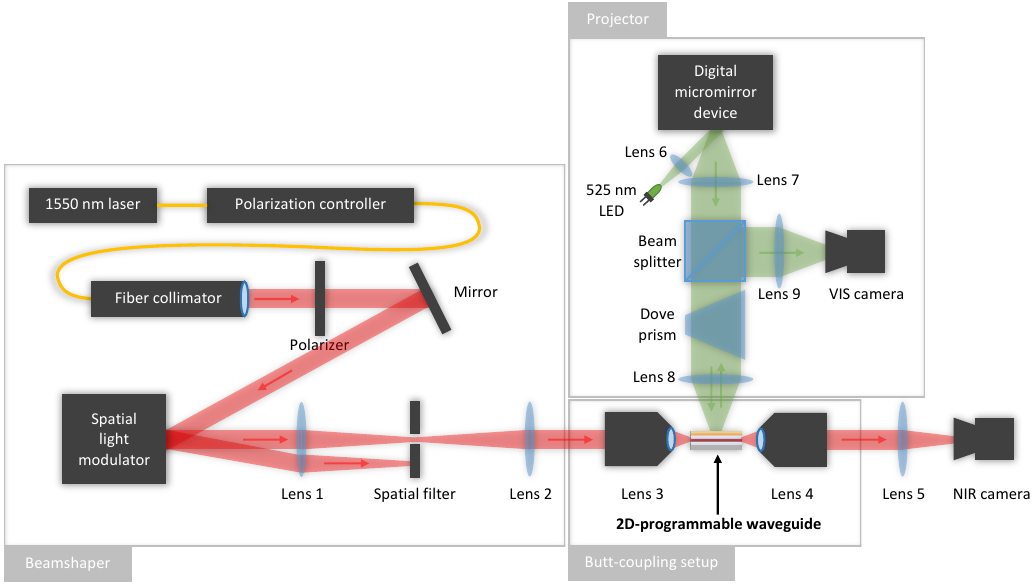}
  \captionsetup{justification=raggedright,singlelinecheck=false}
  \caption{\textbf{Schematic of the experimental setup.} 
  The setup consists of five key units: (1) a beamshaper to generate spatially varying one-dimensional optical inputs to the 2D-programmable waveguide, (2) a digital micromirror projector to impose a programmable illumination pattern controlling the refractive-index distribution, (3) butt-coupling optics to couple light in and out of the device, (4) (not shown) a high-voltage source applying oscillating fields across the electrode and conductive silicon wafer, and (5) an infrared camera to record the output intensity profile. The beamshaper provides full amplitude and phase control of input fields, while the projector enables spatial refractive-index modulation with continuous pixel-values via pulse-width modulation.}

  \label{fig:sketch-setup}
\end{figure*}

As shown in Extended Data Fig. 2, the experimental setup can be roughly divided into five units: 1) An optical beamshaper to create spatially-varying one-dimensional electric field inputs for the 2D-programmable waveguide, 2) a projector to create a programmable illumination pattern that controls the refractive-index distribution inside the waveguide, 3) a butt-coupling setup to couple light in and out of the 2D-programmable waveguide, 4) a high-voltage source to apply an oscillating bias voltage across the electrodes of the 2D-programmable waveguide, and 5) a camera to measure the intensity of the output beam. We note that the experimental setup relies on more free-space optical components than usual for an integrated photonics experiment. This is a direct consequence of our decision to keep the fabrication of the device simple, without lithographically defined structures for this initial proof-of-concept demonstration. We envision that a more compact, fully integrated version of the 2D-programmable waveguide could be built by integrating on-chip lithium niobate modulators and detectors, and a micro-LED display (see Supplementary Section 7D).

In this section, we provide an overview of the key components and functionalities of the experimental setup. For a more detailed description, including photographs of the optical setup and specifics on the components such as part numbers and manufacturers, see Supplementary Section 2.

The free-space beamshaper allows for the realization of arbitrary input optical fields $E(x, z=0)$, up to a spatial resolution of $\SI{2}{\micro\meter}$ and over a distance of $\SI{600}{\micro \meter}$. In this experiment, we used the beamshaper to create both simple input fields, such as a single Gaussian beam for the Y-branch splitter demonstration, and more complex input fields for the machine-learning demonstrations. This flexibility also enabled us to freely vary the encoding of input vectors into the optical field. For instance, we varied the width of the input modes and adjusted their spacing, which is tailored to each machine learning task. Finally, because the beamshaper is capable of shaping both the amplitude and phase of the input field, it was also used to calibrate the 2D-programmable waveguide (see Supplementary Section 4B).

The design of the beamshaper we built closely follows ref.\,\cite{bender-cao2022natphys}, which also programmably shapes the input light that is coupled into slab waveguides. The core working principle of the beamshaper is to create spatially varying phase-gratings on a 2D-phase spatial light modulator \cite{frumker-silberberg2007beamshaper} (SLM, Meadowlark Optics UHSP1K-850-1650-PC8). We varied the amplitude and relative positions of these phase-gratings to control the input optical field $E(x, z=0)$. A lens after the SLM performed a Fourier transform that separates the diffraction maximums of the phase-gratings, and a spatial filter selected the first-order diffraction maximum. Finally, a 4f relay system (comprising lens 2 and lens 3 as shown in Extended Data Fig.~2) demagnified the optical field at the focal plane of the first lens and coupled the light into the 2D-programmable waveguide. Due to the response time of the liquid crystal in the SLM, the beamshaper's fastest update speed is approximately 50~Hz.

The projector setup was designed to create a high-resolution programmable illumination pattern over a large field of view. We used a digital micromirror device (DMD, Vialux V-7000) with a resolution of 1024$\times$768 pixels and a pixel pitch of 13.7 \textmu m. The DMD was illuminated with green light (525 nm) from an LED. We imaged the surface of the DMD onto the surface of the 2D-programmable waveguide via a 4f setup consisting of two tube lenses. The focal length of the tube lenses was chosen to demagnify the image of the DMD by a factor of 1.5, so the projected pattern on the surface of the 2D-programmable waveguide has dimensions of $\SI{9.1}{mm} \times \SI{6.8}{mm}$, with each individual pixel of the projected pattern measuring $\SI{9}{\micro m} \times \SI{9}{\micro m}$. Because the complex wave propagation spans a distance of $\SI{1}{mm}$ in the $x$ direction, in practice, we only use a $\SI{9.1}{mm} \times \SI{1}{mm}$ region of the projected pattern to control the wave propagation in the 2D-programmable waveguide. Finally, although the DMD provides only binary modulation, we achieve continuous refractive-index modulation by pulse-width modulating the illumination pattern. This is feasible because the DMD can be switched on and off at a rate of $\SI{20}{kHz}$, much faster than the RC time constant of the device, which is about $\SI{10}{Hz}$.

In order to maximize the electro-optic effect in lithium niobate, we used high voltages of about \SI{1}{kV}. We created sinusoidal voltages with an arbitrary function generator and amplified the voltage with a Trek 2220 high voltage amplifier, which has a voltage gain of 200$\times$ and is capable of outputting voltages of up to \SI{2}{kV}. We electrically contacted the device using high-voltage-rated probe arms with BeCu probe tips: one probe tip was put in contact with the gold electrode on top of the device, while a grounded probe tip touched the silicon substrate (see Supplementary Fig.~S9).

We used an AC frequency of $\SI{10}{Hz}$ for the experiments shown in Fig.~2, and $\SI{26}{Hz}$ for the experiments shown in Fig.~3 and 4. Because AC voltage is applied, the desired refractive-index distribution is realized at the peak of the sinusoidal modulation. Thus, we trigger the camera to this peak, which explains the use of a higher frequency for the machine learning demonstrations---to maximize the update rate of the experiment. Finally, we opted for AC over DC operation because the resistivity of the silicon dioxide cladding exceeds the dark resistance of the photoconductor; under DC operation, the photoconductor would not modulate the circuit's overall impedance. Future modifications can enable DC operation, such as using an alternative cladding material that is more conductive or by increasing the dark resistivity of the photoconductor (see Supplementary Section 1B).

To measure the output of the computation performed by our device, we imaged the output facet of the device with an infrared camera (Allied Vision Goldeye CL-033). We built a 4f relay with a magnification factor of 5.3, allowing us to image the intensity distribution at the output facet, $I_\text{camera}(x,y)$, with a resolution of $\SI{2.8}{\micro \meter}$ per pixel, and a field of view of $\SI{1.7}{mm}$ in the $x$ direction. We defined a small range of $y$ to be the region of interest and integrated the intensity over this range to obtain the 1-D intensity output of the 2D-programmable waveguide: $I_\text{out}(x) = |E(x, z=L)|^2 = \int_{y_{\text{min}}}^{y_{\text{max}}} I_\text{camera}(x, y) \mathrm{d}y$. We used an exposure time on the order of $\SI{500}{\micro \second}$, chosen to be much shorter than the AC voltage's period, which is approximately $\SI{40}{ms}$.

\section*{Data availability}
All data generated during this work are available at \url{https://doi.org/10.5281/zenodo.10775721}.

\section*{Code availability}
An expandable demonstration code for simulating wave propagation through programmable waveguides is available at
\url{https://github.com/mcmahon-lab/2D-programmable-waveguide}.
All code used for this work is available at \url{https://doi.org/10.5281/zenodo.10775721}.

\end{document}


\title{Arbitrary control over multimode wave propagation for machine learning: supplementary information}
\maketitle
\vspace{-0.2in}
\tableofcontents
\clearpage

\section{Device design and characterization}
\label{section-device-design}

\subsection{Waveguide design}
\label{section-waveguide-design}

\begin{figure*}[!htbp]
    \centering
    \includegraphics[width=0.33\textwidth]{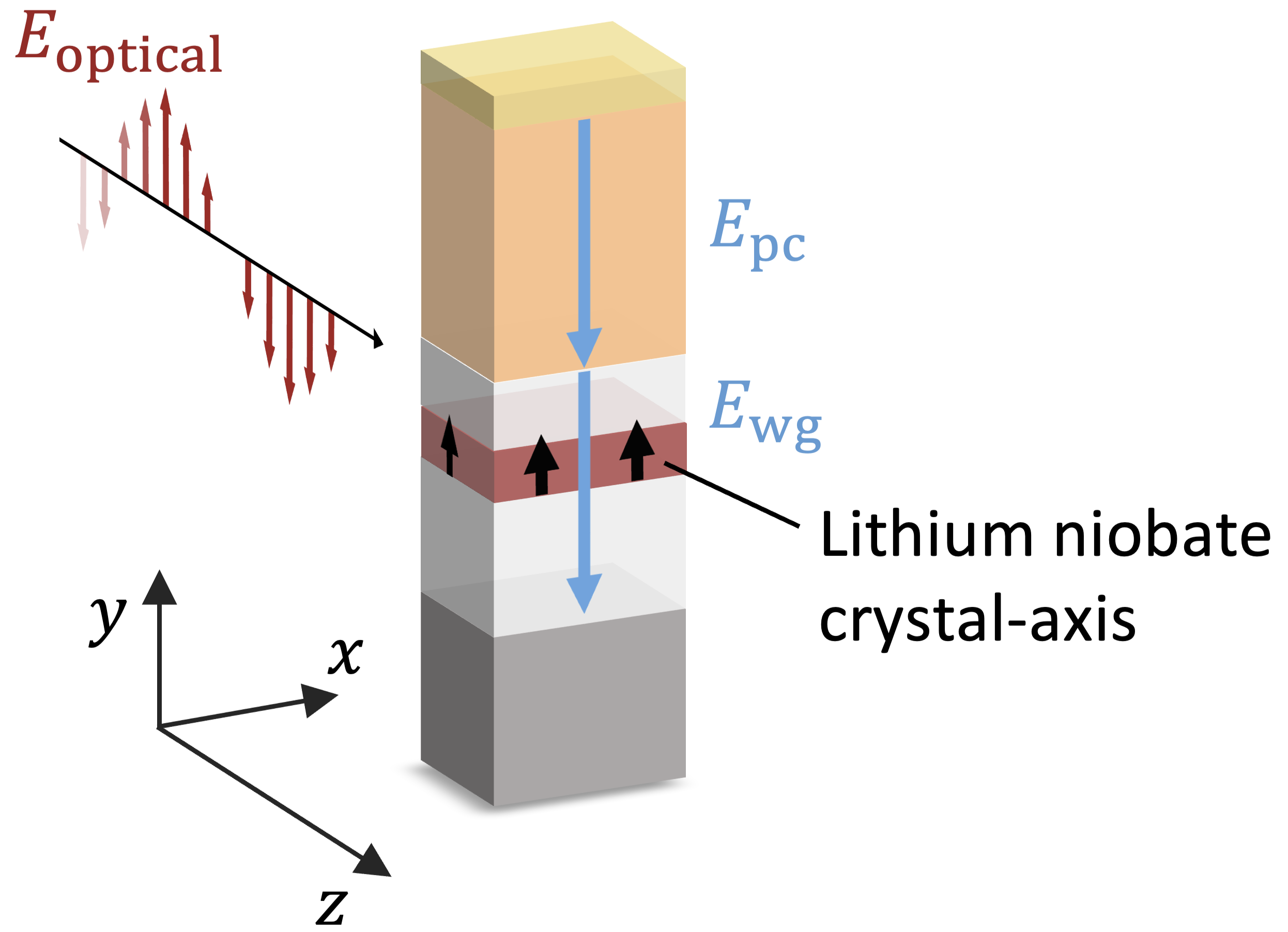}
    \captionsetup{justification=raggedright,singlelinecheck=false}
    \caption{
    \textbf{Orientation of fields and materials with respect to the coordinate system.} Blue arrows: Applied electric fields in the waveguide ($E_\text{wg}$) and in the photoconductor ($E_\text{pc}$). Red arrows: Optical electric field $E_\text{optical}$ of the transverse-magnetic mode. Black arrows: Lithium niobate crystal axis. 
    }
\end{figure*}

\textbf{Choice of core material:} We selected lithium niobate as the material for the waveguide core, as it has a large electro-optic coefficient, low optical loss, and thin-film wafers are commercially available. The change in the refractive index of lithium niobate under an external applied field is given by 
\begin{align}
    \label{eq:e-o-effect}
    {\Delta n_{i} = -n^3_{i} \sum_j r_{ij} E_j/2},
\end{align}
where $r_{ij}$ is the electro-optic tensor, $n_i$ are principal semiaxes of the index ellipsoid, and $E_i$ are the components of the electric field that are induced by the applied voltage.  The strongest component of the electro-optic tensor of lithium niobate is the $r_{33}$ component.
To select the term in the above equation that includes $r_{33}$, we need to ensure that the electric field of the optical mode in the waveguide points in the same direction as the externally applied electric field.
Since the planar electrode and conductive substrate of our device constrain the applied electric field to predominantly point in the (vertical) $y$-direction, we chose to use a transverse-magnetic (TM) mode such that the electric field of the optical mode also predominantly points in the $y$-direction.
Finally, to capitalize on the $r_{33}$ component, the crystal z-axis also needs to be oriented to point in the $y$-direction in this coordinate system, which is the reason we chose a Z-cut lithium niobate film.
\color{white}
\cite{lecun2015deep,patterson2021carbon, shen2017natphoton, Tait2017neuromorphic, wetzstein2020nature, shastri2021natphoton, feldmann2019nature,feldmann2021nature, huang2021silicon, ashtiani2022nature, bandyopadhyay2022single, hamerly2019large, nahmias2019photonic, anderson2023optical, mcmahon2023natrevphysics, larocque2021opticsexp, khoram2019photonres, Hughes-fan2019SciAdv, Nakajima-Hashimoto-NTT2022neuralSE, nikkhah2024inverse, wu-feng2023natphoton, Molesky-vuckovic-rodriguez2018NatPhoton, psaltis1988adaptive, Brady1991holographic, delaney-muskens2021nonvolatile, wu-li2024freeform,delaney-muskens2020family, chiou-wu2005nature, wu2011natphoton, luennemann-buse2003AppliedPhysB, hillenbrand1995acoustic, lecun1998mnist, wright-onodera-mcmahon2022nature, gu-pan2023m3icro, huang2020demonstration, Zhang-Liu2021NatureComms, pai2023experimentally,Marcucci-Conti2020PhysRevLetters,Tegin2021scalable,mohammadi2019science,RoquesCarmes2020heuristic,bogaerts2020nature,xie2017programmable,Feng-wang2024integrated, cheng2018photonic, bender-cao2022natphys, frumker-silberberg2007beamshaper}
\color{black}

\textbf{Choice of cladding material:} For the claddings of the waveguide, we chose silicon dioxide as it is readily available as buried oxide beneath lithium niobate, and simple to deposit via PECVD as top cladding. It also has desirable material properties for our device, including a high dielectric-breakdown field and low optical loss.

\textbf{Choice of waveguide dimensions:} We designed the thickness of each layer to ensure single-mode waveguiding in the $y$-direction, while maximizing the TM$_0$ mode overlap with the lithium niobate core. 
At the same time, to optimize the refractive-index modulation in the lithium niobate core, we minimized the overall thickness of the waveguide (see Sec.~\ref{section-circuit-model}) while keeping propagation loss minimal. 
These needs were balanced with the availability of thin-film lithium niobate wafers from NanoLN. 
We settled on a 700 nm thick film of Z-cut lithium niobate with 2 \textmu m of silicon dioxide as the bottom cladding and 1 \textmu m of silicon dioxide as the top cladding. This slab waveguide technically has two modes, but the higher order $\text{TM}_1$ mode has a significant substrate loss ($>$ \SI{100}{dB/cm}), and thus the waveguide is effectively single mode. The cladding thicknesses are chosen to minimize the loss of the $\text{TM}_0$ mode of the waveguide, which we characterize both numerically and experimentally to be less than $\SI{1}{dB/cm}$ (see Sec.~\ref{section-propagation-loss} for more detail). 

\subsection{Photoconductor design}
\label{sec:photoconductor-design}

\textbf{Choice of photoconductor material:} 
We chose silicon-rich silicon nitride (SRN) as the photoconductor material due to its high dielectric-breakdown field and strong photoconductive response. In addition, the material properties of SRN are easy to tune by varying its silicon-nitrogen ratio \cite{Piccirillo1990}.

\textbf{Equivalent circuit model:} 
\label{section-circuit-model}
Here, we explain the electrical modeling of the device. We consider a simplified model, where each small region of the 2D-programmable waveguide is modeled as multiple layers of homogeneous materials with different electrical resistivity $\rho$ for each layer. This simplified model does not address the fringing fields present when a complex pattern of illumination is applied to the photoconductor (for this, a more complex model is necessary, and is discussed in Sec.~\ref{section-e-field-resolution}), but it is useful for understanding the basic principles of the device. 
We model each layer as a leaky capacitor, i.e. each layer as a capacitor in parallel with a resistor, and multiple layers are connected in series with each other.
The electric field $E_y(y)$ is then a piecewise constant function where the electric field in each material layer is constant and can be found from the corresponding lumped element impedance ($Z_i$) via $E_{y,i} = V_i/d_i$, where $V_i = V_\text{applied}Z_i/Z_\text{total}$ and $d_i$ is the layer thickness for layer $i$.
The gold electrode and strongly-doped silicon substrate are sufficiently conductive that for the purpose of this analysis they can be modeled as perfect conductors and are ignored in the lumped element model.
On the other hand, the lithium niobate core (with $\rho \sim \text{G}\Omega$cm) and both silicon dioxide claddings are sufficiently electrically insulating that we can model the resistors as open circuits. 
Therefore, we simplify the circuit for the waveguide, consisting of top cladding, lithium niobate core, and bottom cladding, as a single capacitor with capacitance $C_{\text{wg}}$.
Hence, a simplified equivalent circuit to model the device is a single capacitor for the lithium niobate waveguide in series with a leaky capacitor for the photoconductor, where the resistance associated with the leaky capacitor can be reduced via external illumination.

\begin{figure*}[!htbp]
    \centering
    \includegraphics[width=0.85\textwidth]{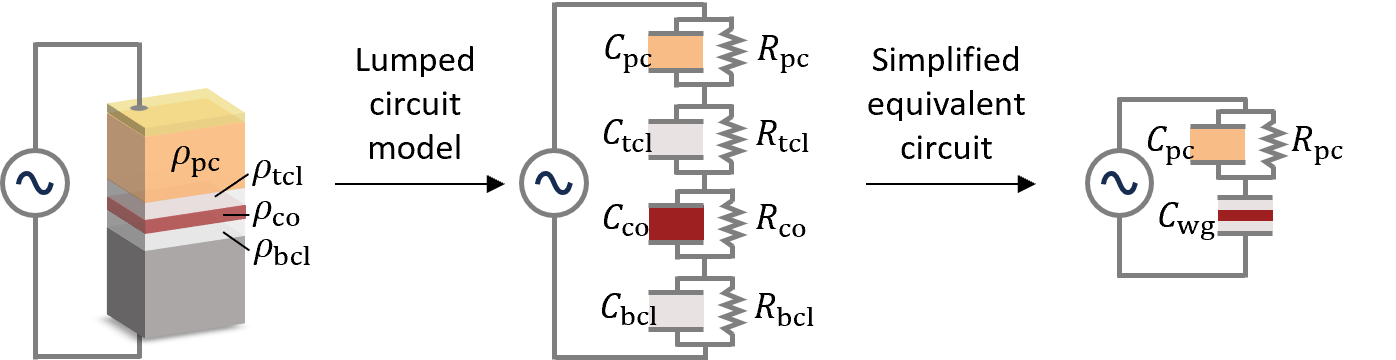}
    \captionsetup{justification=raggedright,singlelinecheck=false}
    \caption{
    \textbf{Different abstractions of the electrical model of the 2D-programmable waveguide.} Left: The device is modeled as a stack of homogeneous layers with different electrical resistivities. Center: A lumped element abstraction models each layer as a leaky capacitor with different resistance and capacitance. Right: A simplified model of the material stack consisting of only one capacitor in series with a leaky capacitor. Abbreviations used in subscripts: \textbf{pc}: photoconductor, \textbf{tcl}: top cladding, \textbf{co}: lithium niobate core, \textbf{bcl}: bottom cladding, \textbf{wg}: waveguide}.
    \label{fig:equivalent-circuit}
\end{figure*}

\textbf{Direct current (DC) vs alternating current (AC):} 
Widely used waveguide claddings are extremely good electrical insulators.
For example, thermally grown silicon dioxide's resistivity at room temperature is reported to be more than $10^{14}$.
This resistivity is far higher than the resistivity of silicon-rich silicon nitride and therefore, at DC voltages, the impedance of the device is dominated by the resistance of silicon dioxide.
This implies that one cannot change the refractive index of the waveguide with a constant applied voltage, as the steady-state electric field would be zero everywhere except across the silicon dioxide layers and the electro-optic effect in lithium niobate would be negligible.
We evade this problem by applying AC voltages to our device with frequencies that roughly satisfy $R_\text{pc} \sim 1/\omega C_\text{pc}$.
As will be explained in the following section, this will allow us to effectively switch the electric field in the lithium niobate core on and off.

DC operation becomes feasible when waveguide claddings are made of materials with lower resistivity, such as doped silicon oxynitride \cite{Janotta2004doping,Piccoli2022silicon}. Such structures are also expected to yield a higher refractive-index modulation, due to the reduced impedance of the waveguide. Additionally, they would require a lower operational voltage to achieve the same refractive-index modulation ($\Delta n$). Another approach towards DC operation is to develop photoconductors with higher dark resistivity than the waveguide core and cladding. 

\textbf{Choice of photoconductor thickness:} 
In order to design the material stack to effectively switch the electric field inside the lithium niobate core on and off, we optimize the parameters of the circuit and the frequency of the applied voltage.
The goal of the design was to find a photoconductor for which the electric field inside the lithium niobate core is as large as possible when the photoconductor is illuminated, and as low as possible when the photoconductor is not illuminated.
The photoconductor's resistance $R_\text{pc}$ varies depending on the illumination. 
In the ideal ``bright'' state (that is, with illumination on the photoconductor), the photoconductor's resistance is so low that all voltage drops across the waveguide and the electro-optic effect is strongest.
In the ideal ``dark'' state (that is, with no illumination on the photoconductor), the photoconductor's resistance is so high that the voltage drop across the waveguide is as low as possible and the electro-optic effect is weakest.
As mentioned above, it is very challenging to design a photoconductor whose dark resistance is higher than that of silicon dioxide.
Therefore, we chose to use AC voltage instead and analyze the complex impedances of the circuit. 
The voltage drop across the waveguide is given by: 

\begin{align}
    \label{eq:voltage-division}
    V_\text{wg} = V_\text{applied} \frac{Z_\text{wg}}{Z_\text{wg} + Z_\text{pc}} = \frac{V_\text{applied}}{1 + \frac{-i\omega C_\text{wg}}{R_\text{pc}^{-1} - i\omega C_\text{pc}}}.
\end{align}

Assuming fixed capacitances, this expression is maximal in the limit $R_\text{pc} \rightarrow R^\text{bright} \ll 1/\omega C_\text{pc}$ (``bright'' state of photoconductor) and evaluates to $V_\text{wg} = V_\text{applied}$, i.e. all voltage drops across the waveguide. 
The expression is minimal in the limit $R_\text{pc} \rightarrow R^\text{dark} \gg 1/\omega C_\text{pc}$ (``dark'' state of photoconductor), and evaluates to $V_\text{wg} = V_\text{applied}/(1 + C_\text{wg}/C_\text{pc})$.
First, this implies that, to maximize the difference of $V_\text{wg}$ in the bright and dark state, the device needs to be operated with an AC applied voltage whose frequency $\omega$ satisfies ${R^\text{bright} \ll 1/\omega C_\text{pc} \ll R^\text{dark}}$.
Second, the difference between $V_\text{wg}$ in the bright and dark state will be larger when the ratio $C_\text{wg}/C_\text{pc} = \epsilon_\text{wg} d_\text{pc}/\epsilon_\text{pc} d_\text{wg}$ is larger.
Since the relative permittivities are determined by the choice of materials and the thickness of the waveguide $d_\text{wg}$ is determined by the considerations presented in Sec.~\ref{section-waveguide-design}, the only freely tunable parameter is the thickness of the photoconductor $d_\text{pc}$.
Therefore, a thicker photoconductor layer minimizes the electric field inside the lithium niobate core in the dark state and maximizes the programmable refractive-index modulation.

\begin{figure*}
    \centering
    \includegraphics[width=0.5\textwidth]{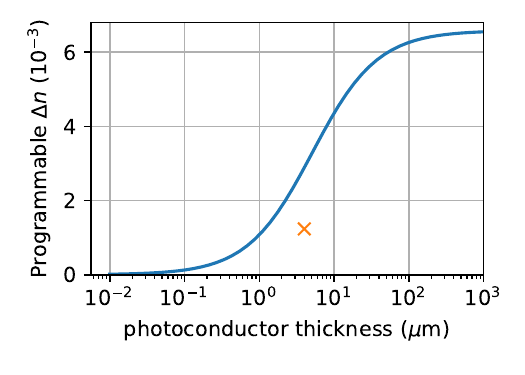}
    \captionsetup{justification=raggedright,singlelinecheck=false}
    \caption{\textbf{Maximal programmable refractive-index modulation as a function of photoconductor thickness.} Blue: Maximal programmable $\Delta n$ as a function of photoconductor thickness at an applied voltage of $1100$ V and $f = 10$ Hz, assuming a photoconductor that switches from a perfectly ``dark'' state to a perfectly ``bright state'' as described in Sec.~\ref{section-circuit-model}. Orange marker: Photoconductor thickness chosen and programmable $\Delta n$ achieved in this work.}
    \label{fig:dn-vs-pc-thickness}
\end{figure*}

Fig.~\ref{fig:dn-vs-pc-thickness} shows the maximal programmable refractive-index modulation $\Delta n$ as a function of the photoconductor thickness, otherwise assuming the waveguide parameters shown in Extended Data Fig.~1 and a maximal field of $50$ V/\textmu m in the lithium niobate core. The programmable refractive-index modulation is defined as the difference between the refractive-index modulation in the bright and dark photoconductor state:

\begin{align}
    \label{eq:programmable-delta-n}
    \Delta n_\text{programmable} = \Delta n_\text{bright} - \Delta n_\text{dark},
\end{align}
where the refractive-index modulation in the dark and bright photoconductor state is calculated using Eq.~\ref{eq:e-o-effect} with the electric field determined from Eq.~\ref{eq:voltage-division}. 
For brevity we refer to the programmable refractive-index modulation $\Delta n_\text{programmable}$ simply as $\Delta n$ throughout the manuscript. In Fig.~\ref{fig:dn-vs-pc-thickness}, the blue line shows the maximal programmable refractive-index modulation assuming an optimal photoconductor with $R^\text{bright} \ll 1/\omega C_\text{pc}$ and $R^\text{dark} \gg 1/\omega C_\text{pc}$.
We selected the thickest photoconductor layer that was reasonable ($d_\text{pc} = 4$ \textmu m), balancing the benefit from a thicker layer with fabrication constraints (layer stress, deposition time, etc.) and the spatial resolution trade-off discussed in Sec.~\ref{section-e-field-resolution}. 
The orange cross shows the programmable refractive-index modulation that we achieve in experiment, which is lower than the maximal value due to the photoconductor not being perfectly insulating in the dark state and not perfectly conductive in the bright state. More details about this measurement are presented below in Sec.~\ref{sec:delta-n-measurements}.

\subsection{Magnitude of refractive-index modulation}
\label{sec:delta-n-measurements}

\begin{figure*}[!tbp]
    \centering
    \includegraphics[width=\textwidth]{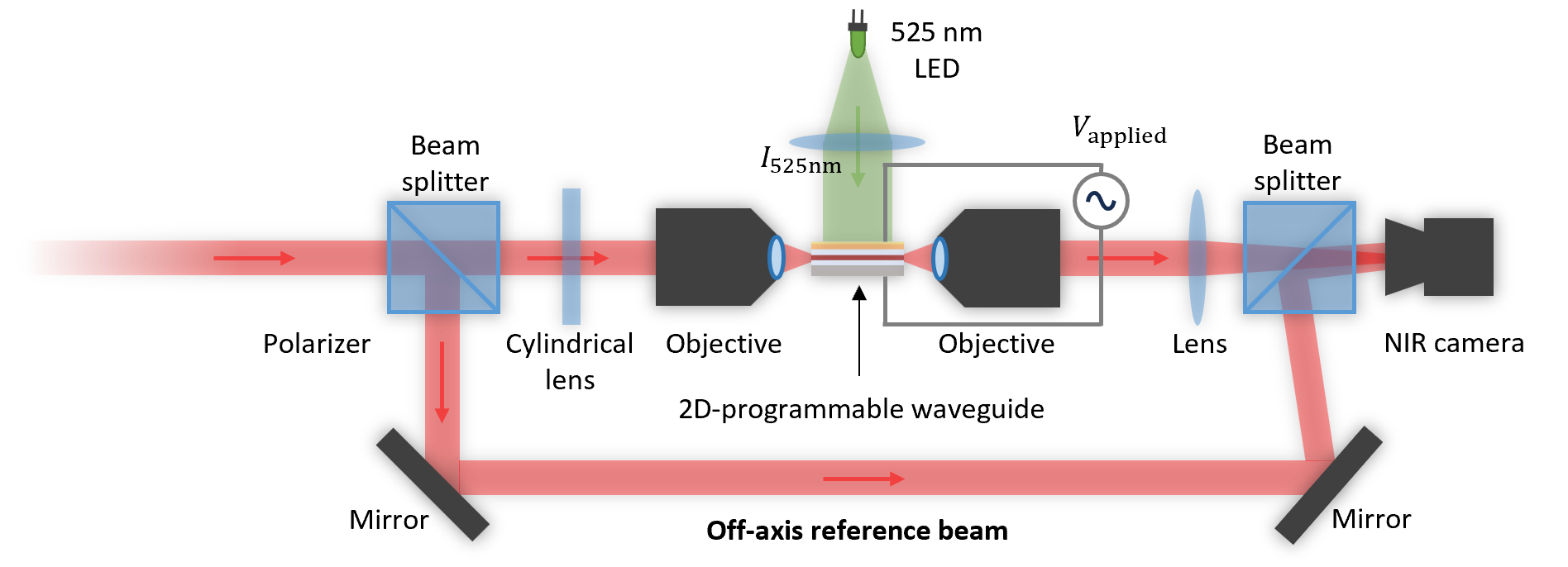}
    \captionsetup{justification=raggedright,singlelinecheck=false}
    \caption{\textbf{Schematic of the off-axis holography setup.} We used this setup to measure the refractive-index modulation under different illumination conditions and applied voltages. 
    We interfered light propagating through the 2D-programmable waveguide with light from the off-axis reference beam. We determined the refractive-index modulation at different applied voltages and illumination intensities from changes of the interference pattern.}
    \label{fig:off-axis-holography}
\end{figure*}

\begin{figure*}[!tbp]
    \centering
    \includegraphics[width=\textwidth]{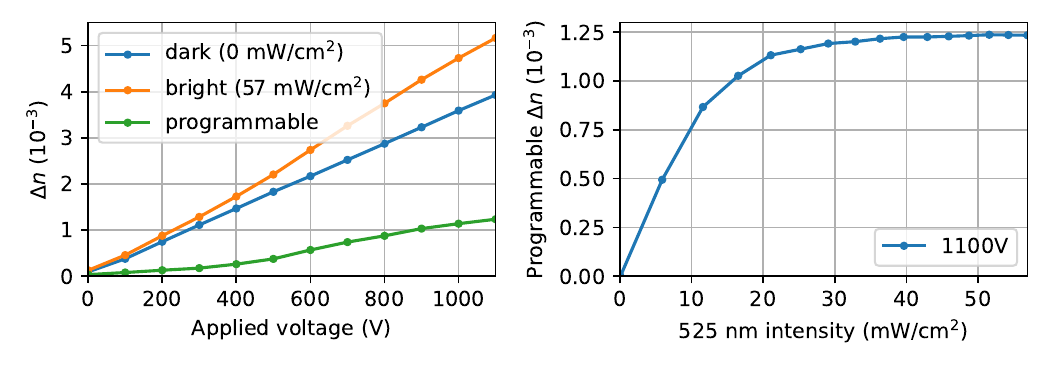}
    \captionsetup{justification=raggedright,singlelinecheck=false}
    \caption{\textbf{Measured refractive-index modulation.} Left: Refractive-index modulation of the TM$_0$ mode as a function of applied voltage for the photoconductor in a bright and dark state, and their difference, which is the programmable refractive-index modulation. Right: Programmable refractive-index modulation of the TM$_0$ mode as a function of illumination intensity.}
    \label{fig:delta_n_measurements}
\end{figure*}

In this subsection, we present measurements of the programmable refractive-index modulation as a function of the applied voltage and illumination intensity on the photoconductor. 
We performed this measurement with an off-axis holography setup, by interfering the light that propagated through the waveguide with an off-axis reference beam (see Fig.~\ref{fig:off-axis-holography}). All measurements presented here are measurements of the effective index of the TM$_0$ mode, and are given in reference to the refractive index of this mode with no external voltage applied.
We measured the refractive index at different applied voltage amplitudes (0 to 1100V) and different illumination strengths (0 - 57 mW/cm$^2$), at a fixed AC voltage frequency of 10 Hz.
As shown in Fig.~\ref{fig:delta_n_measurements}A, without any illumination on the photoconductor (``dark''), the refractive index linearly increases with the applied voltage to about $4\cdot10^{-3}$ at 1100 V.
With the photoconductor being illuminated at about 50 mW/cm$^2$, the refractive-index modulation increases to about $5.2\cdot10^{-3}$ at the highest voltage of 1100V.
We measured the difference between the dark- and bright-state refractive-index modulation, i.e. the programmable refractive-index modulation, to be around $1.2\cdot10^{-3}$ at the highest voltage.
We also present a measurement of the refractive-index modulation as a function of the illumination intensity, interpolating between the ``dark'' and ``bright'' state. 

We attribute the change of the refractive index to the Pockels effect in the thin film lithium niobate layer, due to good agreement with the expected electro-optic coefficient, $d_{33} = 
\SI{27}{\pico\meter/\volt}$. The photorefractive effect is likely negligible as the photorefractive damage threshold of 5\% MgO-doped lithium niobate to green light is more than 6 orders of magnitude higher than the intensities we used in experiment, at $>$\SI{75}{kW/cm^2}\cite{Volk2009,Bryan1984}.

We note that in Fig.~2 of the main manuscript, the largest achievable $\Delta n$ is $1\times 10^{-3}$, as opposed to $1.2\cdot10^{-3}$ presented here. This discrepancy arises because we conducted the experiments for Fig.~2 at \SI{1000}{V}, as opposed to the max voltage applied here of \SI{1100}{V}. Moreover, we note that the off-axis holography measurements were performed on a different device than the one used for the main manuscript's measurements. Variations in the film thickness of the electrode during deposition led to a higher transmission of the illumination through the electrode in the device used for off-axis holography measurements. Thus, more optical power was delivered to the photoconductor, leading to a higher refractive-index modulation.

\subsection{Spatial resolution of refractive-index modulation}
\label{sec:spatial_resolution}
\label{section-e-field-resolution}

\begin{figure*}[!t]
    \centering
    \includegraphics[width=\textwidth]{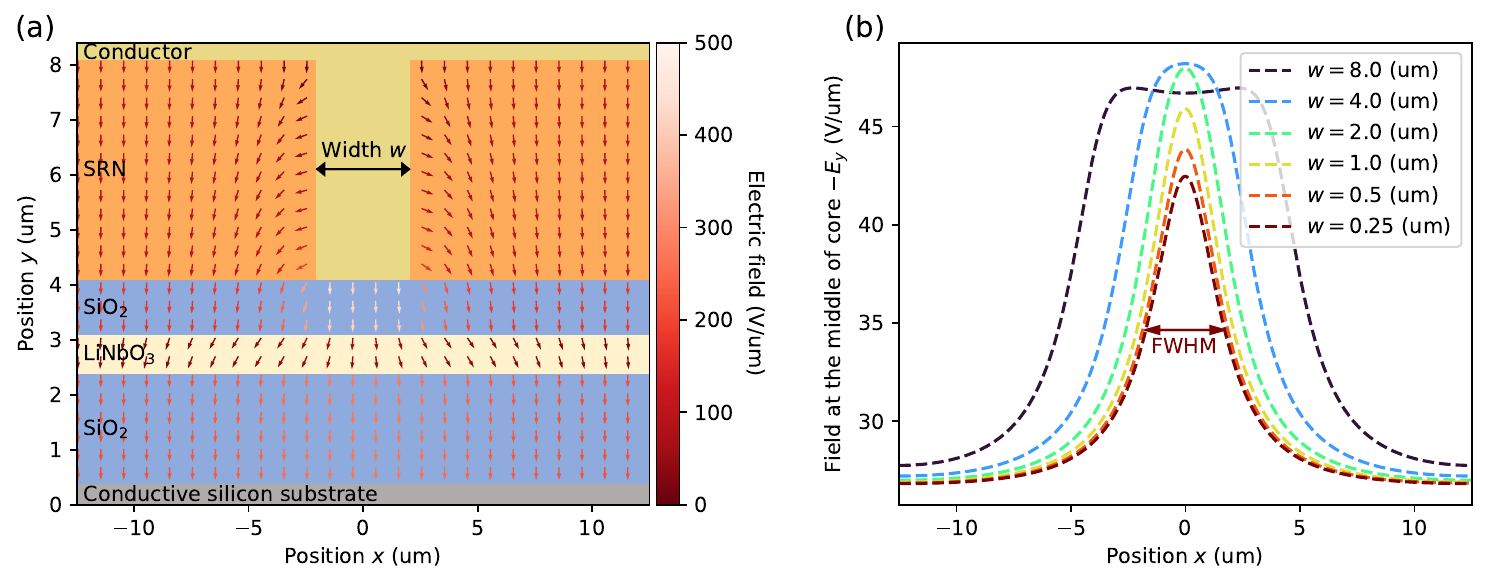}
    \captionsetup{justification=raggedright,singlelinecheck=false}
    \caption{
    \textbf{Electric field distribution in the 2D-programmable waveguide with spatially varying projected illumination.}
    (a) A simplified model to capture the electric-field spreading effects in light-based programming of the refractive-index distribution. We assume dielectric permittivity of $\epsilon=3.9$ for $\mathrm{SiO_2}$ and $\epsilon=7.2$ for SRN. Lithium niobate exhibits anisotropy and the permittivities are $85.2$ and $27.8$ for the $x$- and $y$-directions, respectively. The directions and the colors of the arrows represent the direction and strength of electric field, respectively, for $V_\mathrm{cc}=\SI{1000}{V}$. (b) Distribution of the vertical electric field in the middle of the $\mathrm{LiNbO_3}$ core layer for various conductive strip width $w$. For all the simulations, we assume periodic boundary condition in the $x$-direction.}
    \label{fig:poisson}
\end{figure*}

In this subsection, we discuss limitations on the spatial resolution achievable with the 2D-programmable waveguide.
\begin{enumerate}
    \item The optical diffraction limit dictates the minimal spot size of the pattern projected onto the device.
    \item Independent of the optical diffraction limit, the carrier diffusion length in the photoconductor material dictates the minimal size of high-conductivity areas (i.e. areas that are in the ''bright state'').
    \item The geometry of the waveguide results in an "electrostatic resolution limit" due to the fringing of the electric field inside the waveguide. We estimated this resolution limit to be around $\SI{3.5}{\mu m}$ in our device and in general expect it to be roughly proportional to the thickness of the waveguide.
    \item The largest (but most readily surmountable) spatial resolution limit is currently imposed by the pixel size of the digital micromirror device, which are $9\times\SI{9}{\mu m}$ in size. This can be readily changed by using a projection system with a different magnification.
\end{enumerate}

Comparing outputs from the experimental setup to our simulations suggests that the combination of the first three factors combines to a resolution limit of around \SI{6}{\micro\meter}. This implies that reducing the pixel size below $9\times\SI{9}{\micro m^2}$ would not yield improvements below this spatial resolution.
Our choice not to further reduce the size of projected pixels is also informed by the magnitude of the programmable refractive-index change $\Delta n$: The minimal waist of a laser beam can be determined by $w_0=\lambda_0/(\pi \text{NA})$, where the numerical aperture is $\text{NA}=\sqrt{(n_\text{eff}+\Delta n)^2 - n_\text{eff}^2} \approx \sqrt{2n_\text{eff}\Delta n} = 0.062$. This results in a beam waist of \SI{8.2}{\micro\meter}, which suggests a further reduction of the resolution will not result in significant changes to the capabilities of the 2D-programmable waveguide (until the resolution becomes close to the optical wavelength, which will enable qualitatively new devices such as gratings).

We now discuss an ''electrostatic resolution limit'', arising from an extension to the 1D-model of the electric field inside the waveguide that was introduced in Sec.~\ref{section-waveguide-design}. The model presented here takes into account the 3D-distribution of electric fields inside the dielectric stack when parts of the photoconductor are illuminated to create a spatially varying refractive-index distribution. The considerations in this paragraph show how the spread of the electric field affects the minimal feature size achievable in the programmable refractive-index distribution. 

Inside a dielectric medium with no free charge, the displacement field follows the macroscopic Maxwell equation
\begin{align}
    \nabla\cdot\mathbf{D}(\mathbf{r})=0,
\end{align}
where $\mathbf{D}(\mathbf{r})$ is the displacement field at position $\mathbf{r}=(x,y,z)^\intercal$. Using the constitutive relation $\mathbf{D}(\mathbf{r})=\epsilon_0\mathbf{\epsilon}(\mathbf{r}):\mathbf{E}(\mathbf{r})$, and $\mathbf{E}(\mathbf{r})=-\nabla \phi(\mathbf{r})$ with the scalar potential $\phi$, we obtain
\begin{align}
\label{eq:poisson}
    \nabla\cdot(\mathbf{\epsilon}(\mathbf{r}):\nabla \phi(\mathbf{r}))=0.
\end{align}
Here, $\epsilon_0$ and $\epsilon(\mathbf{r})$ are the vacuum and relative permittivities of the medium, respectively. Solving \eqref{eq:poisson} under appropriate boundary conditions, we can calculate how the electric field spreads inside dielectric materials.

As shown in Fig.~\ref{fig:poisson}(a), as a model for the programmable slab waveguide, we consider a stack of layers composed of conductive Si substrate, \SI{2.0}{\micro m} of $\mathrm{SiO_2}$ bottom cladding, \SI{0.7}{\micro m} of $\mathrm{LiNbO_3}$ core, and \SI{1.0}{\micro m} of $\mathrm{SiO_2}$ top cladding. The layers are stacked in the $y$-direction and are modeled to be infinitely extending in the $x$- and $z$-directions. On top of the top cladding is a \SI{4.0}{\micro m} photoconductive SRN layer, with a T-shaped conductive region of width $w$ in the $x$-direction. The substrate is grounded, and the top conductive region is connected to a voltage source with voltage $V_\mathrm{cc}$. We use such a structure to model the situation in which a limited region on the photoconductive SRN is illuminated by a focused light with spot size $w$, making the material locally conductive. It is our goal to use this simplified model to unravel how the spot size $w$ is related to the distribution of the electric field inside the core material, studying the impact of electric-field spreading on the spatial resolution we can achieve in light-based programming of the refractive-index distribution.

For the structures mentioned above, we solve \eqref{eq:poisson} using the biconjugate gradient stabilized method. In Fig.~\ref{fig:poisson}(a), we show a typical electric field distribution inside the medium for $w=\SI{4}{\micro m}$. To be more quantitative, we show in Fig.~\ref{fig:poisson}(b) the vertical-electric-field distribution $E_y$ at the middle of the lithium niobate core for various conductive strip width $w$. For a large enough $w$, the width of the distribution of $E_y$ is approximately proportional to $w$. On the other hand, as $w$ approaches a value comparable to the thickness of the dielectric stacks, the spreading of electric fields sets a finite lower bound to the width of the distribution of $E_y$. Using the results from $w=\SI{0.25}{\micro m}$, we find the full-width at half maximum (FWHM) of the field distribution to be \SI{3.5}{\micro m}. This sets a limit to the minimum feature size we can achieve, even in the absence of any diffusion of the carriers inside SRN or finite resolution of the illumination profile impinging onto the SRN.

\subsection{Propagation loss}
\label{section-propagation-loss}

\begin{figure*}[!t]
    \centering
    \includegraphics[width=\textwidth]{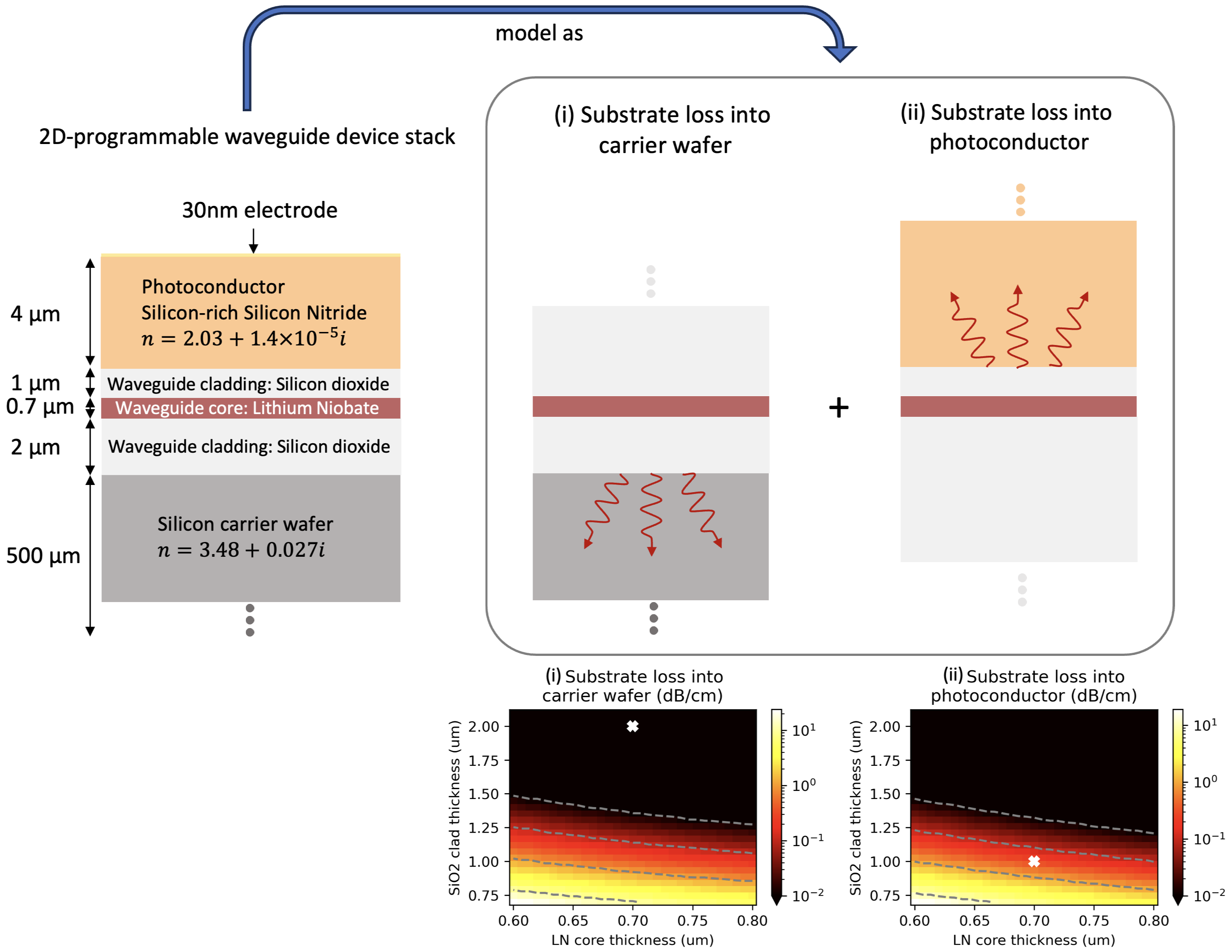}
    \captionsetup{justification=raggedright,singlelinecheck=false}
    \caption{\textbf{Simulation of propagation loss in the 2D-programmable waveguide.} The dominant source of propagation loss is due to radiative substrate loss into both the silicon carrier wafer and the photoconductor. We model this loss as the sum of two independent substrate loss contributions from the carrier wafer and photoconductor. This loss is simulated for different waveguide core and cladding thicknesses with the transfer matrix method \cite{ghatak1987numerical}. The device parameters in this work are represented by the white ``$\times$'' marker in the two subfigures.
    }
    \label{fig:substrate-loss}
\end{figure*}

In this section, we discuss the various sources of optical propagation loss in the 2D-programmable waveguide, along with our numerical and experimental characterization of the loss. There are two primary sources of loss in our device: the first is associated with the lithium niobate slab waveguide, and the second is radiative (substrate) loss into the photoconductor and the conductive silicon carrier wafer.

One source of optical loss in the lithium niobate slab waveguide is due to the ion-sliced thin-film lithium niobate wafer from NanoLN, as well as the silicon dioxide deposited via PECVD. According to Ref. \cite{zhang-loncar2017monolithic}, the optical loss in the lithium niobate slab waveguide is typically less than \SI{1.5}{dB/m} for samples that have not been thermally annealed, as is the case with our device. A key factor contributing to the low loss is the absence of lithographic etching in our process, thereby avoiding side-wall induced loss, which is often the predominant source of loss.

The second source of loss in the 2D-programmable waveguide is associated with the photoconductor and the conductive silicon carrier wafer. We measured the material loss of the photoconductor, silicon-rich silicon nitride, to be \SI{5}{\decibel/\centi\meter} using a Metricon prism coupler. Thus, the material has a refractive index of \(n_\text{pc} = 2.03 + 1.4 \times 10^{-5}i\). The conductive carrier wafer, a p-type silicon wafer, has a refractive index of \(n_\text{Si} = 3.48 + 0.027i\).  Thus, the loss is primarily due to the optical power of the lithium niobate slab waveguide radiating into the photoconductor and the carrier wafer, rather than being absorbed by the materials. This phenomenon, traditionally referred to as substrate loss, typically pertains only to the carrier wafer. However, given the photoconductor's substantial thickness (approximately \SI{4}{\micro\meter}), it can also be approximately modeled as radiative substrate loss. Note that although the refractive index of the photoconductor is smaller than the extraordinary refractive index of lithium niobate ($n_\text{LN,e} = 2.1376$ at \SI{1550}{nm}), the effective index of the TM$_0$ mode in the lithium niobate waveguide only has index $n_\text{eff}=1.94$, which is smaller than $n_\text{pc}$, allowing light from the TM$_0$ mode to leak into the photoconductor. Therefore, we model this loss in the device as the sum of two independent substrate loss contributions from the silicon carrier wafer and photoconductor (see Fig.~\ref{fig:substrate-loss}). The substrate loss is numerically computed using the transfer matrix method \cite{ghatak1987numerical} for different core and cladding thicknesses. For the carrier wafer, the loss is negligible, below \SI{1}{\decibel/\meter}, owing to a thick bottom cladding of \SI{2}{\micro\meter}. For the photoconductor, the loss is estimated at \SI{0.3}{\decibel/\centi\meter}, attributed to a thinner top cladding of \SI{1}{\micro\meter}. We verified these loss estimates using a COMSOL EWFD Mode Analysis, which yields a total loss of 0.23 dB/cm, accounting for all of the effects discussed above. As loss varies sharply with device parameters, we conservatively estimate that the overall propagation loss is below \SI{1}{\decibel/\centi\meter}.

Due to the absence of lithographic etching in our device, precise experimental measurement of this numerically estimated low loss is challenging. We fabricated chips of varying lengths (5mm, 1cm, 2cm) and qualitatively observed that transmission through these chips does not vary significantly with length, consistent with the simulations. 

Finally, we emphasize that for applications like quantum photonics, the loss can be reduced to be less than \SI{1}{\decibel/\meter} by further increasing the top cladding thickness. We also note that it is possible to further enhance device performance, specifically the refractive-index modulation (\(\Delta n\)), without a substantial increase in optical loss, by reducing the thickness of the bottom cladding.

\section{Experimental setup}
\label{sec:experimental_setup}

\begin{figure*}[!t]
    \centering
    \includegraphics[width=0.87\textwidth]{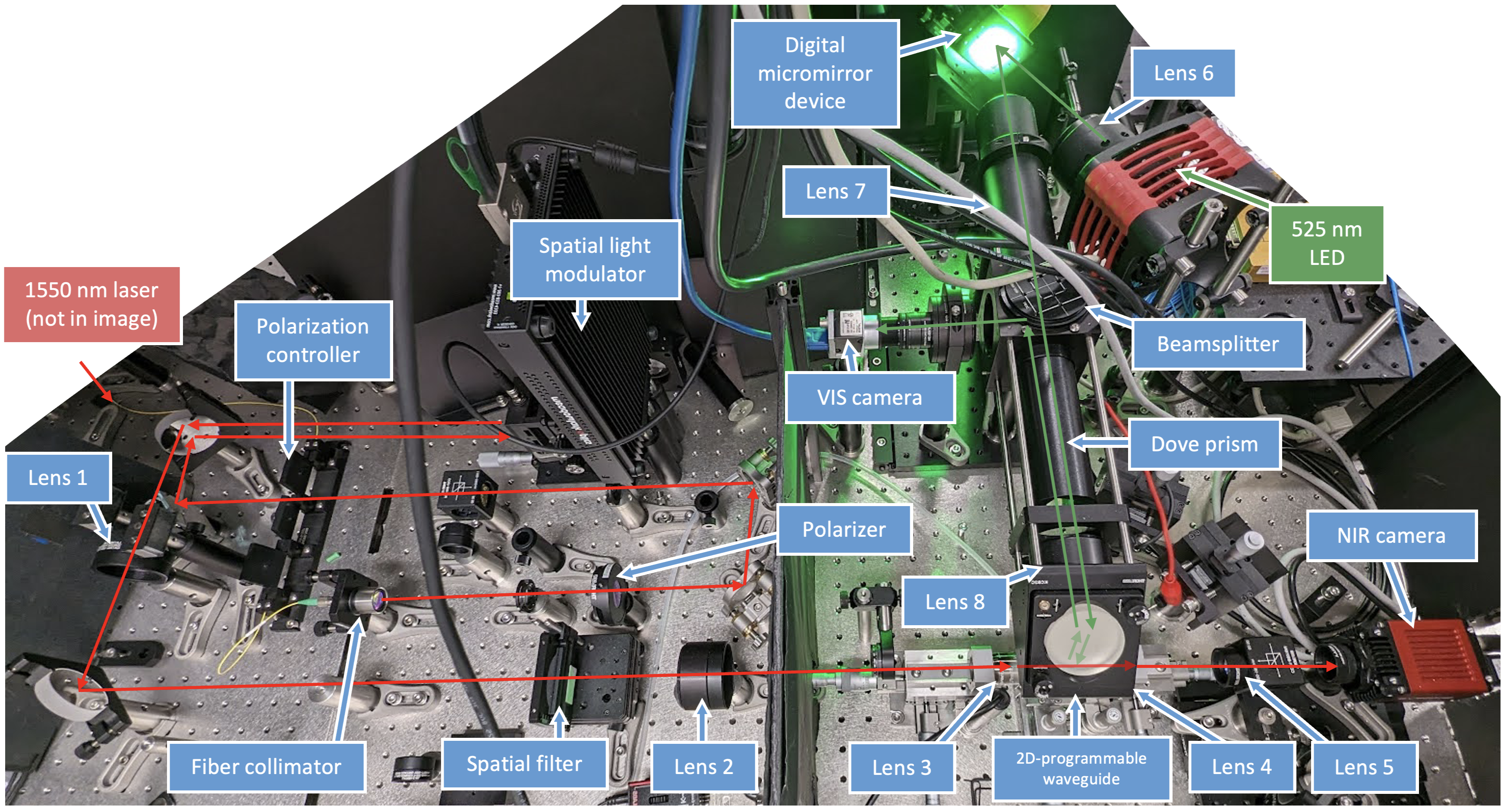}
    \captionsetup{justification=raggedright,singlelinecheck=false}
    \caption{\textbf{Photograph of the experimental setup.} The photograph is overlaid with the optical beam path (red) and the photoconductor illumination path (orange). The laser beam propagating through the 2D waveguide originates on the left and propagates to the camera in the bottom right. The 525nm control light originates in the top right and is projected onto the waveguide in the bottom right.}
    \label{fig:photo-setup-complete}
\end{figure*}

\begin{figure*}[!h]
    \centering
    \includegraphics[width=0.85\textwidth]{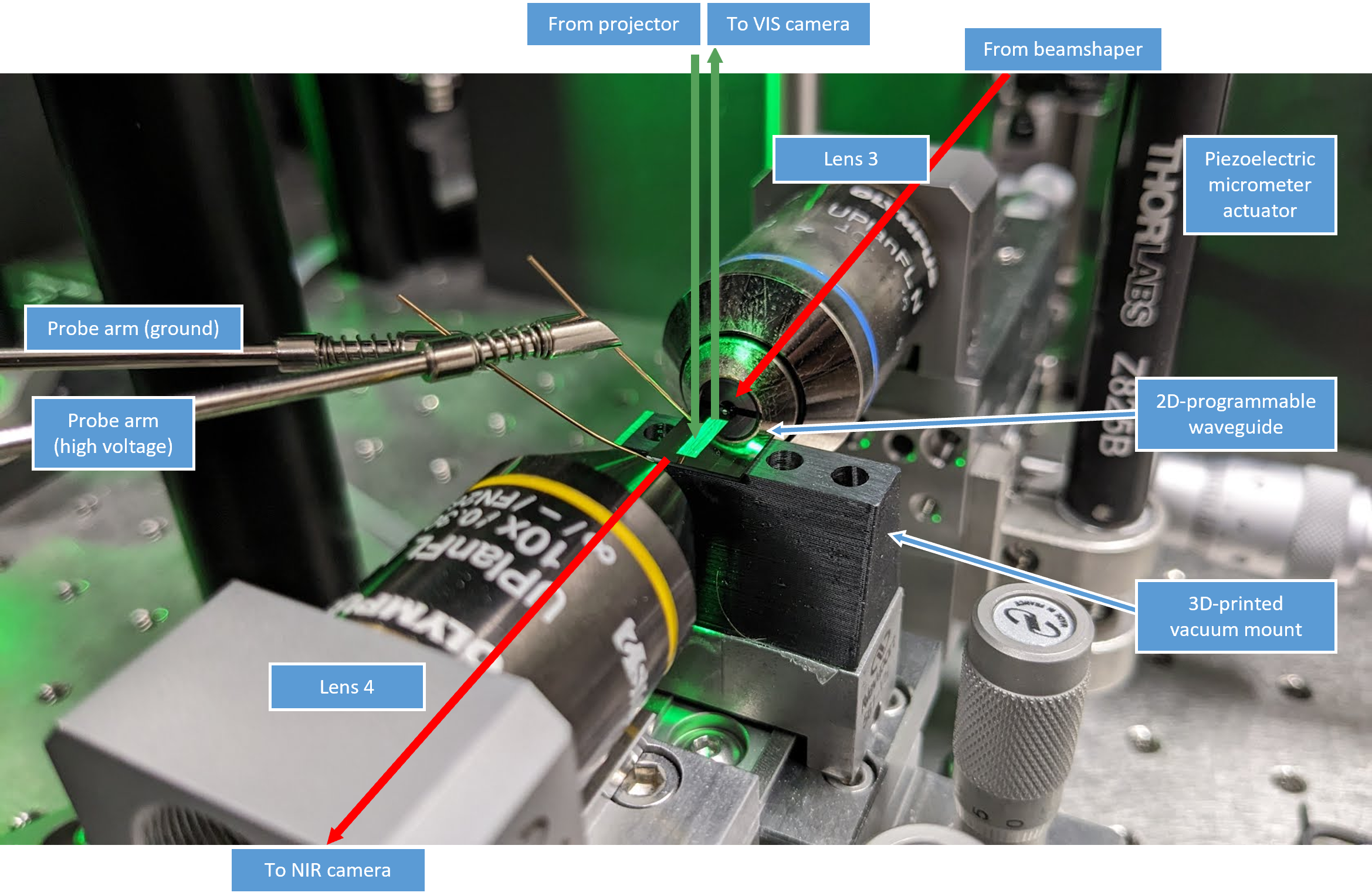}
    \captionsetup{justification=raggedright,singlelinecheck=false}
    \caption{\textbf{Photograph of the butt-coupling setup}. The photograph is overlaid with the optical beam path (red) going from the top right to bottom left and the photoconductor illumination path (orange) coming from the top.}
    \label{fig:photo-setup-buttcoupling}
\end{figure*}

In this section, we further detail the experimental setup described in the Methods, focusing on the beamshaper and the projector.  In the descriptions that follow, we reference Extended Data Fig.~2 and Fig.~\ref{fig:photo-setup-complete} for clarity. To maintain consistency, naming conventions for components within the experimental setup (e.g., ``Lens 3") are aligned with those used in the figures.

\textbf{1D spatial beamshaper to create optical inputs:} The core working principle of the beamshaper is to create spatially varying phase gratings on a 2D-phase spatial light modulator \cite{bender-cao2022natphys, frumker-silberberg2007beamshaper}. More specifically, at every position on the lateral dimension ($x$ direction), a vertical grating with variable amplitude is displayed on the SLM. These gratings are also shifted by a variable amount in the vertical $y$ dimension. By controlling the amplitude and this shift, we controlled the 1D amplitude $A(x)$ and phase $\phi(x)$ distribution of the first-order diffracted light. We shone a collimated beam of vertically polarized, continuous-wave \SI{1550}{nm} light from a JDS VIAVI MAP MTLG-B1C10 Tunable DBR Laser onto a Meadowlark Optics UHSP1K-850-1650-PC8 spatial light modulator (SLM). More specifically, the beam from the laser was collimated to a \SI{7}{mm} waist before it hits the two-dimensional phase-SLM, on which spatially varying gratings with a period of 16 pixels are displayed with a pixel pitch of \SI{17}{\micro m}. Next, the light propagated through a lens with focal length \SI{500}{mm} (Lens 1, Thorlabs A1380-C-ML), which separated the diffraction maxima of the phase gratings displayed on the SLM by approximately \SI{3}{mm}. A spatial filter (a slit) only let the first-order diffraction maximum pass through to another lens (Lens 2, Thorlabs TTL200-A, $f = \SI{200}{mm}$). From there on, the light propagated to an objective (Lens 3, Olympus UPLFLN 40X Objective, NA = 0.75, $f = \SI{4.5}{mm}$) that coupled the light into the 2D-programmable waveguide. The second and third lenses are essentially a 4-f relay system to demagnify the beam at the focal plane of the first lens. Thus, the distribution of light at the input facet of the 2D-programmable waveguide is approximately given by the spatial Fourier transform of the amplitude- and phase-profile corresponding to the gratings displayed on the phase spatial light modulator. We can create arbitrary input electric fields with features as small as $w_0 = \SI{1.6}{\micro m}$ on the input facet, over a distance of \SI{600}{\micro m}.

\textbf{Projector to create a programmable illumination pattern:}
To create programmable illumination patterns, we used a digital micromirror device (DMD, Vialux V-7000) with a resolution of 1024x768 pixels and a pixel pitch of \SI{13.7}{\micro m}.
We illuminated the DMD with green light (\SI{525}{nm}) from a SOLIS-525C high-power LED, collimated by a condenser lens setup (Lens 6).
We imaged the surface of the DMD onto the surface of the 2D-programmable waveguide via a 4-f setup consisting of two tube lenses (Lens 7, Thorlabs TL300-A, and Lens 8, Thorlabs TTL200-A).
To account for the 45-degree rotation of the micromirror array of our DMD, we inserted a Dove prism (Thorlabs PS993M) between the tube lenses, which, appropriately positioned, rotates the image by 45 degrees.
Using the same optical path as for the projection, we imaged the surface of the 2D-programmable waveguide onto a camera by inserting an 8:92 (R:T) Pellicle beamsplitter (Thorlabs BP245B1) between the tube lenses and imaged the waveguide surface via an additional tube lens (Lens 9, TTL100-A) onto a Basler ace acA5472-17um camera.
The projection illumination pattern on the surface of the 2D-programmable waveguide has dimensions of \SI{9.1}{mm} $\times$ \SI{6.8}{mm}, and each individual pixel of the projected pattern is \SI{9}{\micro m} $\times$ \SI{9}{\micro m} in size. 

\section{Training the 2D-programmable waveguide to perform machine learning}

In this section, we explain how the parameters of the 2D-programmable waveguide is trained to perform machine learning. This includes describing the different training algorithms that we considered, and explaining our decision to use physics-aware training for this work. We begin the section with a discussion on how many parameters are present in our device, as it is a dominant factor that determined the choice of training algorithm. 

\subsection{Parameter count of 2D-programmable waveguide} 
\label{sec:parameters}

\begin{figure}[!htbp]
    \centering
    \includegraphics[width=\textwidth]{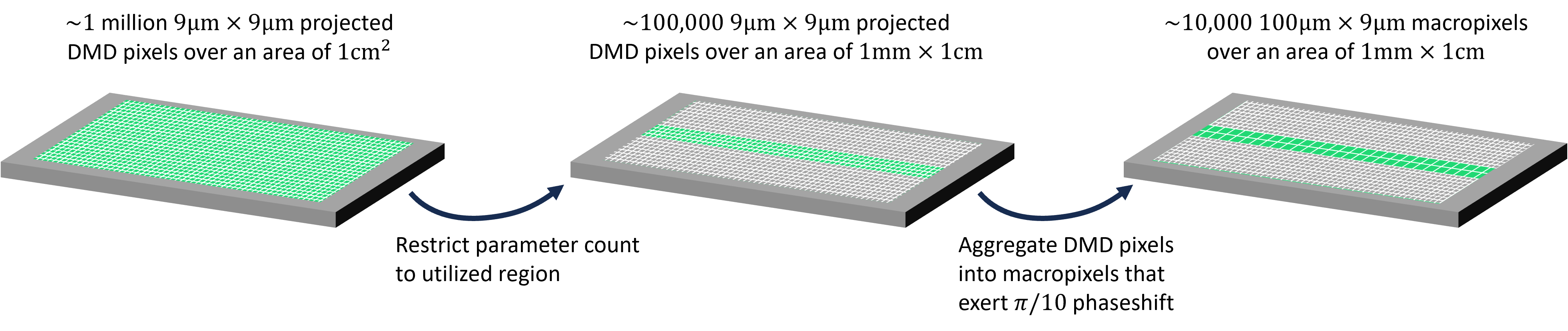}
    \captionsetup{justification=raggedright,singlelinecheck=false}
    \caption{\textbf{Schematic of parameter count.} In principle we can control around 1 million ($1024 \times 768$) refractive-index pixels on the 2D-programmable waveguide. However, most light propagates paraxially in a region of around $1$mm width in the center of the chip and is only affected by projected pixels in this region, reducing the parameter count to around $100,000$. Additionally, because of the weak phase-shift exerted by each individual pixel, we aggregated them into $100$\textmu m long macropixels which exert a significant phase-shift of $\pi/10$, resulting in a parameter count of around $10,000$.}
    \label{fig:parameter-count}
\end{figure}

The parameters of the 2D-programmable waveguide are the refractive-index modulations $\Delta n(x, y)$ of the slab waveguide. While it is conceptually advantageous to reason about the refractive-index modulation as a continuous function over space, it is in practice a discretized quantity as we use a DMD to project different patterns of illumination on the device, which has a discrete number of pixels. Furthermore, as a single pixel does not exert much phase-shift on the propagation of light, we group pixels on the DMD into macro-pixels for practical implementation. In other words, the number of parameters that we train for the device depends on how finely we perform this discretization, which we outline below. 

In the $x$ dimension, which is the direction that is perpendicular to the direction of propagation, we chose to use the finest discretization, which is limited by the imaging setup. This is appropriate as finer features in $x$ directly lead to more control over the wave dynamics. Thus, the number of independent parameters in the $x$ dimension is given by dividing the width of the total area that the wave propagates through ($\SI{1}{mm}$) by the smallest pixel size of the projected pattern on the chip ($\SI{9}{\micro\meter}$).

For the $z$ direction, which aligns with the direction of propagation, the situation differs. To understand this more formally, consider the distance over which a maximal refractive-index modulation can induce a phase shift of $\pi$. This quantity $L_{\pi} = \lambda/(2\Delta n_\text{max})$ is approximately $\SI{1}{mm}$ in our device. Since the length of a single pixel ($\SI{9}{\micro\meter}$) is much smaller than $L_{\pi}$, each pixel only exerts a negligible phase shift on the light that is propagating in the $z$ direction.
Therefore we do not consider each individual pixel to be a freely tunable parameter, but instead only count groups (macropixels) of 11 individual pixels to be one tunable parameter. We chose this number so that the length of macropixels is $L_{\pi}/10$, to ensure that each programmed macropixel can induce a non-negligible phase shift of $\pi/10$. Thus, the number of independent parameters in the $z$ dimension is given by dividing the length of the chip by the length of the macropixel in $z$. 

Consequently, the total number of parameters for the 2D-programmable waveguide is given by the product of the number of parameters in each direction, which is approximately \(\SI{1}{mm} / \SI{9}{\micro\meter}\) in the \(x\) direction and \(\SI{9}{mm}/ \SI{100}{\micro\meter}\) in the \(z\) direction, yielding 10,000 parameters. 

\subsection{Choice of training algorithm}
In this subsection, we describe the algorithms we considered for training the 2D-programmable waveguide. There are two critical factors that determined our choice to use physics-aware training: the number of parameters (10,000) for our device, and update speed of our experiment (\SI{20}{Hz} for inputs and $\SI{3}{Hz}$ for parameters).

\textbf{Model-free training algorithms:} These algorithms do not require a digital model to train the physical system. One approach involves individually perturbing each parameter and computing the gradient of the loss function with respect to each parameter \cite{shen2017natphoton}. Alternatively, random gradient descent can be used, where the parameters are perturbed in a random direction \cite{bandyopadhyay2022single}. Generally, both approaches may result in slow training, as a large number of passes through the setup is required to accurately sample the gradient. Given the large number of parameters and the relatively slow update speed of our experiment, this method would have resulted in impractically long training times for the MNIST machine learning task. We note that these algorithms could be effective for training the 2D-programmable waveguide if the number of parameters is reduced, for instance, by a low-dimensional parametrization of the projected patterns. 

\textbf{In-silico training:} An alternative approach is to use a purely model-based method, termed in-silico training. In this approach, training is performed entirely on a digital computer with the digital model. For differentiable digital models, the physical system can be trained with the backpropagation algorithm, for faster training. After completing the training, the parameters are transferred from the digital computer to the physical system. This requires an excellent digital model of the system. We attempted this approach and found that despite having good digital models (see Sec.~\ref{sec:digital_model}), poor performance was attained on the experiment. For instance, the test accuracy on MNIST handwritten-digit classification is only 60\% with in-silico training.

\textbf{Physics-aware training:} In this work, we use physics-aware training, a hybrid in-silico in-situ training algorithm \cite{wright-onodera-mcmahon2022nature}. This method combines the advantages of both model-free and model-based training. Our reasons for choosing this algorithm are as follows. Firstly, it is a backpropagation algorithm. Thus, approximate gradients are computed in a single backward pass, resulting in faster training. Additionally, since the physical system is used to compute the forward pass, the training is able to mitigate the mismatch between the experiment and digital model as well as the effects of experimental noise. A requirement for this approach is a differentiable digital model of the physical system, which we detail in Sec.~\ref{sec:digital_model}. We will explain physics-aware training in more detail in the following subsection.

\textbf{In-situ backpropagation algorithms:} 
These algorithms use the physical system to obtain the gradient of the loss function  \cite{hughes2018training, pai2023experimentally,spall-lvovsky2023training}. Thus, a digital model of the system is not required, and the training is fast as the algorithm is gradient-based. However, the algorithm requires having bidirectional input of light \cite{pai2023experimentally,spall-lvovsky2023training} and measuring the intensity distribution of light within the chip \cite{bender-cao2022natphys, iluz2024unveiling}. For this reason, the method was not implemented in our current experiment. We note that as we scale up the device to a larger number of parameters, this method may become essential for training the 2D-programmable waveguide.

\subsection{Physics-Aware Training}
\label{sec:PAT}
\begin{figure}[!htbp]
    \centering
    \includegraphics[width=\textwidth]{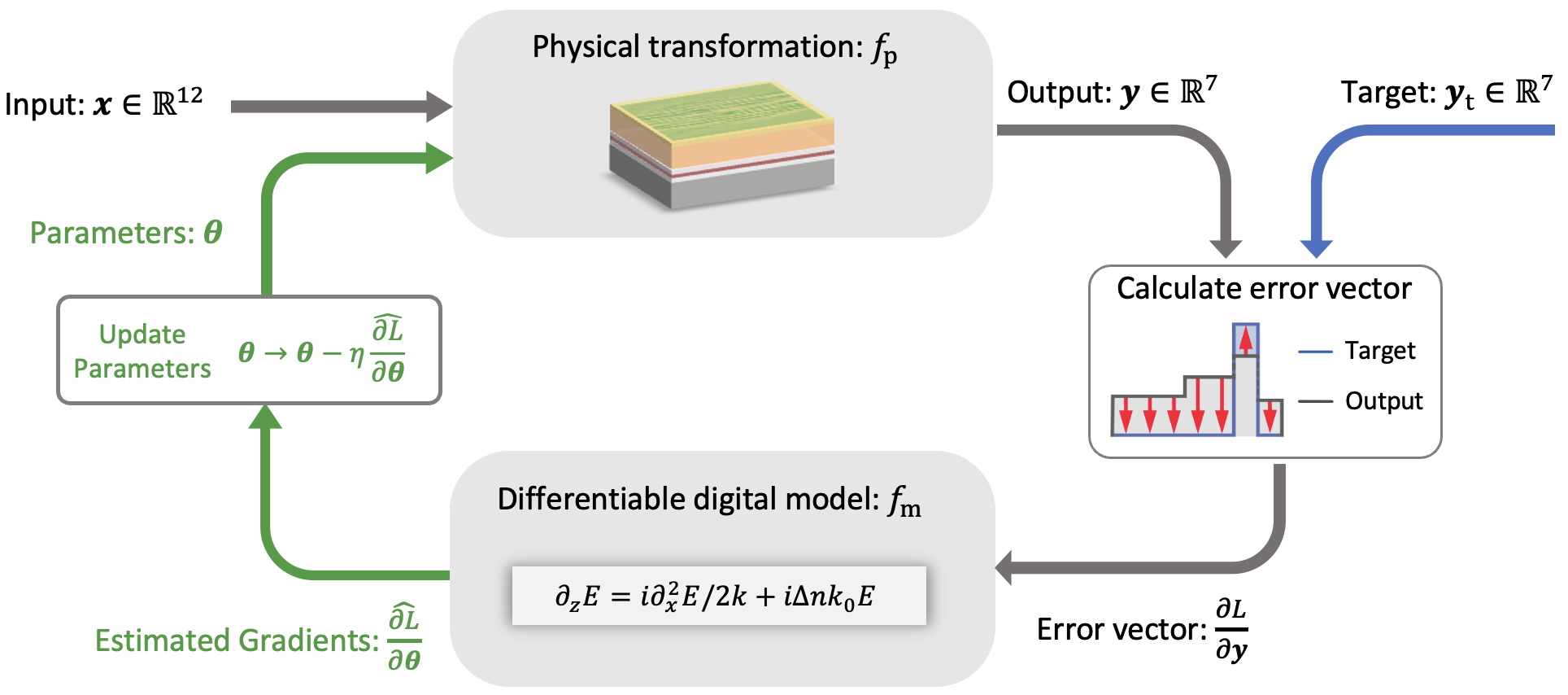}
    \captionsetup{justification=raggedright,singlelinecheck=false}
    \caption{\textbf{Schematic of physics-aware training.} The physical system is used in the forward pass, and the digital model is used in the backward pass, to obtain an approximate gradient of the loss function with respect to the physical parameters.}
    \label{fig:PAT}
\end{figure}

This section will explain the physics-aware training algorithm, which is a hybrid in-silico in-situ training algorithm. It is a summary of the algorithm that is specialized for the 2D-programmable waveguide---a more detailed and general explanation of the algorithm can be found in the original paper \cite{wright-onodera-mcmahon2022nature}. The schematic of the algorithm is shown in Fig.~\ref{fig:PAT}, which outlines the four key steps of the algorithm. The figure also explains the algorithm in the specific context of vowel classification, where the input vector is 12 dimensional and output vector is 7 dimensional. Finally, we note for brevity that the following equations will assume a batch size of one.

\textbf{1. Forward-pass through physical system:} The first step is to perform a forward pass through the physical system, which in our case is the 2D-programmable waveguide. Mathematically, this is given by
\begin{align}
    \label{eq:forward_pass}
    \mathbf{y} = f_{\mathrm{p}}(\mathbf{x}, \boldsymbol{\theta}).
\end{align}
Note that the physical transformation $f_{\mathrm{p}}$ includes both the amplitude encoding of the input machine learning data into an input optical field, and the binning operation that converts the output intensity to the machine learning output. Thus, this is a function that takes in an input vector $\mathbf{x} \in \mathbb{R}^{12}, \boldsymbol{\theta} \in \mathbb{R}^{\sim 10,000}$ and returns an output $\mathbf{y} \in \mathbb{R}^{7}$. 

\textbf{2. Compute the error vector:} Using the output of the physical system $\mathbf{y}$ and the target output $\mathbf{y}_\textrm{t}$, the second step is to compute the error vector with a digital computer. The error vector indicates how the outputs of the physical system should be adjusted to minimize the loss function, and plays a key role in a backpropagation algorithm. Mathematically, this is given by
\begin{align}
    \frac{\partial L}{\partial \mathbf{y}} = \frac{\partial}{\partial \mathbf{y}}H(\mathbf{y_{\mathrm{t}}}, \textrm{softmax}(\mathbf{y})),
\end{align}
where the softmax function is used to turn the outputs from the physical system into a probability, and $H$ is the cross-entropy function that is usually used as a loss function for classification tasks. One key intuition for why physics-aware training works well is that the error vector is accurately computed, as the algorithm has access to the real output of the physical system.

\textbf{3. Compute the estimated gradients:} The third step is to compute the estimated gradients, which is the approximate gradient of the loss function with respect to the physical parameters. This performed via a backward pass through the differential digital model of the physical system $(f_\mathrm{m})$. Mathematically, this is given by
\begin{align}
    \widehat{\frac{\partial L}{\partial \boldsymbol{\theta}}} = \paren{\frac{\partial f_\mathrm{m}(\mathbf{x}, \boldsymbol{\theta})}{\partial \boldsymbol{\theta}} }^\T \frac{\partial L}{\partial \mathbf{y}},
\end{align}
where $\frac{\partial f_m(\mathbf{x}, \boldsymbol{\theta})}{\partial \boldsymbol{\theta}}$ is the Jacobian matrix of the digital model of the physical system. The transposed Jacobian matrix performs a matrix-vector multiplication on the error vector to arrive at the estimated gradients. It should be noted that the estimated gradients $\widehat{\frac{\partial L}{\partial \boldsymbol{\theta}}}$ computed here are approximate rather than exact, as the digital model is only an approximation of the physical system. 
A key reason for the effectiveness of physics-aware training is that as long as the estimated gradient predicts the direction of the true gradient within a 90-degree cone (positive cosine similarity, or $\angle \paren{\widehat{\frac{\partial L}{\partial \boldsymbol{\theta}}},{\frac{\partial L}{\partial \boldsymbol{\theta}}}} < 90^\circ$), a small parameter update will result in a reduction of the loss. 

\textbf{4. Update the parameters:} With the estimated gradients, the final step is to update the parameters of the physical system. This is done by performing a gradient descent step on the parameters. Mathematically, this is given by
\begin{align}
    \label{eq:update_parameters}
    \boldsymbol{\theta} \rightarrow \boldsymbol{\theta} - \eta \widehat{\frac{\partial L}{\partial \boldsymbol{\theta}}},
\end{align}
where $\eta$ is the learning rate. After this, step 1 is repeated, and the algorithm continues to iterate until the loss converges. 

Though the algorithm has been specified mathematically in \eqref{eq:forward_pass}-\eqref{eq:update_parameters}, in practice the algorithm is implemented as a custom autograd function in PyTorch. This custom function can be generated via the package \texttt{Physics-Aware-Training} available at \url{https://github.com/mcmahon-lab/Physics-Aware-Training}. With the package, the user can run the following line of code to define a custom ``physics-aware function'': \\\texttt{f\_pat = make\_pat\_func(f\_physical, f\_model)}. The rest of the code can be written in regular PyTorch to train the system. Thus, the user can access regular PyTorch functionality such as optimizers and schedulers for training.

\section{Digital model of 2D-programmable waveguide}
\label{sec:digital_model}
In this section, we describe the digital model used for the 2D-programmable waveguide. The model is largely physics-based, and is modeled by a partial differential equation. As the calibration and alignment of the chip was complex, we also describe the procedure in this section. Finally, we found that the purely physics-based model only predicts the experimental outputs qualitatively. Thus, to improve the agreement further, we augmented the physics-based model with additional parameters, which were trained on experimental data. We outline this data-driven approach to fine-tune the physics-based model in the final subsection.

\subsection{Physics-based model}
Here, we derive the theoretical model that we use to emulate the 2D-programmable waveguide on a digital computer. To summarize the approach, we begin from the time-independent wave equation (Helmholtz equation) in 2D, and apply the standard beam-propagation method to arrive at the final equation. 

The time-independent wave equation in 2D is given by,
\begin{align}
    \pder{^2\tilde E}{z^2} + \pder{^2\tilde E}{x^2} + n^2(x,y)k_0^2\tilde E = 0,
\end{align}
where $\tilde E$ refers to the electric field in the $y$ dimension as we work with the transverse-magnetic (TM) mode. $n(x, y)$ is the effective refractive index of the slab. We note that the 2D wave equation works well for our setting, as the change in refractive index is generally small \cite{nikkhah2024inverse}. We use the ansatz $\tilde E(x, z) = E(x, z) \exp(\im k z)$ and apply the slowly-varying amplitude approximation, to arrive at, 
\begin{align}
    \label{eq:beam_propagation}
    \pder{E}{z} = \frac{\im}{2k}\pder{^2E}{x^2} + \im \Delta n(x, z) k_0 E,
\end{align}
where $\Delta n(x, z) = n(x, z) - n_0$ is the change in refractive index, which is assumed to be small $\Delta n \ll n_0$. $k_0$ is the wavevector in vacuum, $k = n_0 k_0$ is the wavevector in the medium. 
We note that this equation assumes that the wave is only traveling in the forward direction. This is a good approximation as we did not explicitly train for any resonant structures that could induce significant back reflections.
To numerically solve this partial differential equation,  we use the split-step Fourier method, which is a numerical method that is commonly used to solve PDEs of this form \cite{agrawal2012nonlinear}. 
Eq.~\eqref{eq:beam_propagation} also only describes paraxial waves accurately. We verified in simulation that this is a valid approximation for our experiments by comparing the propagation of inputs from our experiments using Eq.~\ref{eq:beam_propagation} against the ``extended Fresnel approximation''\cite{Feit1978} which handles non-paraxial beams more accurately. We found differences between the two methods to be negligible, but we note that the extended Fresnel approximation is in general preferred for simulations, as it produces more accurate results at the same computational complexity.

\textbf{Input-output relation:} The digital model of the 2D-programmable waveguide is as follows. Recall that the input-output relation is $\mathbf{y} = f_{\mathrm{m}}(\mathbf{x}, \boldsymbol{\theta})$, where $\mathbf{x}$ and $\mathbf{y}$ have dimensionalities associated with the machine learning input data and output data. Thus, the initial condition of the field propagation is given by,
\begin{align}
    \label{eq:mode_basis_input}
    E(x, z=0) = \sum_i x_i \times \underbrace{e^{-(x-\mu_i)^2/w_0^2}}_{E_{\text{mode}, i}(x)},
\end{align}
where $E_{\text{mode},i}(x)$ are the input modes, which we pick to be Gaussian modes with width $w_0$, and whose mean position, $\mu_i$, are translated linearly with respect to the index $i$. By numerically solving \eqref{eq:beam_propagation}, we can obtain the output field at the output facet of the waveguide $E(x, z=L_z)$. The intensity at the output facet is then binned to arrive at the machine learning output data $\mathbf{y}$, which is given mathematically by,
\begin{align}
    y_i = \int_{x_{\textrm{bin},i} - w/2}^{x_{\textrm{bin},i} + w/2} |{E(x, z=L_z)}|^2 \mathrm{d}{x},
\end{align}
where $x_{\textrm{bin},i}$ is the location of the $i$-th bin, and $w$ is the width of the bin.

\textbf{Conversion between projected image and refractive-index modulation:} As shown in Fig.~\ref{fig:delta_n_measurements}, the relation between the intensity of the projected light on the chip $I(x, z)$ and the refractive-index modulation $\Delta n(x, z)$ is nonlinear as saturation effects set in when the intensity is sufficiently high. However, for machine learning, we have found that operating in the linear regime is beneficial to obtain a better digital model. In addition, the spatial resolution of the refractive-index distribution is limited due to the spreading of electric field in the stack, as discussed in Sec.~\ref{sec:spatial_resolution}. Taking both of these facts into consideration, the model we use to convert from the parameters, which is the projected image ($\boldsymbol{\theta} = I(x, z)$) to the refractive-index modulation $\Delta n(x, z)$ is given by,
\begin{align}
    \Delta n(x, z) = \frac{\Delta n_{\text{max}}}{I_{\text{max}}} \paren{g(x, z) \ast I(x, z)},
\end{align}
where $\ast$ is the 2D convolution operation, $g$ is a 2D Gaussian kernel with standard deviation $d_\text{kernel}$, $\Delta n_{\text{max}}$ is the maximum refractive-index modulation, and $I_{\text{max}}$ is the maximum light intensity of the projected pattern that is illuminated on the chip. 

\subsection{Initial calibration of physics-based model}

\textbf{Calibration of beamshaper}: 
In this work, we built a beamshaper that enables us to send arbitrarily programmable input field distributions into the 2D-programmable waveguide. As detailed in Sec.~\ref{sec:experimental_setup}, the beamshaper is able to set both the amplitude as well as the phase of the input field. Thus, we leveraged this capability to calibrate and account for misalignments in the beamshaper. It should be noted that in the future, when on-chip modulators are used for faster input modulation, this calibration will no longer be necessary.

\begin{figure}[!t]
    \centering
    \includegraphics[width=1.0\textwidth]{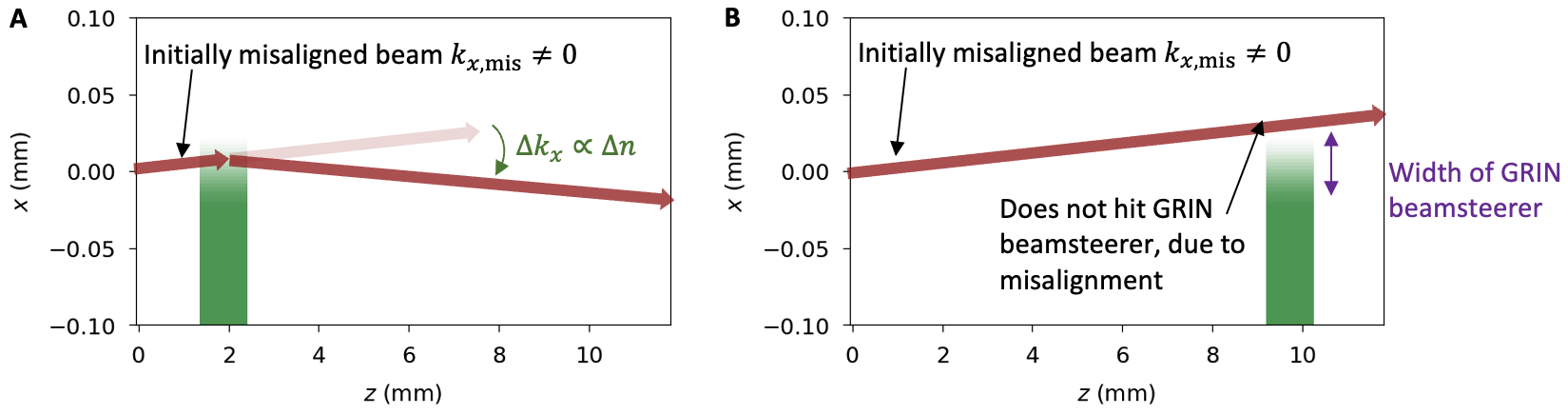}
    \captionsetup{justification=raggedright,singlelinecheck=false}
    \caption{\textbf{Calibration of the programmable 2D-waveguide with a graded index (GRIN) beamsteerer.} As shown schematically in the figure, it can be used to calibrate for misalignments in the beamshaper, as well as to measure system parameters, including the maximal refractive-index modulation $\Delta n_\text{max}$ and the spatial resolution of the chip $d_\text{kernel}$.}
    \label{fig:GRIN_beamsteerer}
\end{figure}

Mathematically, the miscalibrations of the beamshaper can be characterized as follows:
\begin{align}
    \label{eq:beamshaper_miscalibration}
    E_\text{in}(x) = A_\text{mis}(x)\exp\paren{\im{k_{x,\text{mis}} x + \im C_\text{mis} x^2/2}} E_\text{in, set}(x),
\end{align}
where $E_\text{in, set}(x)$ represents the input field that is set to the beamshaper. $A_\text{mis}(x)$, $k_{x, \text{mis}}$ and $C_\text{mis}$ all represents different kinds of misalignments that will be detailed later. Once these misalignments are characterized, \eqref{eq:beamshaper_miscalibration} can be inverted to accurately send in a desired input field. We use the following procedure to calibrate the beamshaper:
\begin{itemize}
    \item The coupling efficiency $A_\text{mis}(x)$ can be characterized by sending in focused Gaussian beam with waist $w_0 \sim \SI{5}{\micro m}$ and different mean input location $x$ to the input facet of the waveguide, and measuring the overall power at the output facet. We found that the transparency window of coupling efficiency (i.e. $A_\text{mis}(x)$ drops when $|x|$ is larger than some value) varies as the slab waveguide is rotated relative to the 1D-axis of the input beam. This misalignment can be corrected by rotating the SLM used in the beamshaper.  Once corrected, $A_\text{mis}(x)$ is mostly flat across the entire field-of-view of the beamshaper, which is about \SI{600}{\micro m} wide for our experiment. 
    \item As the beamshaper involves multiple lenses, there are many distances between lenses that could be misaligned. In practice, we observed that the most relevant misalignment is the distance between lens 2 and 3 not being equal to $f_1$+$f_2$, as this distance changes whenever a new chip is installed into the setup. When this occurs, the input field picks up a quadratic phase front ($C_\text{mis}\neq 0$), which can be understood as a virtual lens being inserted at the input facet. Thus, this lens imparts a wavevector kick that is a linear function of the input location $\Delta k_x = C_\text{mis} x_\text{in}$. To calibrate for this, we send in collimated Gaussian beams ($w_0 \sim \SI{50}{\mu m}$) with different mean input location $x_\text{in}$ and measure how much the mean output location $x_\text{out}$ deviated from the input location. With the formula $x_\text{out} = x_\text{in} + C_\text{mis} x_\text{in} L_z \lambda_0/(2\pi n_0)$, $C_\text{mis}$ can be measured and corrected for. 
    \item To calibrate for $k_{x, \text{mis}}$, which represents the input beams coming in at an angle instead of being straight, we send in a collimated Gaussian beam ($w_0 \sim \si{50}{\mu m}$) and project two different graded index (GRIN) beamsteerer patterns onto the 2D-programmable waveguide. As shown in Fig.~\ref{fig:GRIN_beamsteerer}, when there is a misalignment, the beam will only be steered by the GRIN beamsteerer in the front, and not the one in the back. By comparing with theory, we can measure $k_{x, \text{mis}}$ and correct for it. 
\end{itemize}

\begin{figure}[!t]
    \centering
    \includegraphics[width=\textwidth]{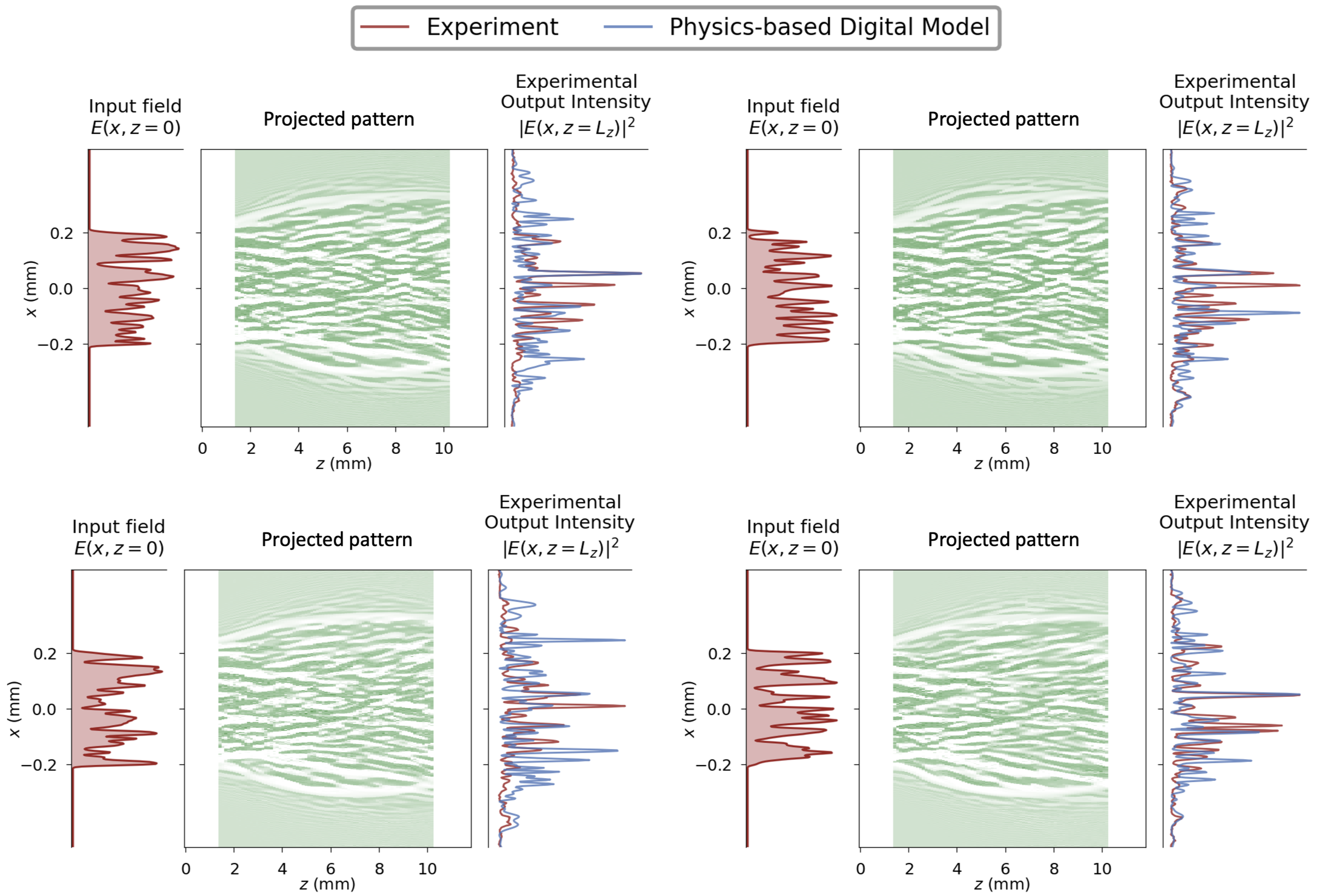}
    \captionsetup{justification=raggedright,singlelinecheck=false}
    \caption{\textbf{Agreement between the output intensity for the experiment and the purely physics-based model.} The agreement is plotted for four different input field distributions and projected patterns. We observe that there is qualitative, but not quantitative agreement between the outputs of the model and experiment.}
    \label{fig:physics_based}
\end{figure}

\textbf{Calibration of programmable 2D-waveguide parameters}:
Once the beamshaper is calibrated, what remains is to characterize the system parameters of the 2D-programmable waveguide. This is also performed by projecting the GRIN beamsteerers onto the device. 
\begin{itemize}
    \item The maximal refractive-index modulation $\Delta n_\text{max}$ can be quantified by measuring the amount of beamsteering exerted by the GRIN beamsteerer. The change in the $x$-component of the wavevector $\Delta k_x$ is linearly proportional to the change in refractive index, and thus $\Delta n_\text{max}$ can be measured by comparing with theory. We characterized that $\Delta n_\text{max}=10^{-3}$ for the experiment in Fig.~2 of the main manuscript, $\Delta n_\text{max}=0.6\times 10^{-3}$ for vowel classification, and $\Delta n_\text{max}=0.8\times 10^{-3}$ for MNIST classification.
    \item The spatial resolution of the chip $d_\text{kernel}$ can be measured by varying the width of the GRIN beamsteerer, measuring the amount of beamsteering, and comparing it to theory. Using this procedure, we found that $d_\text{kernel} = \si{5}{\mu m}$, which is consistent with the theoretical analysis in Sec.~\ref{sec:spatial_resolution}.
\end{itemize}

In Fig.~\ref{fig:physics_based}, we show the agreement between the physics-based model and the experimental data for different input fields and different projected patterns on the 2D-programmable waveguide. While the experimental output and predicted output agrees qualitatively, we observe quantitative differences, especially towards the edges of the output.

\begin{figure}[!t]
    \centering
    \includegraphics[width=\textwidth]{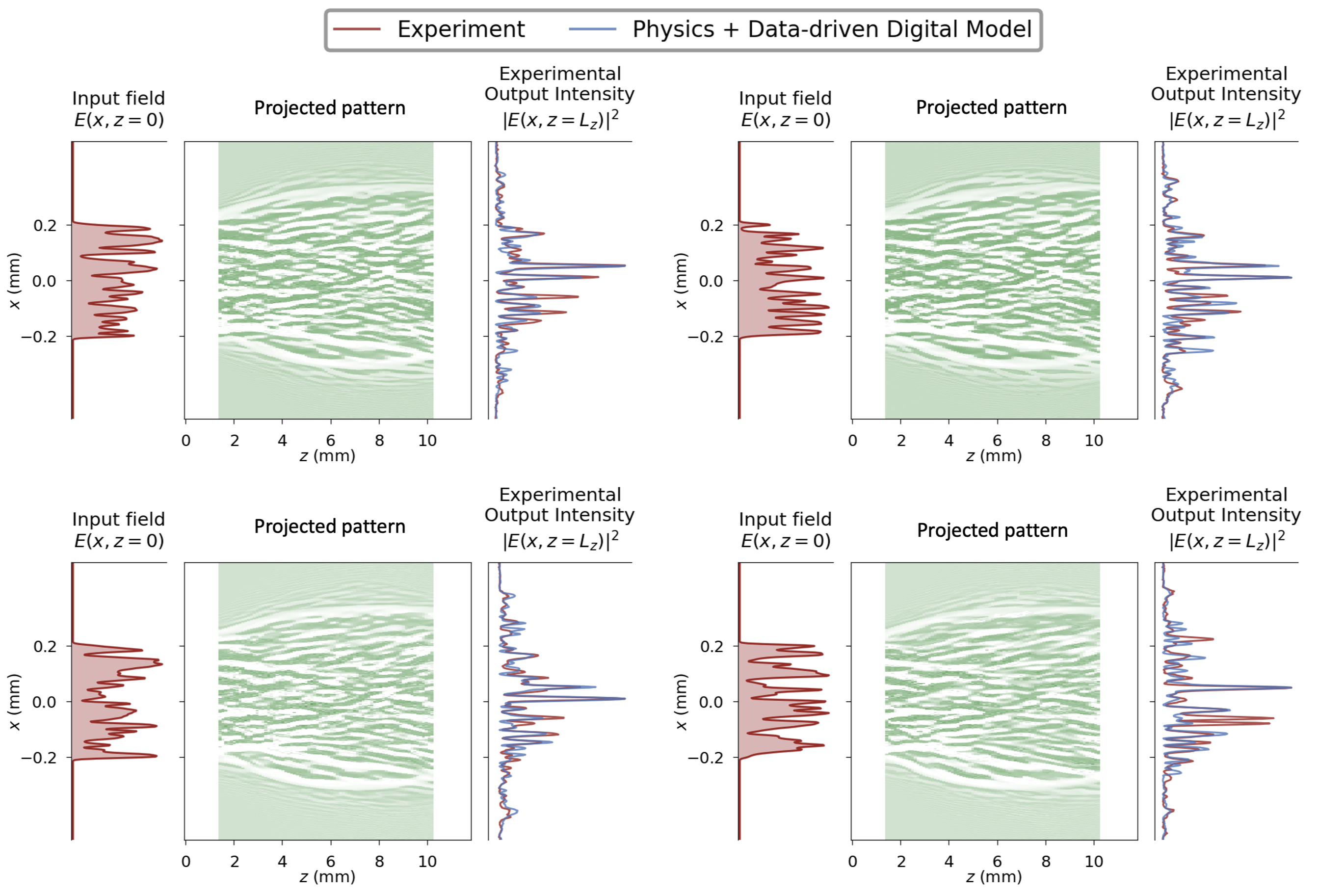}
    \captionsetup{justification=raggedright,singlelinecheck=false}
    \caption{\textbf{Agreement between the output intensity of the experiment and a model that uses data-driven approaches to fine-tune the physics-based model.} The same input field distributions and projected patterns as in Fig.~\ref{fig:physics_based} are shown. We observe that the data-driven refinement improves the agreement between the model and experiment significantly.}
    \label{fig:data_driven}
\end{figure}

\subsection{Data-driven approach to fine-tune the digital model}
To further improve the agreement between the model and the experiment, we used a data-driven approach to fine-tune the physics-based model. To do so, we added additional trainable parameters into the physics-based model and trained them with input-output pairs that are generated from the experimental data. 

The model still solves the beam propagation equation \eqref{eq:beam_propagation}, but now with the refractive-index modulation being
\begin{align}
\label{eq:data_driven}
    \Delta n(x, z) = \frac{\Delta n_{\text{max}}}{I_{\text{max}}} \paren{g(x, z) \ast (I(x,z)\times r(x,z))} + \Delta n_{\mathrm{noise}}(x, z).
\end{align}
Here, $\ast$ represents the 2D convolution operation, $\Delta n_\mathrm{noise}$ denotes a background noise refractive-index distribution, and $r(x,z)$ is another trainable parameter that quantifies the spatial dependence of $\Delta n$ induced by a given amount of illumination. The other variables in \eqref{eq:data_driven} are defined in the text following \eqref{eq:beam_propagation}. We hypothesize that $\Delta n_\mathrm{noise}$ may be attributed to charge noise in the lithium niobate slab waveguide, previously observed in experiments involving light propagation through lithium niobate, which resulted in effects such as wavefront damage \cite{schwesyg2011beam}. The parameter $r(x, z)$ accounts for non-uniform illumination in the imaging setup, specifically referring to the spatial variance in power absorbed by the photoconductor across the field-of-view of the 2D-programmable waveguide, even when $I(x,z)$ is intended to be uniform. Additionally, $r(x, z)$ accounts for film thickness variations in the device stack and the potential presence of small domains within lithium niobate where the crystal axis may be flipped \cite{Paturzo2005LN-defects-internal-field}. In summary, both $\Delta n_{\mathrm{noise}}$ and $r(x,z)$ allow the model to address nonidealities associated with lithium niobate, the experimental setup, and the device fabrication.

Finally, we added a trainable output coupling efficiency term $B_\text{mis}(x)$ to model for imperfections in our measurement of the output intensity  ($I_\text{meas, out}(x) = B_\text{mis}(x)|E(x, z=L_z)|^2$). We also made the previously introduced input coupling efficiency term $A_\text{mis}(x)$ in \eqref{eq:beamshaper_miscalibration} a trainable parameter. 

To collect the training data for the data-driven model, an important detail is to obtain the right distribution of projected patterns that are encountered during machine learning tasks. To do so, we in-silico trained the 2D-programmable waveguide on multiple different variations of the MNIST classification task, with permuted inputs and outputs. This way, we obtained a large number of distinct projected patterns, which we then used to train the data-driven model. For the input fields, we sent in input field distributions with random amplitude and phase in a pixel mode basis. In other words, they were generated with \eqref{eq:mode_basis_input} with $x_i$ being random in both amplitude and phase. Using many experimentally measured input-output pairs, we trained all of these additional parameters ($\Delta n_\mathrm{noise}(x, z)$, $r(x, z)$, $B_\text{mis}(x)$, $A_\text{mis}(x)$) on a digital computer with the backpropagation algorithm.   

In Fig.~\ref{fig:data_driven}, we show the agreement between the experimental data and the physics-based model that is refined with a data-driven approach. We see that the agreement has improved significantly compared to the purely physics-based model, especially at the edges of the output (at large $|x|$).

\section{Machine learning}
In this section, we present additional information regarding the machine learning tasks that we performed on the 2D-programmable waveguide. 

\subsection{Computational model of optical-neural-network demonstrations}
\label{sec:computational_model}

\begin{figure}[!b]
    \centering
    \includegraphics[width=\textwidth]{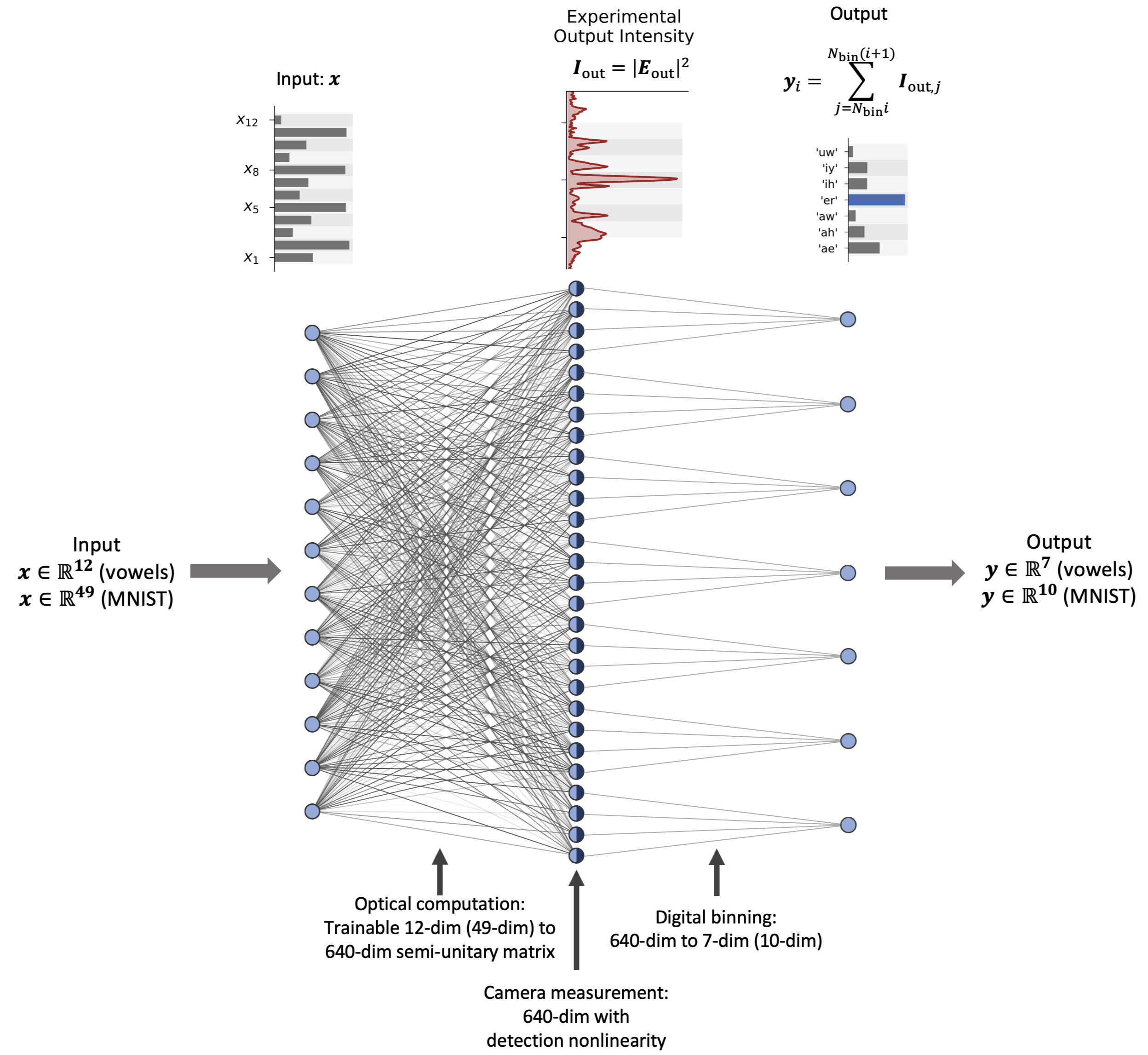}
    \caption{
    \label{fig:equivalent-computational-model}
    \textbf{Schematic for the computational model of the ONN demonstrations in this work.}}
\end{figure}

In this section, we explain the computational model of the optical neural network demonstrations performed by the 2D-programmable waveguide. For concreteness, we focus on the vowel classification task, which has a 12-dimensional input and 7-dimensional output, and the MNIST classification task, which has a 49-dimensional input and 10-dimensional output. At the end of the section, we argue that the computation performed by the 2D-programmable waveguide is similar to that of a 1-layer neural network, and we use this to characterize the computational complexity of the device.

The input vector $\mathbf{x}$ is encoded in the amplitudes of 12 (49 for the MNIST task) spatial Gaussian modes via \eqref{eq:mode_basis_input}. Since each Gaussian mode is normalized, this can be abstracted as a unitary relation: $\mathbf{E}_\text{in} = \mathbf{U}_\text{in} \mathbf{x}$, where $\mathbf{E}_\text{in}$ represents the input field in a pixel basis. Due to the low propagation loss, the field propagation through the device is also a unitary operation, that takes the input field $\mathbf{E}_\text{in}$ and returns the output field $\mathbf{E}_\text{out}$. Thus, the output field is given by $\mathbf{E}_\text{out} = \mathbf{U}_\text{prop} \mathbf{E}_\text{in}$. In this notation, we see that by training the refractive-index distribution, we are able to change the unitary matrix $\mathbf{U}_\text{prop}$ applied by the device $\mathbf{U}_\text{prop}(\Delta n(x, y))$. Moreover, since the first two operations can be collapsed into a single unitary matrix, the device's overall operation can be expressed as $\mathbf{E}_\text{out} = \mathbf{U} \mathbf{x}$, where $\mathbf{U} = \mathbf{U}_\text{prop} \mathbf{U}_\text{in}$. Here, $\mathbf{U} \in \mathbb{C}^{640\times12}$ ($\mathbf{U} \in \mathbb{C}^{640\times49}$ for MNIST), as we discretize the output field into a 640-dimensional vector defined by the number of pixels on the camera used to measure the output.

We model the measurement of the intensity at the output facet by a quadratic nonlinearity applied to the optical field: ${\mathbf{I}_\mathrm{out} \propto |\mathbf{Ux}|^2}$. After the camera measurement, we digitally bin groups of camera pixels---average pooling in the language of machine learning---to create a lower-dimensional output of 7 (10 for MNIST). 
In summary, the computation performed can be expressed as
\begin{align}
    \mathbf{y} = \text{AvgPool}(|\mathbf{Ux}|^2)
    \label{eq:input-output-relation}
\end{align}
where the matrix vector multiplication $\mathbf{Ux}$ is performed optically, the nonlinear activation function is performed by the analog electronics of the camera, and the fixed average pooling is performed by a digital computer. This relation is also shown schematically in Fig.~\ref{fig:equivalent-computational-model}. We note that the entire computation could be performed without a digital computer, for example, by using a lens to optically perform the fan-in operation, and using a photodetector for each class of the classification task.

We emphasize that although we can train the unitary matrix $\mathbf{U}$ by programming the refractive-index distribution, it is not possible to realize an arbitrary $\mathbf{U}$. This limitation is physically intuitive as the number of parameters in $\mathbf{U}$ (125440 for the MNIST task) exceeds the number of parameters available in the 2D-programmable waveguide (10,000).

While the input-output relation defined in \eqref{eq:input-output-relation} is mathematically equivalent to a 2-layer neural network, the computation it performs is similar to that of a 1-layer neural network due to the second layer being sparse and not trainable, and the square nonlinearity not being significantly nonlinear (compared to activation functions such as ReLu and Sigmoid). An alternative argument is that if the square nonlinearity were replaced with an identity operation, \eqref{eq:input-output-relation} would indeed reduce to a 1-layer neural network. Therefore, we characterize the complexity of the computational operation performed by the 2D-programmable waveguide by the number of operations performed by the effective 1-layer neural network, rather than the number of operations in $\mathbf{U}$.

The most complex machine learning task we have undertaken with the 2D-programmable waveguide is the MNIST task, where it achieves an 86\% accuracy, moderately close to the 90\% accuracy achievable by a 1-layer digital neural network (with 49 inputs and 10 outputs). Hence, we characterize the device as capable of performing 490 multiply-accumulate operations in a single pass through the chip.

\subsection{Vowel classification}
\label{sec:vowel-classification}

\textbf{Experiment Settings}: The frequency of the AC voltage drive is set to $\SI{26}{Hz}$, to increase the speed of data collection. The amplitude of the AC voltage was set to $\SI{800}{V}$ instead of $\SI{1000}{V}$ (as in Fig.~2 of the main manuscript), because this experiment was performed at an earlier stage of the project, when we used a more conservative voltage buffer for the device. These two choices led to a decrease in the maximal refractive-index modulation $\Delta n_\text{max}$ to $0.6\times 10^{-3}$, compared to $1\times 10^{-3}$ shown in Fig.~2 of the main manuscript.

\begin{figure}[!t]
    \centering
    \includegraphics[width=1.0\textwidth]{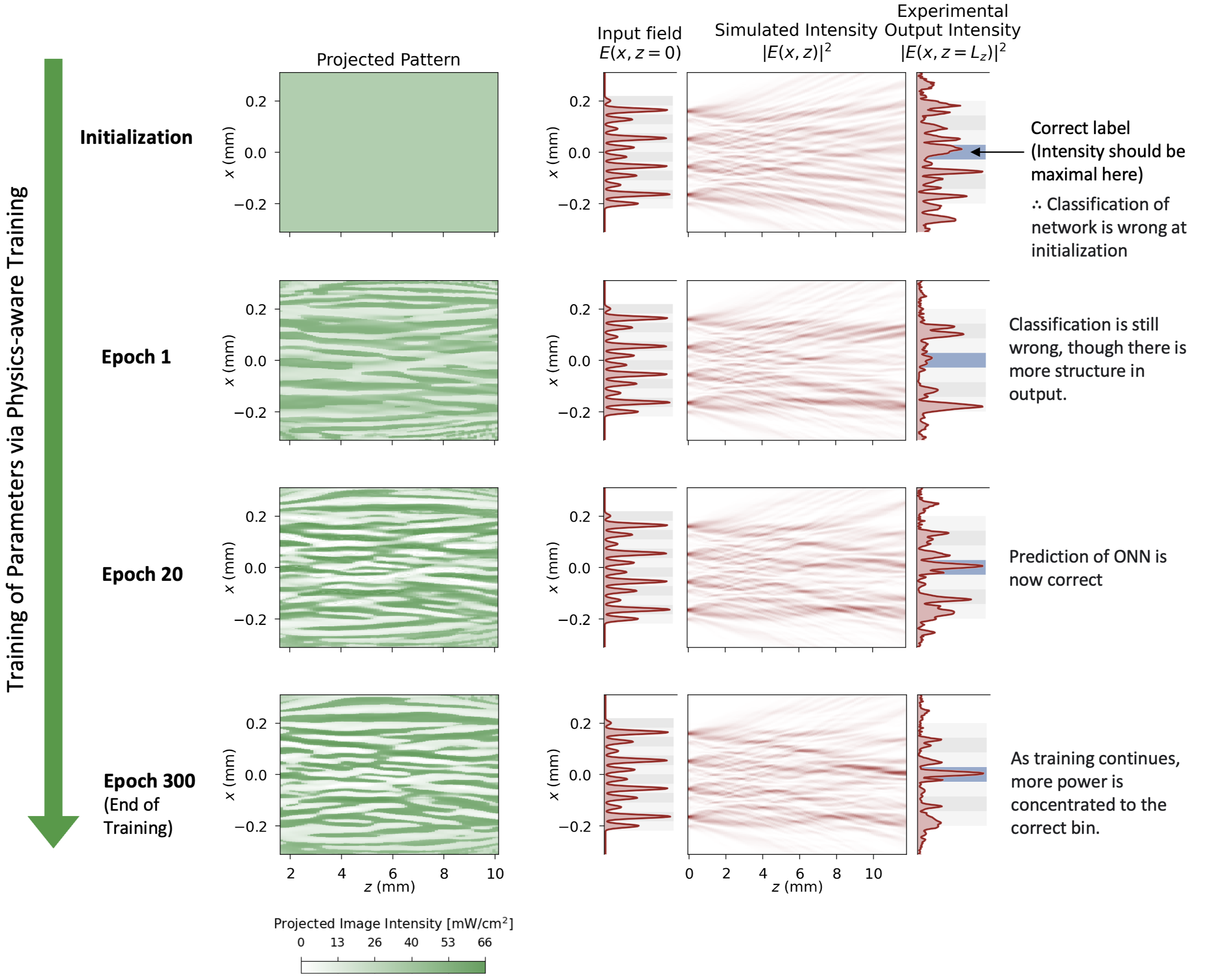}
    \captionsetup{justification=raggedright,singlelinecheck=false}
    \caption{\textbf{Evolution of wave propagation during physics-aware training.}}
    \label{fig:training_propagation}
\end{figure}

\textbf{Additional Results}: In Fig.~\ref{fig:training_propagation}, we show how the wave propagation evolved during the training process. The 4 projected patterns shown on the left correspond to the same projected patterns as shown in Fig.~3 of the main manuscript. The same input vector is also used for consistency, whose correct label is the vowel `er', which corresponds to the fourth bin in the output.  We observe that at initialization, the projected pattern was uniform, and the output intensity was unstructured uniformly spread across all of the 7 bins. After just 1 epoch of training, the wave propagation was already significantly altered, where there is more structure in the output intensity. 

We emphasize that small changes (to the visual eye) in the projected pattern can lead to large changes in the wave propagation. This is particularly evident in the case of the last two projected patterns, at Epoch 20 and Epoch 300. Although these patterns appear similar, the wave propagation differs significantly. More power is concentrated in the bin that corresponds to the correct vowel by Epoch 300. This highlights the importance of an \emph{in-situ} programmable device for controlling complex wave propagation.

\subsection{MNIST classification}
\label{sec:mnist-classification}

\textbf{Experiment Settings}: The frequency of the AC voltage drive is set to $\SI{26}{Hz}$, to increase the speed of data collection. The amplitude of the AC voltage was set to $\SI{1000}{V}$, consistent with Fig.~2 of the main manuscript. Therefore, there is a slight decrease in the maximal refractive-index modulation $\Delta n_\text{max}$ to $0.8\times 10^{-3}$, compared to $1\times 10^{-3}$ shown in Fig.~2 of the main manuscript.

\begin{figure}[!t]
    \centering
    \includegraphics[width=0.85\textwidth]{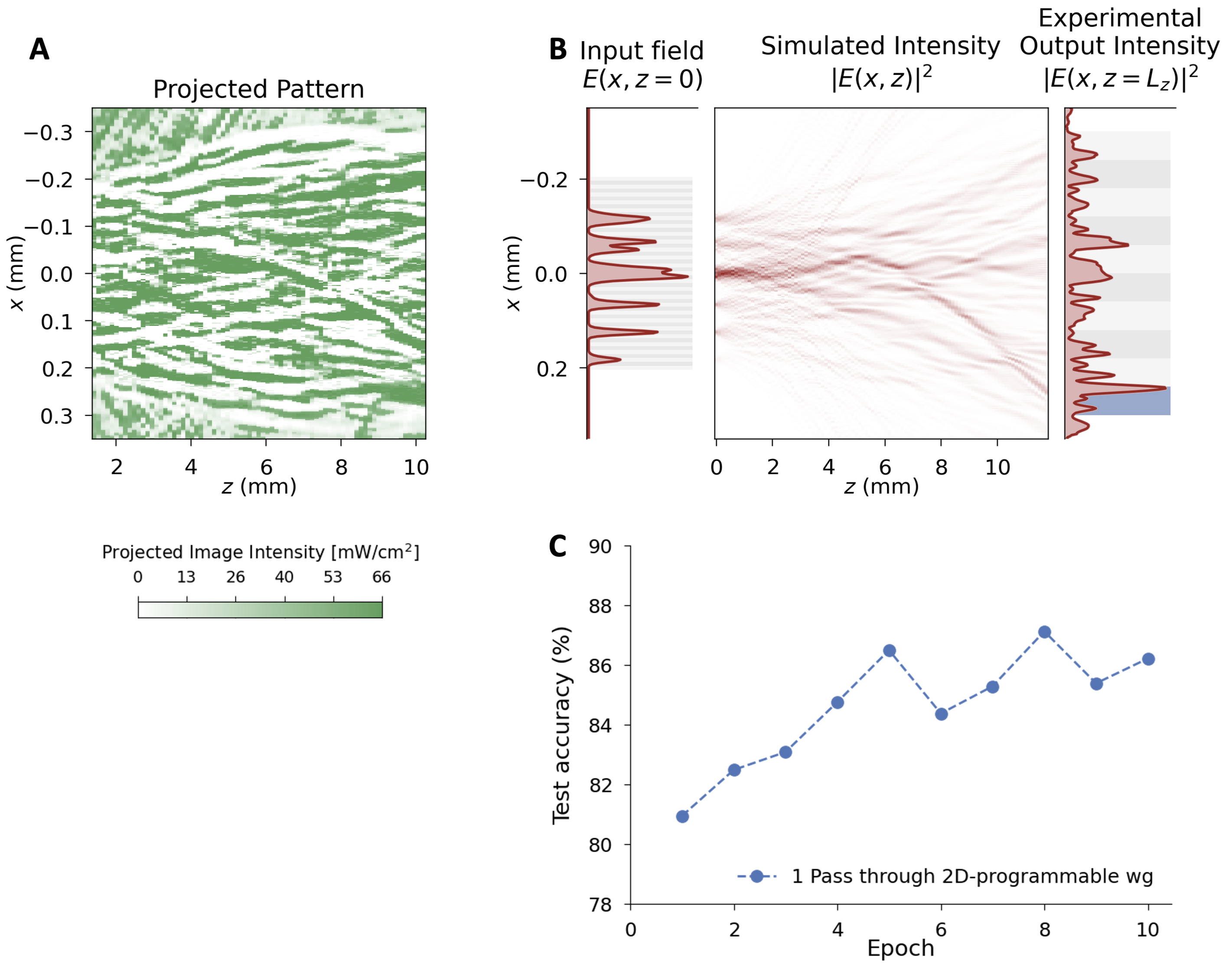}
    \captionsetup{justification=raggedright,singlelinecheck=false}
    \caption{\textbf{Detailed MNIST classification results.} \textbf{A}: Projected pattern after 10 epochs of training. \textbf{B}: Illustration of wave propagation in 2D-programmable waveguide for the particular input MNIST image shown in Fig.4 of main manuscript.  \textbf{C}: The accuracy of the MNIST classification task on the test set as a function of the number of training epochs.}
    \label{fig:MNIST}
\end{figure}

\textbf{Additional Results}: 
In Fig.~\ref{fig:MNIST}, we show additional results regarding the MNIST classification task. Fig.~\ref{fig:MNIST}A shows the parameters of 2D-programmable waveguide, the projected pattern, after training. When compared to the projected pattern for the vowel classification, we observe that more of the pattern takes on more extreme values, where many pixels are at maximum or minimal intensity. This suggests that the MNIST task is more difficult than the vowel classification task, pushing the capabilities of the device closer to its limit.

In Fig.~\ref{fig:MNIST}B, we show the wave propagation for the particular input MNIST image shown in Fig.4 of the main manuscript. The wave propagation is more complex than that in the vowel task. We note that we chose the width of the Gaussian input modes ($w_0$ in \eqref{eq:mode_basis_input}) to be \SI{6}{\micro m}, to limit the diffraction of the modes, while the spacing between the modes is \SI{8.2}{\micro m}. Thus, the input modes used in \eqref{eq:mode_basis_input} overlap, resulting in the input field shown in Fig.~\ref{fig:MNIST}B. This is not a fundamental limitation and will be addressed in future work.

Finally, in Fig.~\ref{fig:MNIST}C, we present the training curve for the MNIST classification task. We observe fluctuations in accuracy; for instance, at epoch 8, it reached 87.2\%. The accuracy is 86.2\% after 10 epochs of training, compared to the 90\% accuracy achieved by a 1-layer digital neural network on the same 49-dimensional MNIST task.

\section{Comparison with other on-chip optical neural network demonstrations}
\label{sec:comparison_ONNs}

In this section, we compare our work with other on-chip optical neural network demonstrations\footnote{We have not included references, such as Ref.~\cite{ramey-lightmatter-2020silicon}, if they don't report machine-learning-inference accuracy.}, focusing particularly on demonstrations that perform matrix-vector multiplication (MVM) with a weight-stationary approach, where the matrix elements are encoded in programmable elements on the chip.

\begin{figure}[!t]
    \centering
    \includegraphics[width=\textwidth]{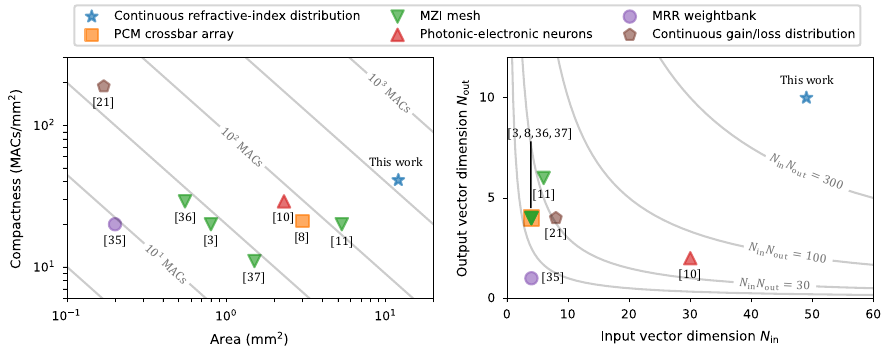}
    \captionsetup{justification=raggedright,singlelinecheck=false}
    \caption{\textbf{Comparison between different on-chip optical neural network demonstrations.} Data from Table~\ref{tab:comparison_ONNs}. Left: Area and compactness of various on-chip ONNs. Grey lines are contours of total MAC operations, which represents the number of multiply-accumulate operations executed in a single pass through the device. Right: Input and output vector dimension of various on-chip ONNs. Grey lines are contours of constant $N_\text{in}\times N_\text{out}$, which corresponds to the number of MAC operations for single-layer ONNs.}
    \label{fig:sm-comparison-plot}
\end{figure}

We acknowledge that various metrics can be used to compare the performance of different on-chip optical neural network demonstrations. Here, we focus on the number of multiply-accumulate operations (MACs) executed in a single pass through the device. For single-layer neural networks, the number of MAC operations is determined by the product of the input and output dimensions of the MVM operation. In multi-layer neural networks, it is the sum of each layer's MAC operations. For ONNs utilizing frequency multiplexing, the number of MAC operations is multiplied by the number of channels, $N_\text{ch}$, used for frequency division multiplexing. Additionally, we compare the devices' compactness, measured by the number of MAC operations per unit area. Here, ``area'' specifically refers to the portion of the chip where light propagates to perform the MVM and excludes regions dedicated to electrical wiring and bond pads. In Table.~\ref{tab:comparison_ONNs}, we summarize the comparison between our work and other on-chip optical neural network demonstrations, and plot some of the key metrics in Fig.~\ref{fig:sm-comparison-plot}.

\begin{table}[!t]
  \centering
  \begin{tabular}{|c|c|c|c|c|c|c|c|c|}
    \hline
    Reference & MVM Scheme & $N_\text{layer}$ & $N_\text{ch}$ & $N_\text{in}$ & $N_\text{out}$ & \makecell{MACs} & \makecell{Area \\ {[\si{mm^2}]}} & \makecell{Compactness \\ {[}MACs/\si{mm^2}{]}}\\ 
    \hline
    This work & Continuous refractive-index distribution & 1 & 1 & 49 & 10 & 490 & 12 & 41 \\ 
    \hline
    \cite{shen2017natphoton} & MZI mesh & 1 & 1 & 4 & 4 & 16 & 0.8 & 20 \\
    \hline
    \cite{Zhang-Liu2021NatureComms} & MZI mesh & 1 & 1 & 4 & 4 & 16 & 0.55 & 29 \\
    \hline 
    \cite{bandyopadhyay2022single} & MZI mesh & 3 & 1 & 6 & 6 & 108 & 5.3 & 20 \\
    \hline 
    \cite{pai2023experimentally} & MZI mesh & 1 & 1 & 4 & 4 & 16 & 1.5 & 11 \\
    \hline 
    \cite{ashtiani2022nature} & Photonic-electronic neurons & 3 & 1 & 30 & 2 & $66^\dagger$ &  2.3 & 29 \\
    \hline 
    \cite{feldmann2021nature} & PCM crossbar array & 1 & 4 & 4 & 4 & 64 & 3 & 21 \\
    \hline 
    \cite{huang2020demonstration} & Microring resonator weightbank & $1^*$ & 1 & 4 & 1 & 4 & 0.2 & 20 \\
    \hline
    \cite{wu-feng2023natphoton} & Continuous gain/loss distribution & 1 & 1 & 8 & 4 & 32 & 0.17 & 188 \\
    \hline
\end{tabular}
  \caption{\textbf{Comparison between different on-chip optical neural network demonstrations.} We focus on neural-network-inference demonstrations that perform matrix-vector multiplication (MVM) on-chip with a weight-stationary approach, where the matrix elements are encoded in programmable elements on the chip. $N_\text{layer}$ is the number of layers (depth) in the ONN. $N_\text{ch}$ is the number of channels used for frequency division multiplexing. $N_\text{in}$ and $N_\text{out}$ are the input and output dimensions of the MVM operation for single-layer neural networks. For multi-layer neural networks with a more complex architecture (such as \cite{ashtiani2022nature}), $N_\text{in}$ and $N_\text{out}$ represent the dimensions of the input and output vectors of the overall ONN. MACs denote the number of multiply-accumulate operations executed in a single pass through the device. MZI and PCM stand for Mach--Zehnder interferometer and phase-change memory, respectively. }
  \label{tab:comparison_ONNs}
\end{table}

\footnotetext[1]{Multilayer ONNs have been demonstrated in a related work \cite{huang2021silicon} on this platform.}
\footnotetext[2]{The first layer of this multilayer ONN performs a convolution operation on the input image. Since it performs dot products on 4 sub-images, each with 12 pixels, we have characterized this first layer as executing 48 MAC operations.}

Table \ref{tab:comparison_ONNs} shows that the 2D-programmable waveguide has demonstrated the largest input and output vector dimensions, as well as the highest number of MAC operations among the on-chip optical neural network demonstrations reviewed. In terms of compactness, the 2D-programmable waveguide is also highly competitive, ranking second only to Ref.~\cite{wu-feng2023natphoton}. This work shares many parallels with ours, notably in its avoidance of discrete programmable photonic elements and in leveraging a continuous, spatially programmable slab waveguide (in their case a spatial gain/loss distribution) to perform neural-network computations. A notable difference between our work and that in Ref.~\cite{wu-feng2023natphoton} is in the type of light used for computation. We use coherent light, enabling complex-valued matrix-vector multiplication. In contrast, their approach uses incoherent light, which is more robust to phase noise, but limits the applied matrix to real and non-negative elements. 

\section{Future device improvements to the 2D-programmable waveguide}

In this section, we discuss potential device improvements to the 2D-programmable waveguide and their implications for neural-network computation. We discuss different directions to improve the device, such as incorporating all-optical nonlinearity, increasing the refractive-index modulation depth, and potential routes to a fully integrated 2D-programmable waveguide. We also discuss how the device can be modified to realize a programmable nonlinear medium.  Finally, we present simulation results of performing unitary matrix operations with a prospective, scaled up 2D-programmable waveguide.

\subsection{Increasing the depth of the neural network computation}
\label{sec:increasing-depth}

In our current work, our chip performs a single-layer neural-network computation (i.e., having a depth of 1). Increasing the depth of the neural network run in a single pass through the chip would require performing nonlinear activation functions on-chip \cite{ashtiani2022nature, bandyopadhyay2022single, huang2021silicon}. Given that our chip's slab waveguide already uses a nonlinear material, lithium niobate, one could periodically pole the lithium niobate in certain regions of the chip, enabling nonlinear interaction between optical waves in those regions, which can be used to engineer all-optical nonlinear activation functions \cite{Li-marandi2022Nanophotonics}. Alternatively, multilayer computation can be performed by propagation of waves in a continuous nonlinear medium \cite{Nakajima-Hashimoto-NTT2022neuralSE, Hughes-fan2019SciAdv, khoram2019photonres}, since programmable nonlinear wave propagation can be mathematically mapped to a deep multilayer neural network \cite{Nakajima-Hashimoto-NTT2022neuralSE, Hughes-fan2019SciAdv, Tegin2021scalable}. A power-efficient realization of this approach in our platform could involve using an appropriately poled lithium niobate slab waveguide with a cascaded second-order nonlinearity \cite{Cui-Fan2022cascaded}.

\subsection{Increasing the refractive-index modulation}
\label{sec:increasing-delta-n}
A weakness of our current demonstration device is the relatively small refractive-index modulation ($\Delta n \sim 10^{-3}$) compared to conventional lithographically defined nanophotonics ($\Delta n \sim 1$): the refractive-index modulation one can achieve in any device with our design is ultimately constrained by the material properties of the waveguide core: its electro-optic coefficient and the maximum electric field it can sustain. With lithium niobate, the modulation we measured, $\Delta n \sim 10^{-3}$, could be improved to $\Delta n = 10^{-2}$ by increasing the thickness and improving the material properties of the photoconductor (see Section~\ref{sec:photoconductor-design}). However, using an alternative electro-optic material for the waveguide, such as barium titanate \cite{tang-wessels2005BaTiO3}, highly nonlinear polymers \cite{koeber-koos2015femtojoule}, or liquid crystals \cite{davis2010liquid}, could allow future realizations to achieve a refractive-index modulation of $\Delta n \sim 5 \times 10^{-2}$ or larger.

\subsection{Realizing a programmable nonlinear medium}
We could also modify our concept of a 2D-programmable waveguide to realize an arbitrarily programmable \textit{nonlinear} medium, where the second-order nonlinear optical susceptibility $\chi^{(2)}(x, z)$ could be programmed in real time. This may be achieved by using a waveguide core material with a large third-order nonlinearity $\chi^{(3)}$ (such as silicon \cite{Leuthold2010silicon}, silicon nitride \cite{blumenthal2018SiN}, or tantalum pentoxide \cite{Jung2021tantala}). Using the principle of photoconductive gain, we can spatially program a strong (quasi-DC) electric field $E_{\text{DC}}(x, z)$ in the waveguide core that can induce an effective second-order nonlinearity $\chi^{(2)}_\text{eff}(x, z)$ \cite{timurdogan2017electric}. Such induced second-order nonlinearities can be quite strong: Ref.~\cite{timurdogan2017electric} demonstrated that the nonlinearity induced in silicon ($\chi_\text{eff}^{(2)} = \SI{41}{pm/V}$) is comparable to that of lithium niobate ($\chi^{(2)} = \SI{50}{pm/V}$). This field-induced second order nonlinearity can be thought of as the higher-order equivalent of the linear refractive-index modulation \eqref{eq:e-o-effect}, and is mathematically described by
\begin{align}
    \chi^{(2)}_\text{eff}(x, z) = 3\chi^{(3)} E_\text{DC}(x, z)
\end{align}
By programming $\chi_\text{eff}^{(2)}(x, z)$, one could reconfigurably phase-match different nonlinear optical processes in real time. This enables the development of programmable nonlinear optical devices, including highly tunable classical and quantum light sources, and more generally, enables programmable nonlinear processing of optical signals.

\subsection{Potential routes to a fully integrated 2D-programmable waveguide}
\label{sec:fully-integrated}

In this section, we discuss potential routes to a fully integrated 2D-programmable waveguide that would be more compact. As discussed in Section~\ref{sec:experimental_setup}, the components occupying the most space in the current experiment are the beamshaper, the detection setup, and the projector. Thus, we address how to integrate these components on-chip in this section.

\begin{figure*}[t!]
    \centering
    \includegraphics[width=0.85\textwidth]{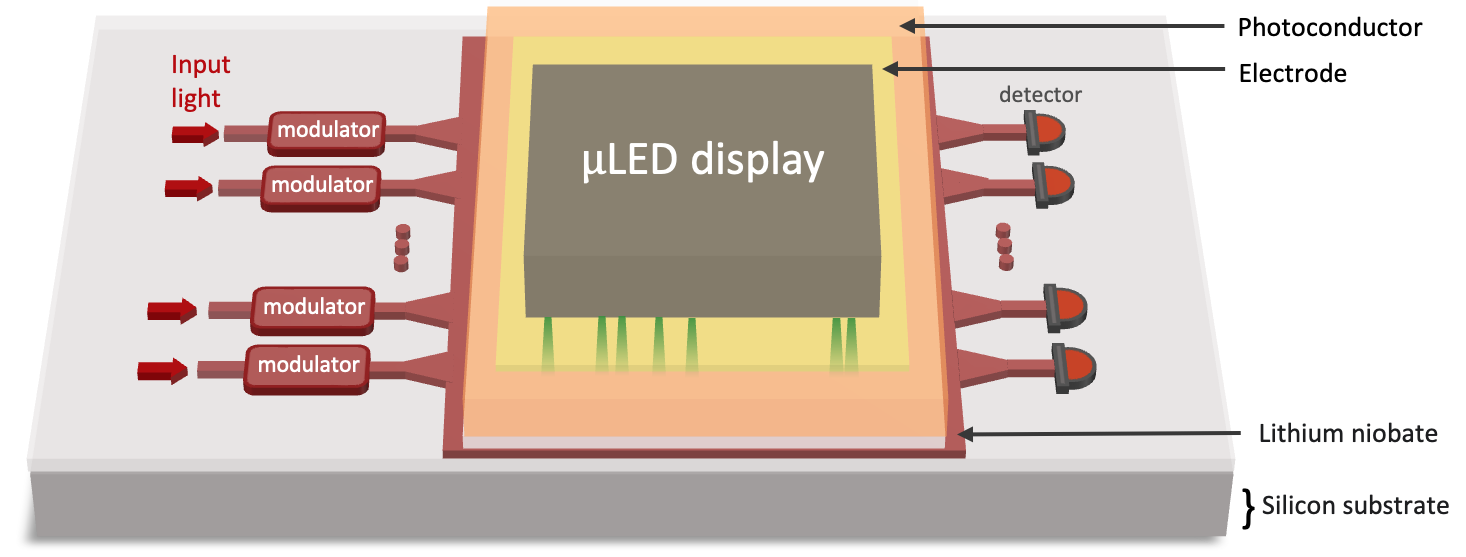}
    \captionsetup{justification=raggedright,singlelinecheck=false}
    \caption{\textbf{Conceptual schematic of a future vision for a fully integrated 2D-programmable waveguide}. The device, which can be built in the future to miniaturize many of the components in the current setup, consists of an array of on-chip lithium niobate modulators to create the shaped input light, a region of the chip with the programmable refractive-index distribution, and an array of on-chip photodetectors to measure the output of the computation. The photoconductor is programmably illuminated by a micro-LED display, which is bonded onto the electrode of the 2D-programmable waveguide.}
    \label{fig:fully_integrated_device}
\end{figure*}

Fig.~\ref{fig:fully_integrated_device} shows a conceptual schematic of a fully integrated 2D-programmable waveguide that could be built in the future. Input fields can be created with an array of on-chip lithium niobate modulators. Due to the strong electro-optic effect in lithium niobate, these modulators have been shown to have high bandwidth at low operating voltages \cite{Wang-loncar2018nature}. These inputs can be directed into a region of the chip featuring the programmable refractive-index distribution. It is important to fan out the waveguide mode to a larger width before entering this region, as the programmable 2D-waveguide region can only control beams with a finite spatial bandwidth $\Delta k_x$, due to the low numerical aperture associated with the maximum refractive-index modulation in the 2D-programmable waveguide.  The detection setup can be replaced with an array of on-chip photodetectors. Although integrating photodetectors on lithium niobate is generally challenging, there has been progress, including heterogeneously integrated III-V materials \cite{zhang2022heterogeneous}, as well as the integration of 2D-material \cite{zhu-wang2023waveguide}, and tellurium thin films \cite{zhu-wang2023waveguide, ahn-jelena2023platform}. High bandwidths of over 40GHz have been demonstrated with these approaches \cite{ahn-jelena2023platform}. For both the modulator and the detector, it is crucial to add the capability to lithographically etch the lithium niobate layer in the 2D-programmable waveguide. This should be feasible by depositing the photoconductive layers (via PECVD) onto a designated region, separate from the region where the lithium niobate is etched for modulators and detectors.

In our current experiment, we use a DMD, illuminated by a green LED to project a programmable illumination pattern onto the 2D-programmable waveguide. As shown in Fig.~\ref{fig:fully_integrated_device}, this programmable illumination can be projected in a more compact manner by bonding a micro-LED display (\textmu-LED) directly onto the electrode of the 2D-programmable waveguide. Among the different display technologies, which include LCDs, organic LEDs (OLEDs), and micro-LEDs, micro-LEDs are the most promising, because they offer high brightness and small pixel pitches. In fact, displays with 30,000 PPI and a brightness of 100,000 nits have been demonstrated \cite{liu2020micro}. This corresponds to a pixel pitch of 0.87 $\si{\micro m}$, and assuming that 50\% of the light emitted is absorbed by the photoconductor, this corresponds to an optical intensity of $45\si{mW/cm^2}$. Thus, such micro-LEDs can deliver sufficient optical power to the photoconductor, for the operation of the 2D-programmable waveguide (see Fig.~\ref{fig:delta_n_measurements}). The key technical challenge will be in getting the LEDs close enough to the waveguide to ensure that the emitted light does not diffract significantly before it is absorbed by the photoconductor. In order to do so, the protective layer (encapsulation) over the LEDs will first need to be either removed or partially etched down to a thin layer. Alternatively, micro-LEDs with monolithically integrated micro-lenses can be used \cite{choi2004GaN}.

\subsection{Simulations of unitary matrices with a prospective, scaled-up 2D-programmable waveguide}
\label{sec:unitary-sim}

\begin{figure*}[!t]
    \centering
    \includegraphics[width=0.95\textwidth]{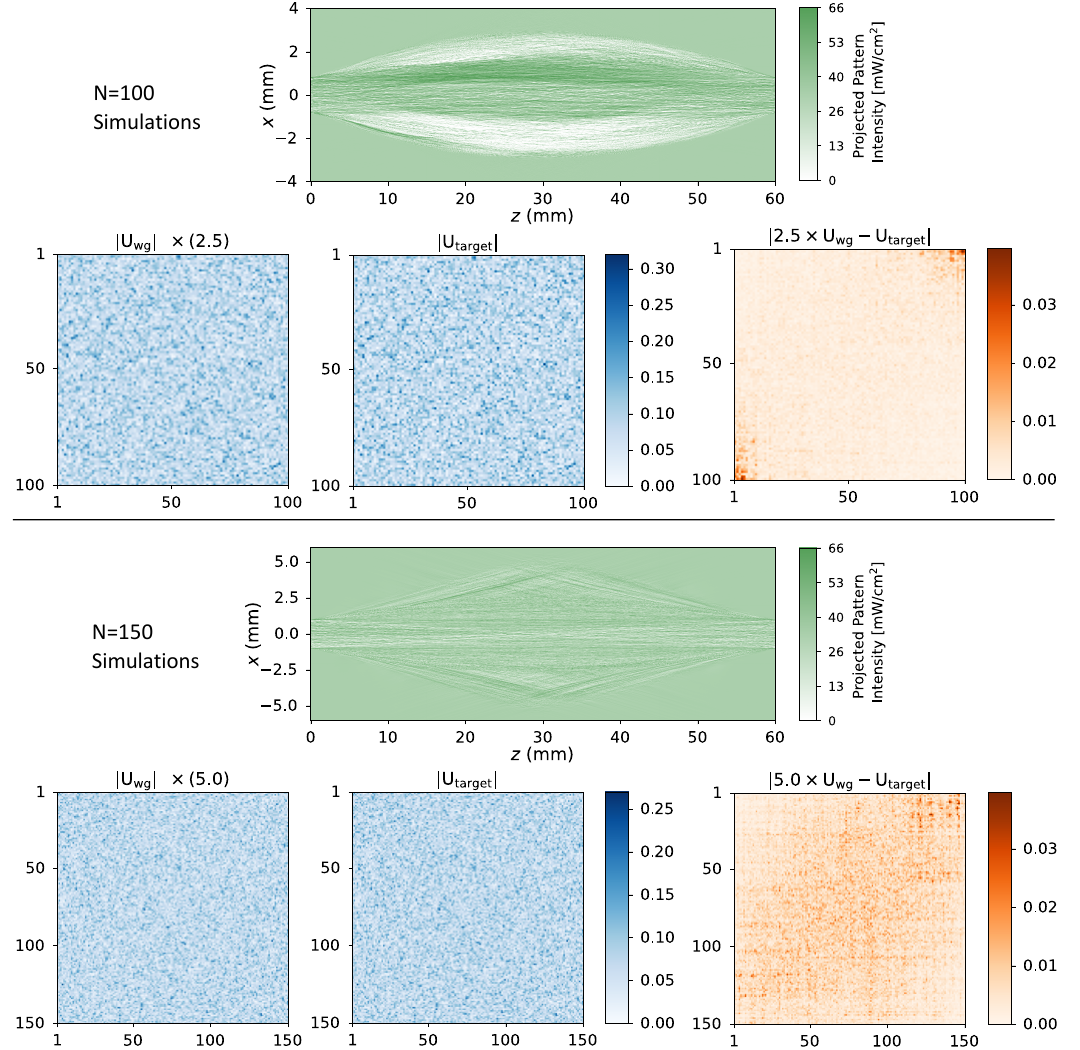}
    \captionsetup{justification=raggedright,singlelinecheck=false}
    \caption{\textbf{Simulations of large-scale unitary matrix operations with a prospective, scaled-up 2D-programmable waveguide}. We train the waveguide to perform unitary matrix operations on input and output vectors of dimensions $N=100$ and $150$. The target unitary matrices are random unitaries sampled from the Haar measure. $U_\text{wg}$ is the matrix operation performed by the 2D-programmable waveguide after training. We note that the multiplicative factor (2.5 for $N=100$, and 5 for $N=150$) is a rescaling factor, indicating that the matrix operation is performed with optical loss (loss of 84\% for $N=100$ and $96\%$ for $N=150$). In these simulations, we assume a $\Delta n_\text{max} = 5\times 10^{-3}$, five times the experiment's current value, and a chip length, $L_z=\SI{6}{cm}$, also five times longer than our current experiment.}
    \label{fig:unitary}
\end{figure*}

In this section, we present simulations of performing large-scale unitary matrix operations with a 2D-programmable waveguide that has a length of $\SI{6}{cm}$ and a maximal refractive-index modulation of $5\times 10^{-3}$. Both of these parameters are five times larger than those of our current experiment ($L_z=\SI{1.2}{cm}$ and $\Delta n_\text{max}=1\times 10^{-3}$).

In order to perform unitary operations, we decompose the output field in discrete output modes. Thus, we do not use the output decoding used in the experiments, where we measure and bin the output intensity distribution, but instead assume that there are output waveguides that filter the output field into discrete modes (this is also the output encoding shown in Fig.~\ref{fig:fully_integrated_device}). We perform simulations for input and output vectors of dimensions $N=100$ and $150$, for random unitaries sampled from the Haar measure. The input and output modes are assumed to be the same. For $N=100$, the width of the Gaussian modes is $w_0=\SI{5}{\micro m}$, while the spacing is \SI{16}{\micro m} (thus the inputs span a distance of \SI{1.6}{mm}). For $N=150$, the width of the Gaussian modes is $w_0 = \SI{4}{\micro m}$, while the spacing is \SI{13}{\micro m} (thus the inputs span a distance of \SI{2}{mm}). We note that the matrix operations are performed up to a rescaling factor, which represents loss. The $N=100$ was performed with a transmission of 16\% (loss of 84\%), while the $N=150$ was performed with a transmission of 4\% (loss of 96\%). Adding this rescaling factor to the training eases the burden on the programmable wave propagation, as it no longer has to funnel all of the power into the discrete and incomplete basis of output modes. 

The results are shown in Fig.~\ref{fig:unitary}. We observe that the 2D-programmable waveguide is able to execute the unitary matrix operations with high fidelity for the $N=100$ case. For $N=150$, the fidelity is lower, but the matrix elements of the target unitary matrix and the matrix performed by the 2D-programmable waveguide still agree to within 15\%. 

\section{Length (and area) scaling for 2D-programmable waveguides}
\label{sec:scaling-argument}

In the main manuscript, we have argued that multimode wave physics provides a favorable length-\textit{scaling} for devices performing matrix-vector multiplications. Specifically, we have claimed that a device with a 2D-programmable index profile embedded in a multimode waveguide can perform a matrix-vector multiplication with an $N\times N$ unitary matrix over a propagation distance that scales as $\sqrt{N}/\Delta n_{\text{prog}}$ and a width that scales as $N$. Here, we present evidence for this claim: An analytical proof (under approximations that make the underlying calculations tractable) and simulations (that suggest that some of the approximations made for the analytical proof are superfluous).

We proceed as follows: In Subsection~A, we analytically construct a continuous refractive-index distribution that tailors multimode wave propagation to perform a desired $N\times N$ unitary. In Subsection~B, we prove that the magnitude of this index distribution scales as $\Delta n_\text{prog}\sim\sqrt{N}/L_z$, or equivalently, the length of the device as $L_z\sim\sqrt{N}/\Delta n_\text{prog}$. 
A key insight is that the mean magnitude of the elements of any $N\times N$ unitary matrix scales as $1/\sqrt{N}$. In Subsection~C, we numerically verify the analytically derived results and compare the performance with an inverse-designed index distribution.
In Subsection~D, we discuss the generality of our analytical finding, present simulations that show that some of our assumptions are expendable, and discuss what would be required for an experimental realization. Finally, Subsection~E summarizes our results.

\subsection{Analytical construction of a refractive-index profile for a given unitary matrix}

We consider a multimode waveguide with background refractive index $\Delta n_{\text{wg}}(x)$ and perturbation due to an additional programmable refractive-index profile $\Delta n_\text{prog}(x,z) \ll \Delta n_{\text{wg}}(x)$.
For this section, we envision a multimode waveguide with a small enough refractive-index contrast between the core and cladding such that we can still accurately describe the propagation in a scalar and two-dimensional model \cite{nikkhah2024inverse} with the beam propagation equation \eqref{eq:beam_propagation}: 
\begin{align}
    \pder{E}{z} = \frac{\im}{2k}\pder{^2E}{x^2} + \im k_0 [\Delta n_{\text{wg}}(x) + \Delta n_\text{prog}(x, z)] E.
    \label{eq:beamprop-wg}
\end{align}
However, all the following arguments are more generally valid for any unidirectional, scalar wave equations of the form: 
\begin{align}
    \pder{E}{z} = \im \hat{H}E + \im\Delta(x,z)E,
\end{align}
where $\hat{H}$ is a Hermitian operator with bounded eigenfunctions and $|\Delta(x,z)|\ll|\hat{H}|$ is a perturbation, for example different approximations to the unidirectional Helmholtz equation \cite{Feit1978}.

\begin{figure*}[!htbp]
    \centering
    \includegraphics[width=\textwidth]{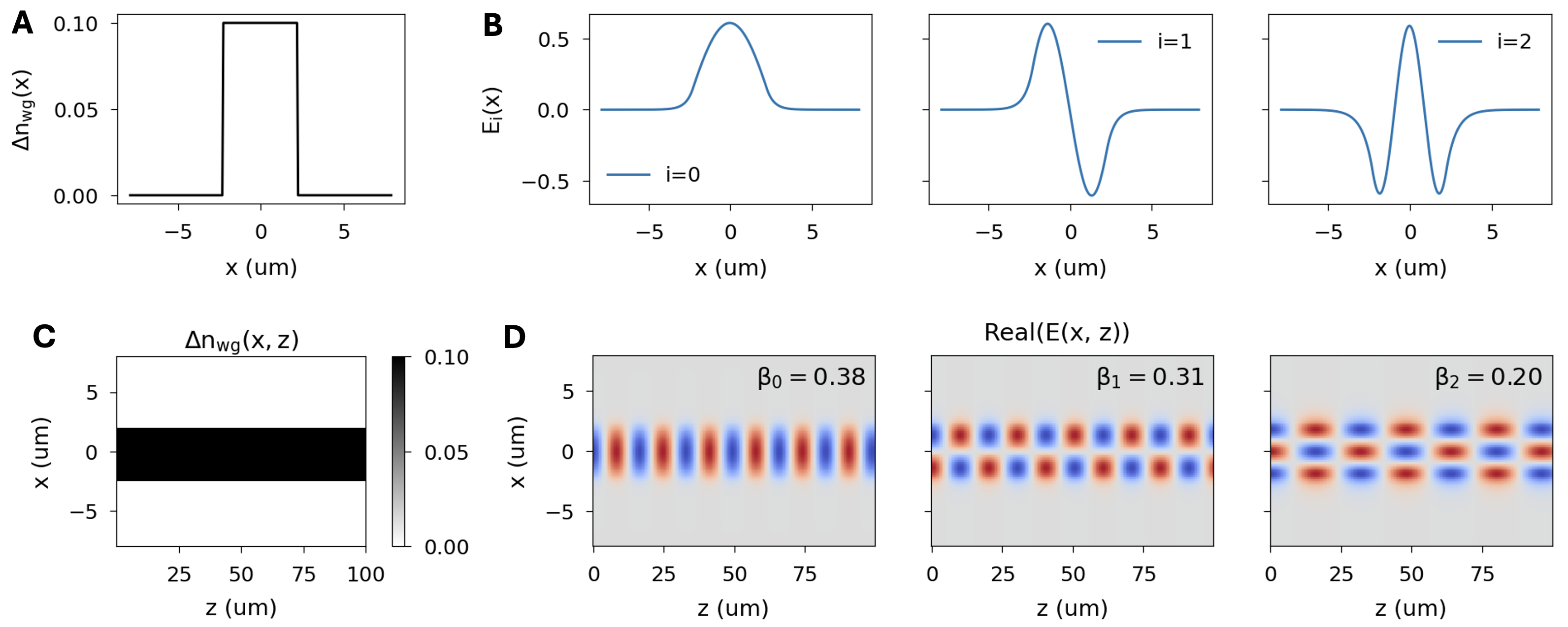}
    \captionsetup{justification=raggedright,singlelinecheck=false}
    \caption{
    \textbf{Modes of a step-index multimode waveguide.} A) The background refractive-index distribution of the multimode waveguide, $\Delta n_{\text{wg}}(x)$. B) The three modes of the waveguide, $E_i(x)$. C, D) Simulations of the different modes propagating through the waveguide (simulated by the beam propagation equation with initial conditions $E(x, z=0) = E_i(x)$ respectively).}
    \label{fig:scaling_modes}
\end{figure*}

We will apply coupled-mode theory to solve Eq. \eqref{eq:beamprop-wg}, therefore we express the electric field in terms of the bound modes of the waveguide:
\begin{align}
    E(x, z) = \sum_i a_i(z) E_i(x),
\end{align}
where $E_i(x)$ denotes the $i$-th mode of the waveguide $\Delta n_\text{wg}(x)$ and satisfies the eigenvalue equation:
\begin{align}
    \frac{1}{2k}\pder{^2E_{i}}{x^2} + k_0 \Delta n_{\text{wg}}(x) E_i = \beta_i E_i.
\end{align}
Here, $\beta_i$ is the propagation constant, and $a_i(z)$ the modal amplitude of the $i$-th mode. For a weakly perturbed waveguide with $\Delta n_\text{prog}\neq 0$, the modal amplitudes are propagation-dependent but the modes remain unchanged (coupled mode theory). Exemplary modes for a step-index waveguide are shown in Fig.~\ref{fig:scaling_modes}. We note that even though the different spatial modes of the waveguides are orthogonal, they do overlap in space. This overlap is what allows the modes to be coupled by the programmable refractive-index distribution, which we will formally introduce and define below. Finally, the different spatial modes also have different propagation constants, as seen in the varying periods of the electric field distributions in Fig.~\ref{fig:scaling_modes}D.

We intend to perform a unitary operation on the amplitudes of these modes by selectively coupling different modes to each other by a carefully chosen refractive-index distribution $\Delta n_\text{prog}(x, z) \ll \Delta n_\text{wg}$. The equations of motion for the mode amplitudes follow from \eqref{eq:beamprop-wg} and are given by:
\begin{align}
    \pder{a_i}{z} = \im \beta_i a_i + \im k_0 \sum_j C_{ij}(z) a_j,
    \label{eq:coupled-mode-propagation}
\end{align}
where, $C_{ij}(z)$ is the coupling matrix, which is given by:
\begin{align}
    C_{ij}(z) = \int dx E_i(x) \Delta n_\text{prog}(x, z) E_j(x).
    \label{eq:Cij}
\end{align}
For guided modes, all $E_i(x)$ can be chosen to be real and hence $C_{ij}$ is a Hermitian matrix. Our goal is to specify a $\Delta n_\text{prog}(x,z)$ that couples modes in precisely the way to realize the desired transformation on the modal amplitudes. This problem is easier tackled in the rotating-wave frame, as we will see in the following paragraph.

In a rotating-wave frame with modal amplitudes $\tilde a_i = a_i \exp\paren{-\im \beta_i z}$, where the fast oscillation associated with the propagation constant of the modes vanishes, the evolution of the modal amplitudes $\tilde a_i$ simplifies to:
\begin{align}
    \label{eq:rotating_frame}
    \pder{\tilde a_i}{z} = \im k_0 \sum_j C_{ij}(z) e^{-i\Delta\beta_{ij}z} \tilde a_j,
\end{align}
where $\Delta\beta_{ij} = \beta_i - \beta_j$. If we would (naively) assume that the $z$-dependence of $C_{ij}(z)$ and the complex exponential cancel out, such that $C_{ij}(z) e^{-i\Delta\beta_{ij}z} = A_{ij}$, integrating this equation over $z$ would lead to a simple solution, which is: 
\begin{align}
    \tilde a(z=L_z) = e^{\im k_0 A L_z} \tilde a(z=0).
    \label{eq:simple-matrix-exp-solution}
\end{align}
Therefore, if one wants to apply a unitary matrix $U$ to the modal amplitudes $\tilde a_i$, one needs to design a coupling matrix A that satisfies:
\begin{align}
    \label{eq:matrix_log}
    U = e^{\im k_0 A L_z} \;\;\; \text{or} \;\;\; A = \text{logm}(U)/\paren{\im k_0 L_z}.
\end{align}

\begin{figure*}[!t]
    \centering
    \includegraphics[width=\textwidth]{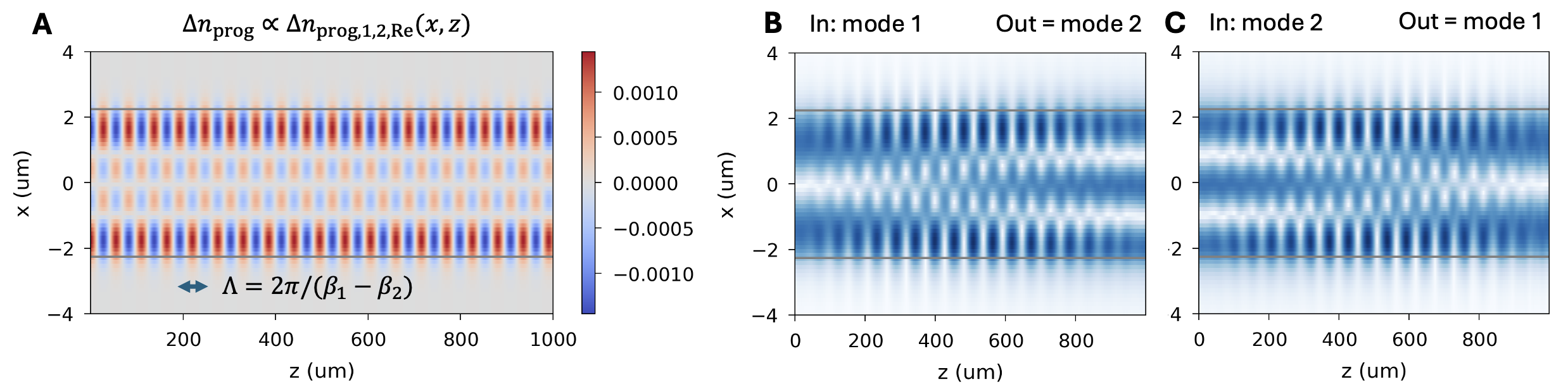}
    \caption{
    \textbf{Coupling modes of a multimode waveguide with gratings.} \textbf{A}: The programmable refractive index $\Delta n_\text{prog}$ chosen to be proportional to $\Delta n_{\text{prog}, 1, 2, \text{Re}}$, to realize a coupling between mode 1 and 2. \textbf{B, C} The intensity distribution of light in the waveguide $|E(x,z)|^2$ for an initial field distribution in mode 1 (panel B) and mode 2 (Panel C). We note that the coefficient of the strength in refractive index $A_{12} = \pi/(2k_0 L_z)$ was chosen to have complete conversion between mode 1 and mode 2. The gray lines denote the waveguide boundaries.
    }
    \label{fig:scaling_modes_coupling}
\end{figure*}

We will now show that for a desired coupling matrix $A_{ij}$, the refractive-index distribution realizing this coupling is given by:
\begin{align}
    \Delta n_{\text{prog}}(x, z)  = \Delta n_{\text{prog, DC}}(x) + \sum_{k \neq l} \mathcal{N}_{kl}E_k(x)E_l(x)\RE(A_{kl}e^{\im\Delta\beta_{kl}z})
    \label{eq:delta_n_prog}
\end{align}

where $\mathcal{N}_{kl} = 1/\int_{-\infty}^{\infty} |E_k(x')|^2 |E_l(x')|^2 dx'$ is a normalization constant. 
The summands are refractive-index distributions that create selective couplings between modes $k$ and $l$. 
As shown in Fig.~\ref{fig:scaling_modes_coupling}, 
each summand only applies to the pair of modes whose difference in propagation constants equals the summand's spatial $z$-frequency. The cross-section of this distribution is chosen to efficiently couple the modes by creating a large overlap.

$\Delta n_{\text{prog, DC}}(x, z)$ is the part of refractive-index distribution which is responsible for the overall phase of the modes, to enforce $C_{ii} = A_{ii}$. We termed it the DC term because it does not vary with $z$. In other words, it is responsible for implementing the diagonal terms of the generator matrix $A$. Since the DC terms do not have a selective $z$-frequency, we solve it numerically by finding the minimal solution to the linear equation:
\begin{align}
    \int |E_i(x)|^2 \Delta n_{\text{prog, DC}}(x) dx = A_{ii}.
\end{align}

Plugging $\Delta n_{\text{prog}}(x, z)$ from Eq. (\ref{eq:delta_n_prog}) into Eq. (\ref{eq:Cij}), we find the coupling matrix $C_{ij}(z)$ to be:
\begin{align}
    C_{ij}(z) = \delta_{ij}A_{ii} + \sum_{k \neq l} \mathcal{N}_{kl} \mathcal{M}_{ijkl} \text{Re}\paren{A_{kl} e^{\im \Delta\beta_{kl}z}}.
\end{align}
where $\mathcal{M}_{ijkl} = \int_{-\infty}^\infty E_i(x)E_j(x)E_k(x)E_l(x) dx$. 

With this coupling matrix, we now derive the equations of motion in the rotating frame:
\begin{align}
    \pder{\tilde a_i}{z} & = \im k_0\sum_j C_{ij}(z)  e^{-\im \Delta\beta_{ij} z}\tilde a_j \nonumber \\
    \
    & = \im k_0\delta_{ij}A_{ii} \tilde a_j  + \frac{\im k_0}{2} \sum_j \sum_{k \neq l} \mathcal{N}_{kl}\mathcal{M}_{ijkl}
    \paren{A_{kl} e^{\im (\Delta\beta_{kl} - \Delta\beta_{ij})z} + 
    A_{kl}^* e^{-\im (\Delta\beta_{kl} + \Delta\beta_{ij})z}} \tilde a_j \nonumber \\
    \
    & \approx \im k_0\delta_{ij}A_{ii} \tilde a_j  + \im k_0 \sum_j \sum_{k \neq l} \mathcal{N}_{ij}\mathcal{M}_{ijkl} A_{kl}\delta_{ik}\delta_{jl} \tilde a_j \nonumber \\
    \
    & = \im k_0 \sum_j A_{ij} \tilde a_j,
    \label{eq:RWA}
\end{align}
which indeed produce the simple solution from Eq. (\eqref{eq:simple-matrix-exp-solution}).
Going from the second to the third line, we used the Hermitian property $A_{kl} = A^{*}_{lk}$ of unitary generators and the rotating wave approximation, assuming that only terms where $\Delta\beta_{kl} = \Delta\beta_{ij}$ and $\Delta\beta_{kl} = -\Delta\beta_{ij}$ will contribute to the dynamics. Since, for a step-index waveguide, the mode spacing is nonlinear w.r.t. the mode index, this only holds when $k=i, l=j$ or $k=j, l=i$. Going from the third to the fourth line, we used $\mathcal{M}_{ijij} \mathcal{N}_{ij} = 1$ to arrive at the final line. 

\begin{figure*}[!t]
    \centering
    \includegraphics[width=\textwidth]{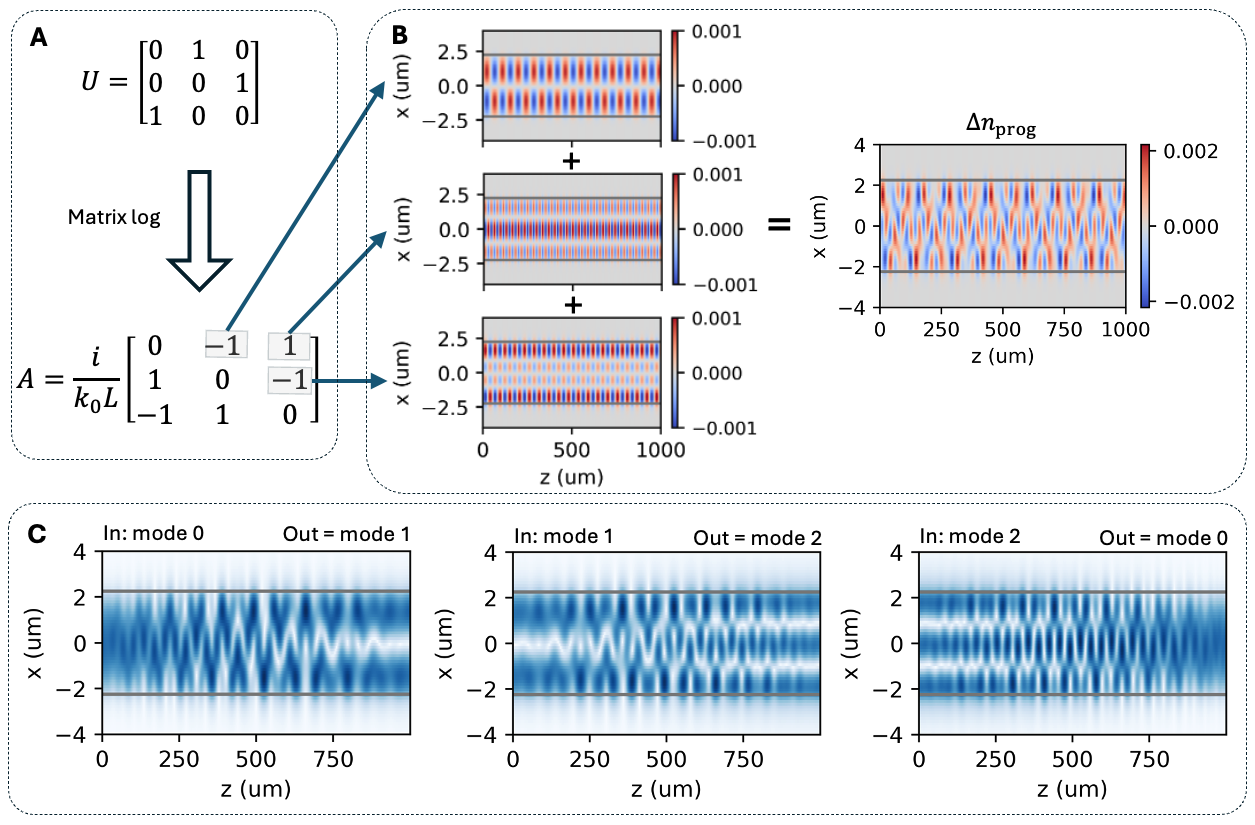}
    \caption{\textbf{
    A case study for implementing unitary operations in a programmable multimode waveguide.} \textbf{(A)}: Given a desired unitary matrix $U$, take its matrix logarithm to arrive at the generator matrix $A$, which describes the mode coupling required to implement it in the waveguide. \textbf{(B)}: Three left panels shows the different constituent refractive-index distributions ($\Delta n_{\text{prog}, k, l}$) required to couple pairs of modes. The full programmable refractive-index distribution $\Delta n_{\text{prog}}$ is the sum of these. \textbf{(C)}: Simulation of the evolution of light intensities in the waveguide for initial conditions corresponding to each waveguide mode, showing that the waveguide with the designed refractive-index distribution performs the desired permutation operation.
    }
    \label{fig:scaling_modes_permutation}
\end{figure*}

To summarize the entire procedure, it is helpful to walk through the example shown in Fig.~\ref{fig:scaling_modes_permutation}. We begin with a desired unitary transformation, here, a simple cyclic permutation matrix, but any unitary could be used. The next step is to calculate the generator matrix \(A\) using \eqref{eq:matrix_log}, which specifies the strength of each pair-wise coupling of modes within the waveguide. As shown in panel B, elements of this matrix specify the magnitude/coefficient for a series of refractive-index distributions, \(\Delta n_{\text{prog}, k, l}\), that each couple a specific pair of modes. All of these constituent refractive-index distributions are then summed to form the total programmable refractive-index distribution, \(\Delta n_{\text{prog}}\). Simulation results, as shown in panel C, confirm that this closed-form refractive-index profile successfully performs the desired permutation of the modes. We reiterate that this procedure applies to any unitary, not just permutations, as in this example.

\subsection{Derivation of the scaling law for the magnitude of $\Delta n_\text{prog}$}

In this section, we derive how the magnitude distribution of $\Delta n_\text{prog}$, as expressed in \eqref{eq:delta_n_prog}, scales with the number of modes $N$ when performing an arbitrarily chosen unitary transformation $U$. We consider the magnitude distribution as measured by its root-mean-square (RMS) value, as the mean value is zero and outliers can be clipped with a small fidelity penalty (or avoided by using an inverse-design procedure with a penalty term for large index values).

The key insight is that the RMS value of the elements of any unitary's generator, which determine the magnitude of the programmable refractive-index distribution according to Eq. \eqref{eq:delta_n_prog}, shrinks with increasing dimensionality.
Formally, the root mean square of the $N^2$ matrix elements is computed as:
\begin{align}
\text{rms}(A) &= \sqrt{\frac{\sum_{i,j} |A_{ij}|^2}{N^2}} = \sqrt{\frac{\operatorname{Tr}(AA^\dagger)}{N^2}}.
\end{align}
We recall that $A = -i\text{logm}(U)$ is Hermitian to simplify the above expression to:
\begin{align}
\text{rms}(A) &= \sqrt{\frac{\operatorname{Tr}(A^2)}{N^2}} = \sqrt{\frac{\sum_i (\lambda_i(A))^2}{N^2}} \leq\sqrt{\frac{\sum_i \pi^2}{N^2}} = \pi/ \sqrt{N},
\end{align}
where $\lambda_i(A)$ are the eigenvalues of $A$. We utilized the property $\operatorname{Tr}(A^2) = \sum_i (\lambda_i(A))^2$, and the diagonalizability of $A$ and $A^2$ implies that $\lambda_i(A^2) = \lambda_i(A)^2$. Since $U$ is unitary and therefore has eigenvalues on the unit circle, the eigenvalues of $A$, being derived from the matrix logarithm of $U$, can always be chosen to lie within the interval $[-\pi, \pi]$, and therefore $|\lambda_i(A)|\leq \pi$.

Equipped with this inequality, we now calculate the RMS value of $\Delta n_\text{prog}$ as given in Eq. (\eqref{eq:delta_n_prog}):
\begin{align}
    \text{rms}(\Delta n_{\text{prog}}) & \approx \sqrt{\frac{1}{L_x L_z}\int \text{d}x\text{d}z \, (\Delta n_\text{prog})^2} \nonumber \\
    & \approx \frac{1}{k_0 L_z}\sqrt{\frac{1}{L_x L_z}\int \text{d}x\text{d}z \, (\sum_{k,l} \mathcal{N}_{kl} E_k E_l \RE{(A_{kl}e^{\im \Delta\beta_{kl}z})})^2} \nonumber \\
    & \lesssim \frac{1}{k_0 L_z}\sqrt{\frac{1}{L_x L_z}\int \text{d}x\text{d}z \, \sum_{k,l} |A_{kl}|^2 + \sum_{k\neq m}\sum_{l\neq n}A_{kl}A^*_{mn}e^{\im (\Delta\beta_{kl} - \Delta\beta_{mn})z})} \nonumber \\
    & \approx \frac{1}{k_0 L_z}\sqrt{\sum_{k,l} |A_{kl}|^2} 
    \quad \leq \frac{\pi \sqrt{N}}{k_0 L_z} 
    = \frac{\sqrt{N}\lambda_0}{2 L_z} \label{eq:rms-sqrt-scaling}.
\end{align}
First, we recognized from the normalization condition $\int |E_i(x)|^2 \, dx = 1$ that the average value of the electric field magnitude is approximately $|E| \sim 1/\sqrt{L_x}$. Consequently, the expression $\mathcal{N}_{kl}E_k(x) E_l(x) \sim 1$ can be dropped going from line two to three. Simultaneously, we separated non-oscillating from oscillating terms in the quadrature of the sum. Integrated over $z$, the oscillating terms can be neglected, just as in the previous section. In the final line, we use the result $\text{rms}(A_{kl}) \leq \pi/\sqrt{N}$, which holds for a suitably chosen generator of \textit{any} unitary, as shown above.

To summarize, the magnitude of $\Delta n_\text{prog}$, as measured by the root-mean-square of its distribution, scales as $\sqrt{N}\lambda_0/2L_z$ when performing $N\times N$-dimensional unitary operations, in contrast to commonly used unitary decompositions for which $\Delta n_\text{prog}$ needs to increase linearly with $N$. 

\subsection{Numerical results and comparison with inverse-designed $\Delta n_\text{prog}$}

\begin{figure}[b!]
    \centering
    \includegraphics[width=\textwidth]{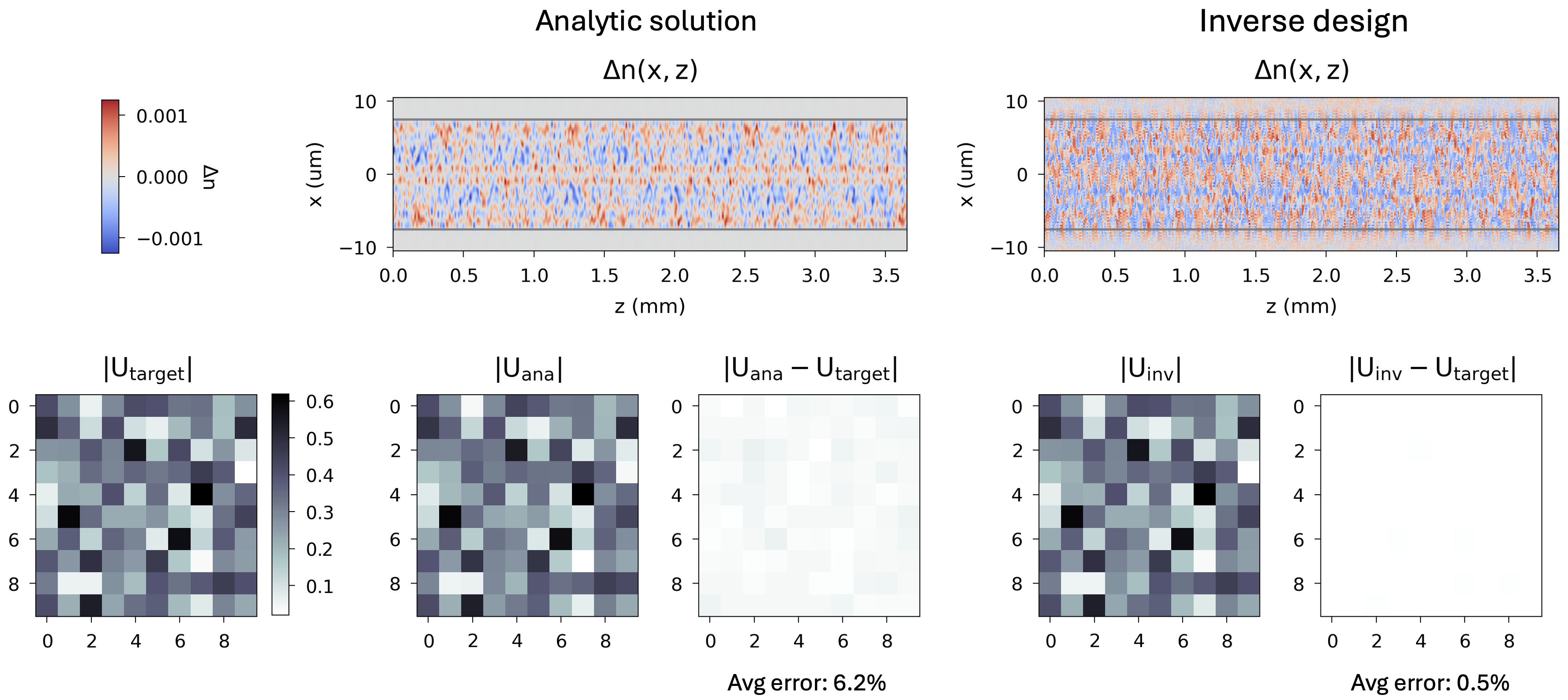}
    \caption{\textbf{Comparison of analytic and inverse design approaches for realizing a target unitary transformation.} A target unitary matrix $U_\text{target}$ of dimension $N=10$ is sampled from the Haar measure, and implemented by a 2D-programmable waveguide. Left: Results using the analytic solution derived from coupled mode theory \eqref{eq:delta_n_prog}. Right: Results using inverse design optimization. The average error of the matrix element is defined as $\|U_\text{target} - U_\text{sim}\|_1/\|U_\text{target}\|_1$, and thus is scaled to the magnitude of the target matrix elements. All plots in the lower row use the same color bar.
    }
    \label{fig:inverse_vs_analytic}
\end{figure}

In this subsection, we numerically verify the solution for the programmable refractive-index distribution $\Delta n_\text{prog}(x,z)$ given in Eq. (\ref{eq:delta_n_prog}).
We take a target random unitary matrix $U_\text{target}$ of dimension $N=10$, sampled from the Haar measure, and quantify the accuracy with which a multimode waveguide realizes the transformation. 
The results are presented in Fig.~\ref{fig:inverse_vs_analytic}.
On the left panel, the analytic solution corresponds to the refractive-index distribution derived in the previous section with coupled mode theory (Eq. \eqref{eq:delta_n_prog}). We see that the fidelity is generally high, with an average error of 6.2\% for the matrix elements. This error is due to the rotating wave approximation not being perfectly accurate, which does become a worse approximation as the number of modes increase and the mode spacing decreases. This can be alleviated by using a longer multimode waveguide with a smaller $\Delta n_\text{prog}$. 

Another approach to reduce the error while keeping the same length is to directly optimize the refractive-index distribution using an inverse design technique. This approach, also employed in the main manuscript, utilizes a gradient-based optimization method to minimize the difference between the desired and realized field distributions. Specifically, we send a batch of $N$ input field distributions (indexed by $i$) $E^\text{in}_i = E_i(x, z=0)$ through the waveguide, and then optimize the refractive-index distribution to minimize the cost function ${\mathcal{L} = \sum_{i=1}^{N_\text{modes}} \| U_\text{target} E_i^\text{in} - E_i^\text{out} \|^2}$. Here, $E_i^\text{out}$ is the output field for the $i$-th input, given by the beam propagation equation \eqref{eq:beamprop-wg}. We included a hard cut-off for $|\Delta n_\text{prog}|$ at $10^{-3}$ in the simulation to ensure that the refractive-index distribution remains within the desired range. We simulated the propagation through the multimode waveguide with an auto-differentiable implementation of the extended Fresnel approximation \cite{Feit1978} using a Runge-Kutta method.

The results of this optimization are shown in the right panel of Fig.~\ref{fig:inverse_vs_analytic}, where we observe a significantly lower average error of 0.5\%. The trained refractive-index distribution also looks qualitatively similar to the analytic solution, maintaining the overall structure. This demonstrates that, while the analytic solution provides a good initial estimate, direct optimization yields more accurate results while preserving the general characteristics of the solution.

We performed two further simulations to verify the scalability of our approach. First, as presented in Fig.~\ref{fig:inverse_vs_analytic}, we sampled Haar-random unitary matrices of different dimensions and created refractive-index distributions according to Eq. \eqref{eq:delta_n_prog}. We calculated the RMS for these distributions and confirmed that the RMS is well described by $\text{rms}(\Delta n_\text{prog}) = \sqrt{N}\lambda_0/2L_z$.
Second, as presented in Fig.~\ref{fig:sm_scaling_different_scenarios}, we sampled Haar-random unitaries and random perturbations, and \textit{inverse-designed} refractive-index distributions to accurately implement these unitaries. The results in Fig.~\ref{fig:sm_scaling_different_scenarios} show the fidelities of unitaries implemented in waveguides of length \SI{3}{mm} with different multimode waveguide profiles, widths that scale linearly with the unitary dimension to accommodate enough modes (${L_x\approx N\lambda_0 /\ 2\text{NA}}$), and varying limits on the maximal $\Delta n_\text{prog}$. The simulations were performed with an autodifferentiable implementation of the extended Fresnel approximation \cite{Feit1978} using a Strang split-step Fourier method, for a favorable speed/accuracy tradeoff. The fidelities were calculated using $F=|\text{Tr}(U_\text{target}^{\dagger}U_\text{realized})|/N$. Very similar simulations are presented in Fig. 5 of the main manuscript, except that we fixed the maximal $\Delta n_\text{prog}$ to $5\cdot 10^{-3}$ and varied the propagation distance $L_z$. For all these simulations, the spatial resolution of the refractive-index was limited to around \SI{1}{\micro\meter} by convolving the refractive-index distribution with a Gaussian kernel of the form $e^{-(k_x^2 + k_z^2) / k_c^2}$ with $k_c=2\pi n_\text{clad}/\lambda_0\approx2\pi /\ \SI{1}{\micro\meter}$.

\begin{figure}[t!]
    \centering
    \includegraphics[width=0.8\textwidth]{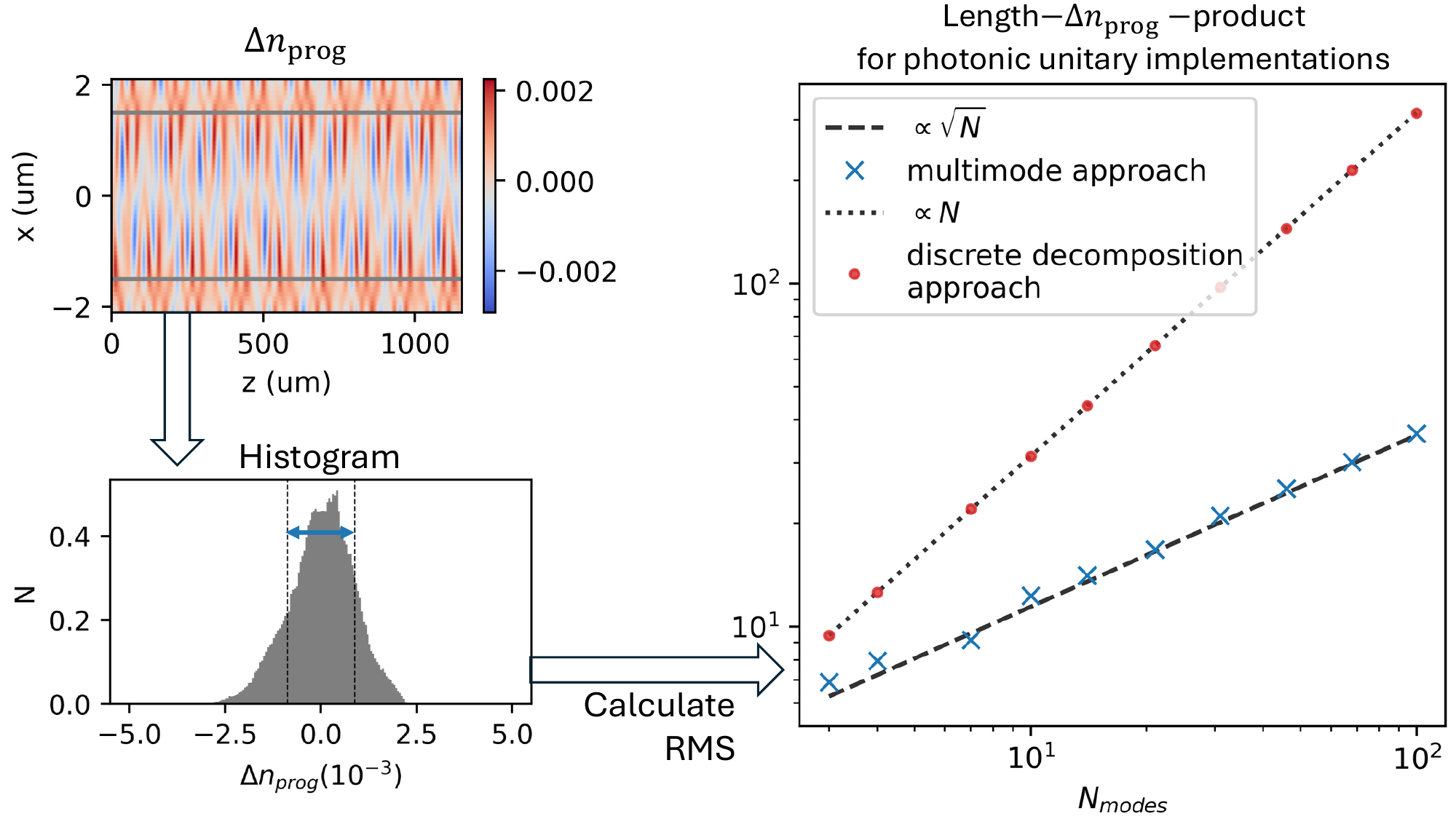}
    \caption{\textbf{Numerical verification of square-root scaling of $\text{rms}(\Delta n_\text{prog})$ of Eq. \eqref{eq:delta_n_prog}.} Target unitary matrices of variable dimension are sampled from the Haar measure and the according $\Delta n_\text{prog}$ distributions are calculated according to Eq. (\ref{eq:delta_n_prog}). The widths of the distributions, as measured by the root-mean-square, are plotted against the number of modes, i.e. the dimension of the unitaries. The plot of the length-$\Delta n_\text{prog}$-product on the right confirms the $\sqrt{N}$ scaling from Eq. (\ref{eq:rms-sqrt-scaling}).
    }
\end{figure}

\subsection{Generality, validity, and experimental implementation of the $\sqrt{N}$-scaling}

\begin{figure}[htb!]
    \centering
    \includegraphics[width=\linewidth]{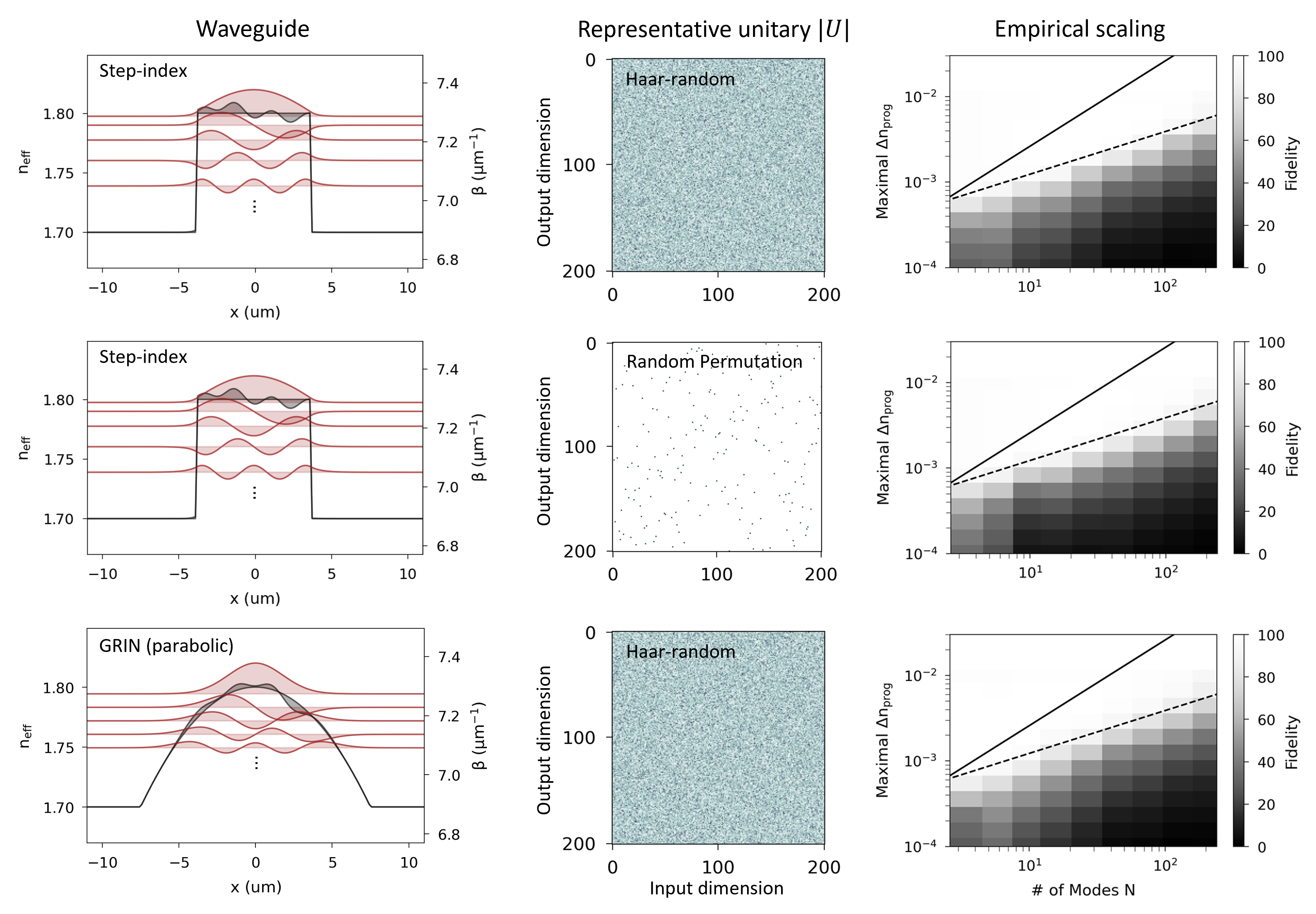}
    \caption{\textbf{Empirical scaling of $\Delta n_\text{prog}$ to realize high-fidelity with inverse-design.} We simulate $L_z=\SI{3}{mm}$ of multimode wave propagation in different waveguide geometries and inverse-design refractive-index distributions $\Delta n_\text{prog}(x,z)$ to obtain high-fidelity realizations of various unitary matrices. In each case, we observe that the maximal $\Delta n_\text{prog}$ to realize the desired unitary with high-fidelity is approximately given by $\sqrt{N} \pi / k_0 L_z$ (dashed line~${\sim\sqrt{N}}$, solid line $\sim N$). \textbf{Top row:} A step-index waveguide with perturbations to realize unitaries uniformly sampled from the Haar-measure. \textbf{Middle row:} To test whether the scaling holds for highly structured unitaries, we simulated randomly sampled permutation unitaries in a step-index waveguide. \textbf{Bottom row:} To test whether degenerate mode spacings, which prevent pair-wise couplings, degrade the scaling we simulated a parabolic graded-index (GRIN) waveguide with perturbations to realize unitaries uniformly sampled from the Haar-measure.
    \label{fig:sm_scaling_different_scenarios}}
\end{figure}

We conclude this section by a discussion of the approximations under which the results were derived, and the generality of the scaling relation.

First, we reiterate that the method from Section~8A can be applied to any PDE of the form
$$\pder{E}{z} = i(\hat{H} + \Delta)E,$$
with a Hermitian propagation operator $\hat{H}$ and a real-valued perturbation $\Delta$.
In principle these could describe systems completely different from optics, e.g. atomic systems. Specifically in optics, this includes more general wave equations such as the \emph{extended Fresnel approximation} from Ref.~\cite{Feit1978}, extending the validity to waves that do not satisfy the paraxial approximation.

Next we consider limitations on the unitary matrices that may be realized with this scaling. The derivation showing that the for the closed-form solution $\text{rms}(\Delta n_\text{prog})\propto \sqrt{N}$ is completely general and strictly holds for \textit{any} unitary matrix $U$. The scaling of $\text{max}(\Delta n_\text{prog})$, however, can in the worst case be linear in $N$ due to outliers in the $\Delta n_\text{prog}$ distribution. Our simulations indicate that inverse-designing refractive-index distributions with a firm limit on $\text{max}(\Delta n_\text{prog})$ also produces unitaries with high fidelity (implemented with a cutoff-step in the optimizer). We showed this in Fig.~\ref{fig:sm_scaling_different_scenarios} for unitaries of dimension up to $N=200$, for Haar-random unitaries and for random permutation unitaries, and in a step-index waveguide and in a parabolic graded-index (GRIN) waveguide. In all the scenarios we explored in numerical simulations, allowing the inverse-designed algorithm to create refractive-index contrasts that scale as $\sqrt{N}\lambda_0 / L_z$ (or larger) resulted in high-fidelity unitaries. We did not inverse-design unitaries of dimension exceeding $N=200$ because of the memory and processing limitations of our simulation computer, but we expect the trends in Fig.~\ref{fig:sm_scaling_different_scenarios} would continue beyond $N=200$.

We turn to a discussion of the approximations made:
we assumed a two-dimensional Helmholtz equation, which is usually derived from an effective-index description stemming from a separable optical potential $n(x,y,z)\approx n(x,z) + n(y)$. This is accurate for shallow index contrasts between the core and cladding, or modes far away from cutoff. However, our method can readily be generalized to a three-dimensional unidirectional Helmholtz equation and produces the same scaling.
We also assumed scalar fields. A generalization to vectorial fields might require anisotropic perturbations (i.e. independent perturbations to the ordinary and extraordinary part of the refractive index). This could potentially double the dimension of the realizable unitary with otherwise same propagation distance and refractive-index changes, but would complicate an experimental implementation.
Lastly, we used coupled-mode theory and the rotating wave approximation to only consider pair-wise couplings in the analytic solution of Eq.~\eqref{eq:RWA}. Our simulations in a GRIN waveguide (bottom row of Fig.~\ref{fig:sm_scaling_different_scenarios}) clearly violate the rotating-wave approximation (RWA), as mode spacings are degenerate, but still exhibit a $\Delta n_\text{prog}\sim\sqrt{N}/L_z$ scaling. This suggests that the rotating-wave approximation was only necessary to produce a closed-form solution to Eq.~\eqref{eq:RWA} and that the same $\sqrt{N}$ scaling holds even when the RWA is violated. The use of coupled-mode theory and the implicit assumption that $|\Delta|\ll |\hat{H}|$ (or $\Delta n_\text{prog} \ll \Delta n_\text{wg})$ appears more pertinent to the scaling relation. We suspect that, as this approximation breaks down for larger perturbations, so does the $\sqrt{N}$ scaling, and a more general bound including the magnitude of both $\Delta$ and $\hat{H}$ will arise\cite{Arenz2017}.

We were initially surprised by the $\sqrt{N}$ length-scaling because fundamental considerations suggest that in linear optical devices the length required to perform arbitrary operations on $N$ modes is also proportional to $N$ \cite{Miller2023, Li2022WadeHsu}. Specifically, Miller poses that $L_z$ needs to be $\gtrsim N\lambda_0/2\alpha n_\text{max}$ for a universal mode converter, where $n_\text{max}$ is the largest refractive index in the optical device and $\alpha=1-\cos{\theta}$ for an optical device in which light can propagate at angles between $0$ and $\theta$ with the optical axis.
On the other hand, our theory implies that $L_z\approx\sqrt{N}\lambda_0/2\Delta n_\text{prog}$ is sufficient to perform universal optical operations on $N$ modes. Comparison of the two bounds yields that the $\sqrt{N}$ scaling would violate the bound of Miller if $N\gtrsim(\Delta n_\text{wg}/\Delta n_\text{prog})^2$, where we have used that $\alpha\approx\text{NA}^2/2n_\text{core}^2$ and $\text{NA}\approx\sqrt{2n_\text{core}\Delta n_\text{wg}}$ for a weakly guiding waveguide with $\Delta n_\text{wg}\ll n_{core}$. This limitation of the $\sqrt{N}$-scaling may also be understood in terms of limited grating bandwidth: Our proposed refractive index distribution (Eq.~\eqref{eq:delta_n_prog}) is a superposition of many gratings of length $L$, each with a bandwidth of $\Delta \beta_g=2\pi/L$. At best, the propagation constants $\beta_i$ of the $N$ waveguide modes are spaced such that $\min(\Delta\beta_{ij}) \geq \Delta n_\text{wg} / N$. In order for each grating to contribute an independently controllable degree of freedom to the mode conversion, the grating bandwidth needs to be smaller than the mode spacing. Consistent with the bound by Miller, $\Delta\beta_g<\Delta n_\text{wg}/N$ imposes that $N\lesssim(\Delta n_\text{wg}/\Delta n_\text{prog})^2$.

We conclude with a discussion on what would be necessary for an experimental implementation of devices as we simulated them above.
The main difference between the 2D-programmable waveguide device we used for the experiments presented in the main text, and the waveguides discussed in this section is that the latter assumes an additional (relatively) strong multimode waveguide---in other words, there are boundaries on either side of the slab waveguide in the $x$-direction imposed by a refractive-index contrast $\Delta n_\text{wg}$. There is a sense in which our experimental prototype device already has this character because the slab waveguide is bounded by air on both sides in the $x$ direction, but in practice we used a sufficiently small region in the middle (in the x direction) of the waveguide that the air-waveguide boundaries were effectively infinitely far away. However, one could etch a wide (in the x direction) waveguide that is sufficiently wide to support $N$ modes in the $x$-direction, but sufficiently narrow to give the device well-defined boundaries so that the coupled-mode theoretical analysis assuming confined modes directly applies to it. Our method of programming the refractive index in the waveguide (involving adding a layer of photoconductor and biasing the device etc.) is compatible with the presence of an etched multimode waveguide.
The biggest obstacle to an experimental realization stems from the necessity to programmably create potentially high-frequency gratings.
For gratings in the propagation ($z$) direction, the highest frequency is required for gratings that couple modes with the largest momentum mismatch to each other, i.e., the fundamental and highest-order modes. The momentum mismatch can be bounded by $\Delta\beta_\text{max}<k_0 (n_\text{core}-n_\text{clad})$ and the associated upper bound on the grating frequency is $\Lambda_z>\lambda_0/(n_\text{core}-n_\text{clad})$. For the waveguides we simulated above, this evaluates to $\Lambda_z>\SI{15}{\micro m}$, which is readily attainable on our device.
It is also necessary to write gratings in the lateral ($x$) dimension of the waveguide. A waveguide of a given numerical aperture supports beams with a transverse momentum $k_x = 2\pi\text{NA}/\lambda_0 $ . The worst-case scenario requires coupling of two counter-propagating beams with this transverse momentum, requiring a grating of period $\Lambda_x=2\pi/2k_x=\lambda_0/2\text{NA}$. For the waveguides we simulated above, with a numerical aperture of around $0.6$ and $\lambda_0=\SI{1.55}{\micro m}$, this requires grating periods as small as $\Lambda_x=\SI{1.3}{\mu m}$, which is currently unattainable in our device (see Supplementary Section 1D), but might be attainable with improvements to our device, or other refractive-index modulation methods such as acousto-optic modulation, plasma-dispersion with green-light excitation, phase-change materials, and of course, lithography---although this last option would then impose strong restrictions on the programmability of such a device.

\subsection{Conclusion}

To summarize the results of this section: We have described a method that analytically derives a refractive-index distribution that implements a desired $N\times N$-dimensional unitary operation using multimode wave propagation. We have shown numerical evidence that the magnitude of said index distribution, or equivalently the length of a device implementing it, scales as $L_z\Delta n_\text{prog}\sim\sqrt{N}$. This is in contrast to widely used unitary decompositions for which the refractive-index magnitude or the device length needs to increase linearly with $N$. We do not have conclusive proof that this scaling is possible for all unitaries in all waveguide geometries, but we have shown analytical and numerical evidence that it holds for multiple relevant classes of waveguides and unitaries. These scaling relations crucially relax the need for long propagation distances that hamper many of the devices presented in Table~\ref{tab:comparison_ONNs}.